\documentstyle[12pt,epsfig]{article}
        \newdimen\eqskip
        \newdimen\txtskip
        \baselineskip=\txtskip
\begin{document}

  \newcommand{\ccaption}[2]{
    \begin{center}
    \parbox{0.85\textwidth}{
      \caption[#1]{\small{{#2}}}
      }
    \end{center}
    }
\newcommand{\BS}{\bigskip}
% MATH SYMBOLS
\def    \be             {\begin{equation}}
\def    \ee             {\end{equation}}
\def    \ba             {\begin{eqnarray}}
\def    \ea             {\end{eqnarray}}
\def    \nn             {\nonumber}
\def    \=              {\;=\;}
\def    \frac           #1#2{{#1 \over #2}}
\def    \ret            {\\[\eqskip]}
\def    \ie             {{\em i.e.\/} }
\def    \eg             {{\em e.g.\/} }
\def    \lsim           {\raisebox{-3pt}{$\>\stackrel{<}{\scriptstyle\sim}\>$}}
\def    \bentarrow      {\:\raisebox{1.1ex}{\rlap{$\vert$}}\!\rightarrow}
\def    \rd             {{\mathrm d}}    
\def    \Im             {{\mathrm{Im}}}  
\def    \bra#1          {\mbox{$\langle #1 |$}}
\def    \ket#1          {\mbox{$| #1 \rangle$}}

% UNITS                 
\def    \kev            {\mbox{$\mathrm{keV}$}}
\def    \mev            {\mbox{$\mathrm{MeV}$}}
\def    \gev            {\mbox{$\mathrm{GeV}$}}

% KINEMATICAL VARIABLES 

\def    \mq             {\mbox{$m_Q$}}  
\def    \mqq            {\mbox{$m_{Q\bar Q}$}}
\def    \mqqsq          {\mbox{$m^2_{Q\bar Q}$}}
\def    \pt             {\mbox{$p_T$}}
\def    \ptsq           {\mbox{$p^2_T$}}

% QUARKONIUM STATE MACROS
\def    \o              {\ifmmode {\cal{O}} \else ${\cal{O}}$ \fi}
\def    \q              {\ifmmode {\cal{Q}} \else ${\cal{Q}}$ \fi}
\def    \oo              {\ifmmode \overline{\cal{O}} \else 
                            $\overline{\cal{O}}$ \fi}
\def    \oneSzero       {\ifmmode {^1S_0} \else $^1S_0$ \fi}
\def    \threeSone      {\ifmmode {^3S_1} \else $^3S_1$ \fi}
\def    \onePone        {\ifmmode {^1P_1} \else $^1P_1$ \fi}
\def    \threePJ        {\ifmmode {^3P_J} \else $^3P_J$ \fi}
\def    \threePzero     {\ifmmode {^3P_0} \else $^3P_0$ \fi}
\def    \threePone      {\ifmmode {^3P_1} \else $^3P_1$ \fi}
\def    \threePtwo      {\ifmmode {^3P_2} \else $^3P_2$ \fi}

% QUARKONIUM PARAMETERS                                     
\def    \heightp         {\mbox{$H_8^{\prime}$}}
\def    \vevpsi         {\mbox{$\langle {\cal O}_8^{\psi}(^3S_1) \rangle$}}
\def    \vevpsp         {\mbox{$\langle {\cal O}_8^{\psp}(^3S_1) \rangle$}}
% QCD PARAMETERS                                      
\newcommand     \MSB            {\ifmmode {\overline{\rm MS}} \else 
                                 $\overline{\rm MS}$  \fi}
\def    \muf            {\mbox{$\mu_{\rm F}$}}
\def    \mufsq          {\mbox{$\mu^2_{\rm F}$}}
\def    \mur            {{\mbox{$\mu$}}}
\def    \mursq          {\mbox{$\mu^2$}}
\def    \mul            {{\mu_\Lambda}}
\def    \mulsq          {\mbox{$\mu^2_\Lambda$}}
     
\def    \as             {\mbox{$\alpha_s$}}
\def    \asb            {\mbox{$\alpha_s^{(b)}$}}
\def    \assq           {\mbox{$\alpha_s^2$}}
\def    \ascube         {\mbox{$\alpha_s^3$}}
\def    \asfour         {\mbox{$\alpha_s^4$}}
\def    \asfive         {\mbox{$\alpha_s^5$}}

\def    \eps            {\ifmmode \epsilon \else $\epsilon$ \fi}
\def    \epsbar         {\ifmmode \bar\epsilon \else $\bar\epsilon$ \fi}
\def    \epsir          {\ifmmode \epsilon_{\rm IR} \else $\epsilon_{\rm IR}$ \fi}
\def    \epsuv          {\ifmmode \epsilon_{\rm UV} \else $\epsilon_{\rm UV}$ \fi}

% VARIOUS NOTATIONS

\def    \Atot           {{\rm A_{tot}}}
\def    \aeight         {A^{[8]}}
\def    \aeightb        {A^{[8]}_{\rm Born}}
\def    \aeights        {A^{[8]}_{\rm soft}}
\def    \ovaeight         {\overline{A^{[8]}}}
\def    \ovaeightb        {\overline{A^{[8]}}_{\rm Born}}
\def    \ovaeights        {\overline{A^{[8]}}_{\rm soft}}
\def    \aone           {A^{[1]}}
\def    \asins          {A^{[1]}_{\rm soft}}
\def    \aoneb          {A^{[1]}_{\rm Born}}
\def    \aones          {A^{[1]}_{\rm soft}}
\def    \aoneo          {A^{[8]}_{\rm soft}}

\def    \C              {\frac{\Phi_{(2)}}{2 M} \, \frac{N}{K}}
\def    \Cggg           {\frac{1}{2M} \frac{\Phi_{(2)}}{3!}\frac{N}{K}}
\def    \Cggg2          {\frac{1}{2M} \frac{\Phi_{(2)}}{2!}\frac{N}{K}}
\def    \m              {\mbox{${\cal{M}}$}}
\def    \mbar           {\mbox{${\overline{\cal{M}}}$}}
\def    \mborn          {\mbox{${\cal{M}}_{\rm Born}$}}
\def    \gborn          {\mbox{$\Gamma_{\rm Born}$}}
\def    \gbh            {\mbox{$\Gamma_{\rm Born}^{H}$}}
\def    \sborn          {\mbox{$\sigma_{\rm Born}$}}
\def    \sborno         {\mbox{$\sigma^0_{\rm Born}$}}
\def    \sborno         {\mbox{$\sigma_0$}}
\def    \pgg#1          {P_{gg}(#1)}
\def    \cpgq#1         {{\cal{P}}_{gq}(#1)}
\def    \cpgg#1         {{\cal{P}}_{gg}(#1)}
\def    \cpqq#1         {{\cal{P}}_{qq}(#1)}
\def    \dk             {\mbox{${\cal D}_k$}}
\def    \fk             {\mbox{$f_k$}}
\def    \sp#1#2         {#1#2}       
\def    \eik#1          { \frac{#1 \epsilon_c}{#1 k} }
\def    \hone   {\mbox{$H_1$}}
\def    \height {\mbox{$H_8$}}
%%%%%%%%%%%%%%%%%%%%%%%%%%%%%%%%%%%%%%%%%%%%%%%%%%%%%%%%%%%%%%%%%%%

\def\jpsi{\mbox{$J\!/\!\psi$}}
\def\chic{\mbox{$\chi_c$}}
\def\chij{\mbox{$\chi_J$}}
\def\chicj{\mbox{$\chi_{cJ}$}}
\def\psp {\mbox{$\psi'$}}
\def\ups {\mbox{$\Upsilon$}}
\def\mups {\mbox{$M_\Upsilon$}}

\def \oacube {\mbox{$ O(\alpha_s^3)$}}
\def \oatwo {\mbox{$ O(\alpha_s^2)$}}
\def \oas   {\mbox{$ O(\alpha_s)$}}

\def \chiz {\mbox{$\chi_{0}$}}
\def \chio {\mbox{$\chi_{1}$}}
\def \chit {\mbox{$\chi_{2}$}}

\def \chiqz {\mbox{$\chi_{Q0}$}}
\def \chiqo {\mbox{$\chi_{Q1}$}}
\def \chiqt {\mbox{$\chi_{Q2}$}}

\def \chicz {\mbox{$\chi_{c0}$}}
\def \chico {\mbox{$\chi_{c1}$}}
\def \chict {\mbox{$\chi_{c2}$}}

\def \chibz {\mbox{$\chi_{b0}$}}
\def \chibo {\mbox{$\chi_{b1}$}}
\def \chibt {\mbox{$\chi_{b2}$}}

\def \QQ {Q \overline Q}
\def \qq {\mbox{$q \overline q$}}
\def \cc {\mbox{$c \overline c$}}

\def\lqcd{\mbox{$\Lambda_{QCD}$}}

\def \chizgg {\mbox{$\Gamma(\chiz \to \gamma\gamma)$}}
\def \chitgg {\mbox{$\Gamma(\chit \to \gamma\gamma)$}}
\def \chijgg {\mbox{$\Gamma(\chi_J \to \gamma\gamma)$}}
\def \chizlh {\mbox{$\Gamma(\chiz \to LH)$}}
\def \chiolh {\mbox{$\Gamma(\chio \to LH)$}}
\def \chitolh {\mbox{$\Gamma(\chij \to LH)$}}
\def \chijlh {\mbox{$\Gamma(\chi_J \to LH)$}}
\def \chijqqg{\mbox{$\Gamma(\chi_J \to q \overline q g)$}}
\def \lh{{\mathrm LH}}

\def \ebind {\mbox{$E_{\mathrm bind}$}}
\def \rprime {\mbox{${\cal R}^{\prime}_{1P}(0)$}}
\def \rprimes {\mbox{$\vert {\cal R}^{\prime}_{1P}(0) \vert^2$}}
\def \rr {\mbox{${\cal R}_S(0)$}}
\def \rros {\mbox{$\vert{\cal R}_{1S}(0)\vert^2$}}
\def \rrts {\mbox{$\vert{\cal R}_{2S}(0)\vert^2$}}
\def \rrns {\mbox{$\vert{\cal R}_{nS}(0)\vert^2$}}
\def \ei {\mbox{$ \epsilon_{ \!\!\mbox{ \tiny{IR} } } $}}
\def \eu {\mbox{$ \epsilon_{ \!\!\mbox{ \tiny{UV} } } $}}
\def \s0 {\mbox{$\sigma_{0} $}}
\def \se {\mbox{$\sigma_{0}(\epsilon) $}}
\def \ep {\mbox{$\epsilon $}}

\def    \eps            {\ifmmode \epsilon \else $\epsilon$ \fi}
\def \cf {{\ifmmode C_F{}\else $C_F$ \fi}}
\def \ca {{\ifmmode C_A {}\else $C_A$ \fi}}
\def \caf {\mbox{$ C_F-\frac{1}{2}C_A $}}  
\def \tf {{\ifmmode T_F {}\else $T_F$ \fi}}
\def \nf {{\ifmmode n_f {}\else $n_f$ \fi}}
\def \nc {{\ifmmode N_c {}\else $N_c$ \fi}}
\def \da {{\ifmmode D_A {}\else $D_A$ \fi}}
\def \Bf {{\ifmmode B_F {}\else $B_F$ \fi}}
\def \df {{\ifmmode D_F {}\else $D_F$ \fi}}
                                           
\def \fe{\mbox{$f(\ep)$}}                  
\def \de{\mbox{$F(\ep)$}}
\def \feps#1 {f_{\epsilon}(#1)}

\def\der{\mbox{$\stackrel{\leftrightarrow}{\bf D}$}}
\def\nder{\mbox{$\stackrel{\leftrightarrow}{D}$}}
\def\tr{{\mathrm Tr}}

\def\opchizh {\mbox{$\langle 0\vert {\cal O}_8^{\psi}(^3P_0)\vert 0 \rangle$}}
\def\opchioh {\mbox{$\langle 0\vert {\cal O}_8^{\psi}(^3P_1)\vert 0 \rangle$}}
\def\opchith {\mbox{$\langle 0\vert {\cal O}_8^{\psi}(^3P_2)\vert 0 \rangle$}}

\def\opchijh {\mbox{$\langle 0\vert {\cal O}_8^{\psi}(^3P_J)\vert 0 \rangle$}}
\def\opetah  {\mbox{$\langle 0\vert {\cal O}_8^{\psi}(^1S_0)\vert 0 \rangle$}}
\def\oppsih  {\mbox{$\langle 0\vert {\cal O}_8^{\psi}(^3S_1)\vert 0 \rangle$}}
\def\oppsis  {\mbox{$\langle 0\vert {\cal O}_1^{\psi}(^3S_1)\vert 0 \rangle$}}
\def\pppsisq  {\mbox{$\langle 0\vert {\cal P}_1^{\psi_Q}(^3S_1)\vert 0 \rangle$}}

\def\opchijhq {\mbox{$\langle 0\vert {\cal O}_8^{\psi_Q}(^3P_J)\vert 0 \rangle$}}
\def\opetahq  {\mbox{$\langle 0\vert {\cal O}_8^{\psi_Q}(^1S_0)\vert 0 \rangle$}}
\def\oppsihq  {\mbox{$\langle 0\vert {\cal O}_8^{\psi_Q}(^3S_1)\vert 0 \rangle$}}
\def\oppsisq  {\mbox{$\langle 0\vert {\cal O}_1^{\psi_Q}(^3S_1)\vert 0 \rangle$}}
\def\ppchijh  {\mbox{$\langle 0\vert {\cal O}_8^{\tiny\chij}(^3S_1)\vert 0 \rangle$}}
\def\opchizhq {\mbox{$\langle 0\vert {\cal O}_8^{\psi_Q}(^3P_0)\vert 0 \rangle$}}
\def\opchiohq {\mbox{$\langle 0\vert {\cal O}_8^{\psi_Q}(^3P_1)\vert 0 \rangle$}}
\def\opchithq {\mbox{$\langle 0\vert {\cal O}_8^{\psi_Q}(^3P_2)\vert 0 \rangle$}}

\def\ppchizh  {\mbox{$\langle 0\vert {\cal O}_8^{\tiny\chiz}(^3S_1)\vert 0 \rangle$}}
\def\ppchioh  {\mbox{$\langle 0\vert {\cal O}_8^{\tiny\chio}(^3S_1)\vert 0 \rangle$}}
\def\ppchith  {\mbox{$\langle 0\vert {\cal O}_8^{\tiny\chit}(^3S_1)\vert 0 \rangle$}}

\def\ppchiczh  {\mbox{$\langle 0\vert {\cal O}_8^{\tiny\chicz}(^3S_1)\vert 0 \rangle$}}
\def\ppchicoh  {\mbox{$\langle 0\vert {\cal O}_8^{\tiny\chico}(^3S_1)\vert 0 \rangle$}}
\def\ppchicth  {\mbox{$\langle 0\vert {\cal O}_8^{\tiny\chict}(^3S_1)\vert 0 \rangle$}}

\def\oppsph  {\mbox{$\langle 0\vert {\cal O}_8^{\psi^{\prime}}(^3S_1)\vert 0 \rangle$}}
\def\oppsps  {\mbox{$\langle 0\vert {\cal O}_1^{\psi^{\prime}}(^3S_1)\vert 0 \rangle$}}

\def\spectrc {\mbox{$^{2S+1}L^{[c]}_J$}}
\def\spectr {\mbox{$^{2S+1}L_J$}}
\def\spectrs {\mbox{$^{2S+1}L_J^{[1]}$}}
\def\spectrh {\mbox{$^{2S+1}L_J^{[8]}$}}
\def\szh    {\mbox{$\sigma_0^{H}$}}
\def\sbh    {\mbox{$\sigma_{\rm Born}^{H}$}}
\def\szpsi {\mbox{$\sigma_0^{\psi}$}}
\def\szpsiq {\mbox{$\sigma_0^{\psi_Q}$}}
\def\spectro {{^{2S+1}L^{[8]}_J}}

\def\etah {\mbox{$^1S_0^{[8]}$}}
\def\etas {\mbox{$^1S_0^{[1]}$}}
\def\psih {\mbox{$^3S_1^{[8]}$}}
\def\psis {\mbox{$^3S_1^{[1]}$}}
\def\chizh {\mbox{$^3P_0^{[8]}$}}
\def\chioh {\mbox{$^3P_1^{[8]}$}}
\def\chith {\mbox{$^3P_2^{[8]}$}}
\def\chijh {\mbox{$^3P_J^{[8]}$}}
%%%%%%%%%%%%%
\def\chizs {\mbox{$^3P_0^{[1]}$}}
\def\chios {\mbox{$^3P_1^{[1]}$}}
\def\chits {\mbox{$^3P_2^{[1]}$}}
\def\chijs {\mbox{$^3P_J^{[1]}$}}
\def\chijo {\mbox{$^3P_J^{[8]}$}}
\def\opchizs {\mbox{$\langle 0\vert {\cal O}_1^{\tiny\chiz}(^3P_0)\vert 0 \rangle$}}
\def\opchios {\mbox{$\langle 0\vert {\cal O}_1^{\tiny\chio}(^3P_1)\vert 0 \rangle$}}
\def\opchits {\mbox{$\langle 0\vert {\cal O}_1^{\tiny\chit}(^3P_2)\vert 0 \rangle$}}
\def\opchijs {\mbox{$\langle 0\vert {\cal O}_1^{\tiny\chij}(^3P_J)\vert 0 \rangle$}}
\def\opchizh {\mbox{$\langle 0\vert {\cal O}_8^{\psi}(^3P_0)\vert 0 \rangle$}}
\def\opchioh {\mbox{$\langle 0\vert {\cal O}_8^{\psi}(^3P_1)\vert 0 \rangle$}}
\def\opchith {\mbox{$\langle 0\vert {\cal O}_8^{\psi}(^3P_2)\vert 0 \rangle$}}
\def\opchijh {\mbox{$\langle 0\vert {\cal O}_8^{\psi}(^3P_J)\vert 0 \rangle$}}
\def\opetah  {\mbox{$\langle 0\vert {\cal O}_8^{\psi}(^1S_0)\vert 0 \rangle$}}
\def\oppsih  {\mbox{$\langle 0\vert {\cal O}_8^{\psi}(^3S_1)\vert 0 \rangle$}}
\def\oppsis  {\mbox{$\langle 0\vert {\cal O}_1^{\psi}(^3S_1)\vert 0 \rangle$}}
%%%%%%%%
\def\ophtpzs {\mbox{$\langle 0\vert {\cal O}_1^{H}(^3P_0)\vert 0 \rangle$}}
\def\ophtpos {\mbox{$\langle 0\vert {\cal O}_1^{H}(^3P_1)\vert 0 \rangle$}}
\def\ophtpts {\mbox{$\langle 0\vert {\cal O}_1^{H}(^3P_2)\vert 0 \rangle$}}
\def\ophtpjs {\mbox{$\langle 0\vert {\cal O}_1^{H}(^3P_J)\vert 0 \rangle$}}
\def\ophoszs {\mbox{$\langle 0\vert {\cal O}_1^{H}(^1S_0)\vert 0 \rangle$}}
\def\ophtsos {\mbox{$\langle 0\vert {\cal O}_1^{H}(^3S_1)\vert 0 \rangle$}}
%%%%%%%%%
\def\ophtpzo {\mbox{$\langle 0\vert {\cal O}_8^{H}(^3P_0)\vert 0 \rangle$}}
\def\ophtpoo {\mbox{$\langle 0\vert {\cal O}_8^{H}(^3P_1)\vert 0 \rangle$}}
\def\ophtpto {\mbox{$\langle 0\vert {\cal O}_8^{H}(^3P_2)\vert 0 \rangle$}}
\def\ophtpjo {\mbox{$\langle 0\vert {\cal O}_8^{H}(^3P_J)\vert 0 \rangle$}}
\def\ophoszo {\mbox{$\langle 0\vert {\cal O}_8^{H}(^1S_0)\vert 0 \rangle$}}
\def\ophtsoo {\mbox{$\langle 0\vert {\cal O}_8^{H}(^3S_1)\vert 0 \rangle$}}

\def\opchih {\mbox{$\langle 0\vert{\cal O}_8^{\tiny\chij}(^3S_1)\vert 0\rangle$}}

%%%%%%%%%%%%%
\def\b0{\mbox{$b_0$}}
\def\dsh  {\mbox{$\sigma^{H} $}}  
\def\dspq  {\mbox{$d\sigma^{\psi_Q} $}}
\def\dsp  {\mbox{$d\sigma^{\psi_Q} $}}
\def\qf     {\mbox{$\mu^2_{\rm F}$}}
\def\asopi{\mbox{$\frac{\as}{\pi}$}}
\def\szchiz{\mbox{$\sigma_0^{\tiny\chiz}$}}
\def\szchio{\mbox{$\sigma_0^{\tiny\chio}$}}
\def\szchit{\mbox{$\sigma_0^{\tiny\chit}$}}
\def\szchij{\mbox{$\sigma_0^{\tiny\chij}$}}
\def\dschiz  {\mbox{$d\sigma^{\chiz} $}}
\def\dschio  {\mbox{$d\sigma^{\tiny\chio} $}}
\def\dschit  {\mbox{$d\sigma^{\tiny\chit} $}}
\def\dschij  {\mbox{$d\sigma^{\tiny\chij} $}}

\def\vd{\mbox{${\bf D}$}}
\def\ve{\mbox{${\bf E}$}}
\def\vb{\mbox{${\bf B}$}}
\def\vs{\mbox{${\bf \sigma}$}}

\def\rj {{\mathrm J}}

\def \chizgg {\mbox{$\Gamma(\chiz \to \gamma\gamma)$}}
\def \chitgg {\mbox{$\Gamma(\chit \to \gamma\gamma)$}}
\def \chijgg {\mbox{$\Gamma(\chi_J \to \gamma\gamma)$}}
\def \chizlh {\mbox{$\Gamma(\chiz \to \lh)$}}
\def \chiolh {\mbox{$\Gamma(\chio \to \lh)$}}
\def \chitlh {\mbox{$\Gamma(\chit \to \lh)$}}
\def \chijlh {\mbox{$\Gamma(\chi_J \to \lh)$}}

\def\aem{\mbox{$\alpha_{{\mathrm\tiny EM}}$}}
\def\coll{\mbox{$ \vert\vert$}}
\def\Qb{\mbox{$\overline Q$}}
\def\ko{\mbox{$k_1$}}
\def\kt{\mbox{$k_2$}}
\def\mqs{\mbox{$m_Q^2$}}
\def\ks{\mbox{$k^2$}}
\def\ne{\mbox{$N_\epsilon$}}
\def\mzs{\mbox{$m_0^2$}}
\def\mos{\mbox{$m_1^2$}}
\def\mds{\mbox{$m_2^2$}}
\def\mts{\mbox{$m_3^2$}}
\def\mfs{\mbox{$m_4^2$}}
\def\tu{\mbox{$t_1$}}
\def\uu{\mbox{$u_1$}}
\def\li{\mbox{$\mathrm{Li}_2$}}
\def\meas{\mbox{$\frac{\rd^D k}{(2\pi)^D}$}}
% !!***^^**$$###  \def\tt{\mbox{$\leftrightarrow$}}
\def\fc{\mbox{${\mathrm F} \chi$}}
\def\nfc{\mbox{${\mathrm NF} \chi$}}

\def\mufrag{\mbox{$\mu_{\mathrm frag}$}}
\def\mufrags{\mbox{$\mu^2_{\mathrm frag}$}}

\def\mct{\mbox{$m_c^3$}}
\def\mcf{\mbox{$m_c^5$}}

\def\msb{\mbox{${\overline{MS}}$}}
\def\shad{\mbox{$S_{\mathrm had}$}}
\def\psiq{\mbox{$\psi_Q$}}
\def\pspq{\mbox{$\psi^\prime_Q$}}
\def\chijq{\mbox{$\chi_{QJ}$}}
\def\nj{\mbox{$N_J^\epsilon$}}

\def\bfp{{\bf p}}
\def\bfpp{{\bf p}'}
\def\bfk{{\bf k}}

\def\slash#1{{#1\!\!\!/}}

%%%%%%%%%%%%%%%%%%%%%%%%%%%%%%%%%%%%%%%%%%%%%%%%%%%%%%%%%%%%%%%%%%%%%%
\begin{titlepage}
\nopagebreak
{\flushright{
        \begin{minipage}{5cm}
        CERN-TH/97-142\\
        DESY 97-090\\ 
        hep-ph/9707223\\
        \end{minipage}        }

}
\vfill
\begin{center}
{\LARGE { \bf \sc NLO Production and Decay of Quarkonium}}
\vfill                
\vskip .5cm
{\bf Andrea PETRELLI\footnote{Address after Oct. 1st: Theory Division, 
Argonne National Laboratory, Argonne, IL, USA.} } \\
{INFN, Sezione di Pisa, Italy} \\                      
\verb+petrelli@ibmth.difi.unipi.it+\\
\vskip .5cm
{\bf Matteo CACCIARI\footnote{Address after Oct. 1st: LPTHE, Universit\'e Paris
XI, France}} \\
{Deutsches Elektronen-Synchrotron DESY, Hamburg, Germany} \\
\verb+cacciari@desy.de+\\
\vskip .5cm
{\bf Mario GRECO} \\
{Dipartimento di Fisica E. Amaldi, Universit\`{a} di Roma III, \\
and INFN, Laboratori Nazionali di Frascati, Italy} \\
\verb+greco@lnf.infn.it+\\
\vskip .5cm
{\bf Fabio MALTONI\footnote{Permanent address:
     Dipartimento di Fisica dell'Universit\`{a} and Sez. INFN, Pisa, Italy} 
and Michelangelo L. MANGANO\footnote{On leave of absence from 
    INFN, Pisa, Italy}}\\
{CERN, TH Division, Geneva, Switzerland} \\
\verb+fabio.maltoni@cern.ch, mlm@vxcern.cern.ch+\\
\end{center}
\nopagebreak
\vfill
%\vskip 3cm
\begin{abstract} We present a calculation of next-to-leading-order (NLO) QCD
corrections to total hadronic production 
cross sections and to light-hadron-decay rates of 
heavy quarkonium states. Both colour-singlet and colour-octet contributions  
are included. We discuss in detail the
use of covariant projectors in dimensional regularization, the
structure of soft-gluon emission and the overall finiteness of radiative
corrections. We compare our approach with the NLO version of the
threshold-expansion technique recently introduced by Braaten and Chen.
Most of the results presented here are new. Others represent the
first independent reevaluation of calculations already known in the literature.
In this case a comparison with previous findings is reported. 
\end{abstract}                                                
\vskip 1cm
CERN-TH/97-142\hfill \\
\today \hfill 
\vfill 
\end{titlepage}

\section{Introduction}
Since the discovery of the  $\jpsi$~\cite{Aubert74} and
its interpretation within QCD as a charm-anticharm bound
state~\cite{Appelquist75}, the
study of quarkonium has received much attention from both a 
theoretical and an experimental point of view, providing a good testing ground
for studies of Quantum Chromodynamics (QCD) in both its
perturbative and non-perturbative regimes.
                                          
Decay rates of heavy quarkonium states into photons and
light hadrons have first been calculated at leading-order (LO)
and compared to experimental data
(see, e.g., \cite{Appelquist78}) 
under the assumption of  a factorization between a short-distance 
part describing the annihilation of the heavy-quark pair in a colour-singlet
state and a non-perturbative long-distance    
factor, related to the value at the origin 
of the non-relativistic wave-function or its derivatives.
Calculations of decay rates at full next-to-leading order (NLO) have also been 
performed in this framework
since the early days of quarkonium physics~\cite{Barbieri1S0}. 
This approach can be extended to the case of quarkonium production.
The prescription to evaluate the short-distance coefficients is
very simple: the $\QQ$ pair has to be produced during 
the  short-distance interaction in a colour-singlet state, 
with the same spin and angular momentum quantum numbers
of the quarkonium state we are interested in~\cite{Kuehn79}. 
A single non-perturbative 
parameter, the same appearing  in the quarkonium decay and provided by the
bound state Bethe-Salpeter  wave-function, accounts for the hadronization
of the $\QQ$ pair into the physical quarkonium state.                     
Applications of this approach, usually referred to as
the Colour Singlet Model (CSM), led to the calculation of LO matrix elements
for the production of total cross sections and \pt\ distributions in
hadronic collisions~\cite{csm}.
Phenomenological successes and failures of the CSM in describing the data
available up to 1993 on
charmonium production and decay are nicely reviewed in ref.~\cite{Schuler94}.
                                                      
In the case of $S$-waves the simple factorization hypothesis underlying
the CSM was confirmed by the calculation of
one loop corrections~\cite{Barbieri1S0,Hagiwara}.
The appearance of a logarithmic  infrared divergence in
the case of NLO $P$-wave decays 
into light hadrons~\cite{Barbieri3PJ,BarbieriIR}
violated instead factorization explicitly.
Phenomenologically, this singularity could be handled by       
relating it to the binding energy of the bound quarks.                 
Nevertheless, it has since then been clear that
in spite of its simplicity and physical transparency, the CSM suffers from 
serious theoretical limitations. The principal one being the absence of a
general  theorem assessing its validity in higher orders of perturbation theory,
as  already indicated by  the explicit example of $P$-states decay. The
striking observation by CDF 
of large-$\pt$ $\jpsi$  and $\psi'$ states produced at
the Tevatron~\cite{cdf,d0} at a rate
more than one order of magnitude larger than the 
theoretical prediction, and the serious discrepancies between fixed-target
data and predictions for the relative production rates of \threeSone
\threePone\ and \threePtwo\ states (for recent total cross-section
measurements, see~\cite{e705,e672}), 
provided equally compelling evidence that
some important piece of physics was missing from the CSM.
                                    
The road to solve these formal and phenomenological problems  has been
indicated by the work of  Bodwin, Braaten and Lepage (BBL)~\cite{bbl}, which
provided a new framework for the study of quarkonium production and decay
within QCD.  In this work, perturbative factorization is retained by allowing
the quarkonium production and decay to take place via intermediate $\QQ$ states
with quantum numbers different than those of the physical quarkonium state
which is being produced or which is decaying.  In the case of production, for
example,  the general expression for a cross section is given by:
\be                                                                  
d\sigma(H + X) = \sum_n d\hat\sigma(\QQ[n] + X)\langle{\cal O}^H[n]\rangle\, .
\label{eq-fm}
\ee
Here $d\hat\sigma(\QQ[n] + X)$ describes the short distance production of
a $\QQ$ pair in the colour, spin and  angular momentum state $n$, and 
$\langle{\cal O}^H[n]\rangle$, the vacuum expectation value of a
four-fermion operator defined within Non-Relativistic QCD 
(NRQCD)~\cite{CL86,LMNMH92},  describes the
hadronization of the pair into the observable quarkonium state $H$. 
$\QQ$ states with quantum numbers other than $H$
arise from the expansion of the $H$ Fock-space wave function 
in powers of the heavy quark velocity $v$.               
The relative importance of the various contributions in eq.~(\ref{eq-fm})  can
be estimated by using NRQCD velocity scaling rules \cite{LMNMH92}, which allow
the truncation of the series in eq.~(\ref{eq-fm}) at any
 given order of accuracy.
If one only retains the lowest order in $v$, the description of $S$-wave
quarkonia production or annihilation reduces to the CSM one. In the case of $P$
waves, instead, contributions from colour-octet, $S$-wave $\QQ$ states are of
the same order in $v$ as those from the
 leading colour-singlet $P$-wave state. Infrared
singularities which appear in some of the  short-distance coefficients of
$P$-wave states can then 
be shown~\cite{bbl} to be absorbed into the long-distance
part of colour-octet $S$-wave terms,  thereby  ensuring a well defined overall
result. This framework for the calculation of quarkonium production and decay
is often referred to as the colour-octet model (COM), perhaps with an abuse of
language, as the COM pretends to be a direct outcome of QCD, rather than just a
``model''.
Good reviews of the underlying physical principles and applications 
can be found in refs.~\cite{Braaten96a,Beneke97a}.                
                                      
The effect of colour-octet contributions can be extremely important
even in the case of $S$-wave production. In fact, while 
their effects are predicted by the scaling rules to be suppressed 
by powers of $v$ with respect to the leading colour-singlet ones, 
their short-distance coefficients can receive contributions at lower orders of 
\as, thereby enhancing significantly the overall production
rate~\cite{Braaten95}.            
The inclusion of these colour-octet processes, in conjunction with the
observation that gluon fragmentation provides the dominant contribution to the
short-distance coefficients at large-\pt~\cite{Braaten93}, leads to a very
satisfactory description of the Tevatron                
data~\cite{Cacciari94,Cacciari95,Cho96}.
                                                 
One of the most important consequences of factorization for quarkonium
production is the prediction that the value of the non-perturbative parameters
does not depend on the details of the hard process, so that parameters
extracted from a given experiment can be used in different ones. For
simplicity, we will refer to this concept as ``universality''. 
Several studies of experimental data coming 
from different kind of reactions have been performed to assess the validity
of universality. For example  calculations of inclusive quarkonia production
in $e^+e^-$ \cite{ee}, fixed target experiments \cite{coft}, 
$\gamma p$  collisions \cite{Cacciari96,gammap} and $B$ decays 
\cite{bdecays} have  been carried out within this framework. 
The overall agreement of theory and data is satisfactory, but there are clear
indications that large uncertainties are present. The most obvious one is the
discrepancy~\cite{Cacciari96}
between  HERA data~\cite{HERApsi} and the large amount  of
inelastic \jpsi\ photoproduction predicted by applying the colour-octet matrix
elements extracted from the Tevatron large-\pt\ 
data~\cite{Cho96,Beneke97}.
                                                                       
It becomes therefore important to assess to which extent is universality
applicable. Several potential sources of universality violation are indeed
present, both at the perturbative and non-perturbative level.
On one hand there are potentially large corrections to the factorization
theorem itself. In the case of charmonium production, for example, the mass of
the heavy quark is small enough that non-universal power-suppressed corrections
can be large.  Furthermore, some higher-order corrections in the velocity
expansion are strongly enhanced at the edge of phase-space~\cite{Beneke97b}.
For example, the alternative choices of using as a mass parameter for the
matrix elements and for the phase-space boundary
the mass of a given quarkonium state or twice the heavy-quark mass $2m$,
give rise to a large uncertainty in the production rate near threshold.
These effects, which are present both in the total cross-section and in the
production at large-\pt\ via gluon fragmentation, violate universality. This is
because the threshold behaviour depends on the nature of the hard process under
consideration.

Another source of bias in the use of universality comes from purely
perturbative corrections. Most of the current predictions for quarkonium
production are based on the use of leading-order (LO) matrix elements. Possible
perturbative $K$-factors are therefore absorbed into the non-perturbative matrix
elements extracted from the comparison of data with theory. Since the size of
the perturbative corrections varies from one process to the other, an
artificial violation of universality is introduced. Examples of the size of
these corrections are given by the large impact of $k_T$-kick effects and
initial-state multiple-gluon emission in open-charm~\cite{Frixione97} and
charmonium production~\cite{e789,CanoColoma97}.

In this paper we focus on the evaluation of the \oacube\ corrections to
quarkonium total hadro-production cross sections. Of all the efforts needed to
improve the quantitative estimates of quarkonium production, this is probably
the less demanding one, as the formalism and the techniques to be used are
a priori rather well established. Nevertheless there are some subtleties
related to the regularization of the infrared (IR) and ultraviolet (UV)
singularities which need to be addressed with some care. Furthermore,
as pointed out in a preliminary account of part of this work~\cite{Mangano96}, 
the impact of NLO corrections can be significant and a general study of their
effects is necessary.

To carry out our calculations, we need a framework for calculating NLO
inclusive production cross sections and inclusive  annihilation decay rates. As
well known, in perturbative calculations of the short distance coefficients
beyond leading order in  $\as$, the most convenient method for  regulating both
UV and IR divergences is dimensional regularization. On  the
other  hand most calculations of production cross sections and decay rates for 
heavy quarkonia  have been performed using the covariant projection method
\cite{Kuehn79}, 
which involves the  projection of the $\QQ$ pair onto states with definite 
total angular momentum $J$, and which is specific to four dimensions.

In this paper we generalize the
method of covariant projection  to $D = 4 - 2\epsilon$ dimensions and perform 
all our calculations within this framework. 
Very recently a different method for calculating production cross sections 
and decay  rates, which fully exploits the NRQCD factorization framework and
bypasses the need for projections -- the
so  called ``threshold expansion method'' -- has been generalized to
$D$
dimensions so that  dimensional  regularization can be used \cite{Braaten96,
Braaten97}.
We will show that, whenever common calculations exist,
the results obtained using the two techniques coincide.

The knowledge of perturbative corrections to quarkonium production is rather
limited. The only examples of full NLO calculations we are aware of are the
following:
\begin{itemize}
\item total hadro-production cross-sections for \oneSzero\
      states~\cite{Kuehn93,Schuler94};
\item inclusive \pt\ spectrum of \jpsi\ in photo-production, within the
CSM~\cite{Kraemer95};
\item corrections to the polarization of \jpsi\ produced via gluon
fragmentation~\cite{Beneke96}.                  
\end{itemize}       

In this work we present the first calculation of the \oacube\ total
cross-sections for hadroproduction of several $\QQ$ states of
phenomenological relevance: $\oneSzero^{[8]}$, $\threeSone^{[8]}$ and
$\threePJ^{[1,8]}$, where the right upper index labels the colour configuration
of the $\QQ$ pair. In addition, we repeat the calculation for
$\oneSzero^{[1]}$ states within our formalism, in order to establish
consistency with the previous calculations of this
quantity~\cite{Kuehn93,Schuler94}. We also calculate the \oacube\ light-parton
decay rates for \oneSzero, \threeSone\ and \threePJ\ colour-singlet and
colour-octet states\footnote{Notice that in the case of colour-singlet
\threeSone\ and \threePone\ states the \oacube\ result is however LO only.}.
Some of these results are new, some have been known for a                 
long time. In the case of $\oneSzero^{[1]}$ decays we find agreement with
previous results~\cite{Barbieri1S0,Hagiwara}. In the case of $\chijs$
and $\etah$ decays we find a disagreement with the previous         
calculations, reported in~\cite{Barbieri3PJ} and in \cite{Huang96},
respectively\footnote{After this paper was released as a preprint, the authors
of ref.~\cite{Huang96} reviewed their calculation.
We have been informed that their final result now coincides with ours.}. 
While these discrepancies have a negligible numerical impact, it
would be interesting to see in the future new independent calculations
performed, to confirm either of the results.
                                         
This paper is structured as follows. In Section~\ref{sec:projectors} we 
introduce our formalism, with emphasis on the extension to $D$ dimension of
the covariant projection method and on its link with the NRQCD factorization 
approach. Section~\ref{sec:general} gives a brief general description of the 
NLO calculation. In particular, we describe the technique used to identify the
residues of the IR and collinear singularities and to allow their cancellation
without the need for a complete $D$-dimensional calculation of the
real-emission matrix elements.
                              
Section~\ref{sec:soft} describes the behaviour of the soft 
limit of the NLO real corrections, and Section~\ref{sec:decay} presents the 
results for the various components (real and virtual corrections) of the
NLO result for the decay widths. Section~\ref{sec:opren} describes how, within 
the NRQCD formalism, the Coulomb singularity of the virtual corrections and 
the aforementioned infrared singularity which appears in NLO corrections to 
$P$-wave decays can be disposed of. Finally, Section~\ref{sec:production} 
presents the NLO results for the production processes. 

Appendix~\ref{appA} collects symbols and notations and Appendix~\ref{appLO}
collects the  results for the Born cross sections and decay rates in $D$
dimensions.
A summary of all 
results is provided in Appendix~\ref{appNLO}, where hadroproduction cross
sections and decay widths into light hadrons  are presented in  their final
form, after the cancellation of all singularities. In this Appendix we also
comment on the discrepancies observed when comparing 
the results for $\chizs$, $\chits$ and
$\etah$ decays with those presented in previous calculations.
Results for  quarkonia    
photoproduction and decay into one photon plus light hadrons can be easily
extracted from these calculations: explicit results will be presented in a
forthcoming publication.

Appendix~\ref{appc}  presents, for comparison, a calculation of the NLO virtual
corrections performed using
the threshold expansion method. The results agree with those obtained using the
$D$-dimensional projector technique.
Appendix~\ref{appFF} presents the results 
for the $g\to\chi_J$ fragmentation function obtained within our formalism,
showing again agreement with the results of the threshold expansion
technique~\cite{Braaten97}.
                           
A preliminary account of some of the results contained in this paper, together
with a first phenomenological study of NLO $\chi_{c,b}$ total production cross
sections at fixed target and collider energies, was presented in
ref.~\cite{Mangano96}. A more complete study, including the additional
processes calculated in the present work, will be the subject of a separate
publication.                                          
                                                        
\section{Introduction to the formalism}
\label{sec:projectors}
\begin{figure}[t]
\begin{center}
\epsfig{file=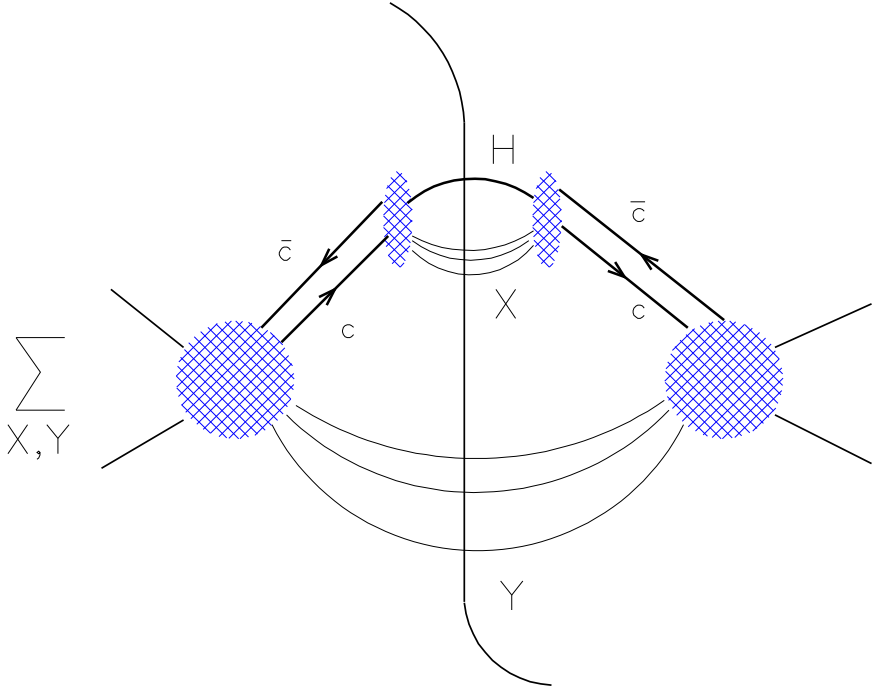,
            width=12cm,clip=}
\caption{\small \label{fig:fact} Diagrammatic representation of factorization
in parton + parton $\to  H + X$  . }
\end{center}                                        
\end{figure}

The calculation  of cross sections and decay rates for heavy quarkonia deals
with two kinds of contributions: short distance ones, related to the production
or annihilation of the heavy-quark pair, and long distance ones,
related to non-perturbative transition between the quarks and the 
observable quarkonium state.                         
Rigorous theorems exist~\cite{bbl} proving the factorization of these two
stages in the case of inclusive     
quarkonium decays. Less rigorous, but widely accepted, proofs have also been
given of the
factorization between the perturbative and non-perturbative phases in the
hadroproduction case~\cite{bbl}. 
In order to lay down the strategy of our calculation, and
establish some notation, we briefly review here the essence of these
factorization statements in the case of hadroproduction.
Figure~\ref{fig:fact} shows a sketch of a typical diagram contributing to the 
inclusive production of a quarkonium state $H$. Factorization implies that we
can identify an intermediate perturbative state composed of a pair of on-shell
heavy quarks, whose non-perturbative evolution will lead to the formation of a
physical hadron $H$ plus possibly a set $X$ of light partons. The set of
light partons $Y$ consists of additional states produced during the hard
collision that created the $\QQ$ pair. Whether a light parton belongs to
the set $X$ or to the set $Y$ depends on our (arbitrary) choice of
factorization scale. The possibility to maintain the freedom to select this 
scale at all orders of perturbation theory is a consistency check of the
factorization hypothesis, and leads to renormalization-group relations
between the non-perturbative matrix elements which describe the long-distance
physics~\cite{bbl}. The factorization theorem states that the effect of soft
gluons exchanged between the sets $X$ and $Y$ cancel out in the production
rate, and their effect can therefore be neglected. The emission of hard gluons
between $X$ and $Y$, furthermore, 
is suppressed by powers of the ratio between the scales of
the soft and hard processes. 
                           
At the amplitude level, and assuming the validity of the factorization
theorem, the processes shown in fig.~\ref{fig:fact} can
then be represented as follows:                
\be
{\cal M} \= {\cal T}_{ij}(\QQ + Y) \;\langle H\,X\vert\overline{\psi_i} 
      \psi_j\vert 0\rangle
\ee
where ${\cal T}_{ij}$ is the
matrix element for the production of the $\QQ+Y$
final state with the heavy quark spinors removed, 
$i$ and $j$ are spinorial indices,
and $\psi$ is the fermion field operator. 
Summing over a complete set of states containing the heavy quark pair, one then
obtains:                                                               
\be \label{eq:3}
{\cal M} \=
\sum_n\langle H\,X\vert \QQ[n]\rangle \langle \QQ[n]\vert\overline{\psi_i}
\psi_j\vert 0\rangle{\cal T}_{ij} 
\,\equiv \; \sum_n\langle H\,X\vert \QQ[n]\rangle \;
     \tr\left[\,{\cal T}\, {\mit \Pi}^{(n)}\right] \; ,
\ee                                                
where ${\mit \Pi^{(n)}}$ selects the quantum numbers of the quark pair
relative to the state $n$. Its specific form for $S$ and $P$ wave states will
be given in the following.
The first term on the right-hand side is
interpreted as the overlap between the perturbative $\QQ$ state and the 
final hadronic state. It can be written as the transition matrix element
of an operator ${\cal O}^{(n)}_2$, 
bilinear in the heavy-quark field, between the vacuum and the
final state $\langle HX\vert$. 
We can therefore define the following quantity, 
following the notation introduced in ~\cite{bbl}:
\be           
\label{eq:OH}
\langle{\cal O}^H(n)\rangle \= \sum_X 
  \langle 0\vert {{\cal O}_2}^{(n)\dagger}\vert H\, X\rangle
\langle H\, X\vert {{\cal O}_2}^{(n)}\vert 0\rangle \=
\langle 0\vert {{\cal O}_2}^{(n)\dagger}{\cal P}^{H} 
{{\cal O}_2}^{(n)}\vert 0\rangle \; ,
\ee
where 
\be
{\cal P}^H = \sum_X \vert H\,X\rangle\langle H\,X\vert \;.
\ee                                                       
We finally get the standard result~\cite{bbl}:
\be                                              
d\sigma(H + X) = \sum_n d\hat\sigma(\QQ [n] + X)\langle{\cal O}^H(n)\rangle
\label{eq:xsect}
\ee
The quantity $\langle{\cal O}^H(n)\rangle$ is proportional to the 
inclusive transition probability of the perturbative state $\QQ[n]$ into the
quarkonium  state $H$. Notice that in principle we could have specified some
detail of the inclusive state $X$, for example we could have specified the
fraction $z$ of light-cone momentum of the pair $\QQ[n]$ carried away by $X$,
in which case we could have defined, starting from eq.~(\ref{eq:OH}), the
equivalent of a non-perturbative fragmentation function (as opposed to an
inclusive transition probability). Such a fragmentation function provides a
more detailed description of the hadronization process, which can be used to
parametrize potentially large higher-order $v^2$ corrections, such as those
appearing near the boundary of phase-space. The factorization
theorem allows us to extract it from some set of data, and its renormalization
group properties should allow its use in different contexts.  
These more general non-perturbative quantities have also been introduced
and discussed recently in ref.~\cite{Beneke97b}, where several interesting
applications have been explored.
                                
We now come to the discussion of the set of rules which define the perturbative
part of eq.~(\ref{eq:xsect}), namely the projection over the perturbative
states $\QQ[n]$. These projections have been known for some time~\cite{Kuehn79}
and have been used since then for the evaluation of quarkonium production and
decays~\cite{csm}. We will extend these rules here for use in dimensions
$D=4-2\ep$, in view of our applications to NLO corrections, where
dimensional regularization techniques are most useful to handle the appearance
of infrared (IR) and ultraviolet (UV) divergences. 
With the exception of the normalization of spin and colour factors, discussed 
below, our choice of normalization is consistent with the standard                    
non-relativistic normalizations used in \cite{bbl} to define the
operators entering the definition  of the non-perturbative matrix elements
$\langle {\cal O}^H(n)\rangle $.

The spin projectors 
for outgoing heavy quarks momenta $Q = P/2+q$ and ${\overline Q }= P/2-q$, are 
given by:
\ba                                                                
&&{\mit \Pi}_{0}
     = {1\over{\sqrt{8m^3}}} \left({{\slash{P}}\over 2} - \slash{q} - m\right)
       \gamma_5 \left({{\slash{P}}\over 2} + \slash{q} + m\right)\, , 
\label{proj_00}                                          
\\                                                       
&&{\mit \Pi}_{1}^{\alpha} 
     = {1\over{\sqrt{8m^3}}} \left({{\slash{P}}\over 2} - \slash{q} - m\right)
       \gamma^\alpha \left({{\slash{P}}\over 2} + \slash{q} + m\right)\, ,
\label{proj_1Sz}                                              
\ea
for spin-zero and spin-one states respectively. In these relations, $P$ is the
momentum of the quarkonium state, $2q$ is the relative momentum between the
$\QQ$ pair, and $m$ is the mass of the heavy quark $Q$. 
The above normalization of the projection operators corresponds to a
relativistic normalization of the state projected out. 
Higher-order powers of
$q$, not required for the study of $S$ and $P$-wave states, have been neglected
in eqs.~(\ref{proj_00}) and (\ref{proj_1Sz}).
                                             
The use of these operators, equal to those used in $D=4$ dimensions, requires
some justification. In $D \neq 4$ dimensions, the product of two
spinorial representations gives rise in fact to other representations, in
addition to the spin-singlet (scalar) and spin-triplet (vector) ones. For
example, the equivalent of a quarkonium state in $D=6$ can have a total spin
corresponding to a scalar, 
a vector, and a fully-antisymmetric three-index tensor of $SO(5,1)$. The
total dimension of these representations is indeed 1+5+10=16. Once we move away
from $D=4$, therefore, we should in principle take into account the existence of
other states in addition to the $S=0$ and $S=1$ ones found in $D=4$. 
For non-integer $D$, in particular, we should deal with an infinite set of
states, since the Clifford algebra becomes infinite-dimensional. The projectors
defined in eqs.~(\ref{proj_00},\ref{proj_1Sz}) are still, nevertheless, 
the right operators needed to define the scalar and vector states. One can
neglect the presence of higher-spin states in the infinite-mass limit, since
spin-flip transitions, which would mix scalars and vectors 
to higher spins, vanish.                                   
With finite mass one can neglect the mixing among these
operators provided one works at sufficiently low orders in $v$. At higher
orders in $v$ spin-flip transitions occur, and in principle mixing with
higher-spin states should be handled. In the calculations illustrated in this
paper this never occurs, and we can safely work with just the states we are
interested in, namely the $D$-dimensional scalar and vector.

The problem of dealing with the higher-spin states goes beyond the
scope of this work, and will not be analyzed here in any detail. Nevertheless
we want to provide some argument to suggest that even at high
orders in $v$ no problems are expected. The argument proceeds as follows:
higher-spin operators should vanish in 4 dimensions, and therefore their effect
should be suppressed by at least one power of $\ep$. In order to contribute to a
cross-section in 4 dimensions, they must be accompanied by some $1/\ep$ pole.
However IR and collinear poles cannot appear: IR poles arise from the emission
of soft gluons, which cannot change the spin, and collinear poles admit a
factorization in terms of lower-order amplitudes times universal splitting
functions. Therefore no new operators can appear in the collinear limit, and
higher-spin operators should decouple. As for UV poles, in the cross-section
calculations presented here they are all reabsorbed
by the standard renormalization procedure. 
We believe it should be possible to set the above arguments
on a more solid footing. We point out that a proof of the decoupling of
higher-spin evanescent operators is also required when 
using the $D$-dimensional threshold expansion technique introduced in
ref.~\cite{Braaten97}, since in its current formulation the explicit assumption
is made that only spin-0 and spin-1 operators are relevant.
                                                           
The $D$-dimensional character of space-time is implicit in
eqs.~(\ref{proj_00},\ref{proj_1Sz}), and appears explicitly when  performing the
sums over polarizations, as shown later. The manipulation of expressions
involving $\gamma_5$ is carried out by using standard techniques, as discussed
in the following.  
The colour singlet or octet state content of a given state will  be
projected out by contracting the amplitudes with the following
operators :
\ba
&&{\cal C}_1 = {{\delta_{ij}}\over{\sqrt{\nc}}}\qquad{\rm for~the~singlet}
\label{proj_sing}
\\
&&{\cal C}_8 = \sqrt{2} T_{ij}^c\qquad{\rm for~the~octet}
\label{proj_oct}
\ea

The projection on a state with orbital angular momentum $L$ is obtained by 
differentiating $L$ times the spin-colour projected amplitude with respect to 
the momentum $q$ of the heavy quark in the $\QQ$ rest frame, and then 
setting $q$ to zero. We shall only
deal with either $L=0$ or $L=1$ states, for which the amplitudes take
the form:
\ba      
&&{\cal A}_{S=0,L=0} = \tr\left.\left[{\cal C}\,{\mit  \Pi}_0\, {\cal A}\right]
\right|_{q=0}\qquad\qquad\qquad\,{\rm Spin}\;{\rm  singlet }~S~{\rm states}\\
&&{\cal A}_{S=1,L=0} = \epsilon_\alpha
          \tr\left.\left[{\cal C}\,{\mit  \Pi}_1^\alpha\, 
     {\cal A}\right]\right|_{q=0}\qquad\qquad\;\;\,\,\,{\rm Spin} 
     \;{\rm triplet }~S~{\rm states}\\                                        
&&{\cal A}_{S=0,L=1} = \epsilon_\beta
{{\rd}\over{\rd q_\beta}} \tr\left.\left[{\cal C}\, {\mit \Pi}_0 
{\cal A}\right]\right|_{q=0}\quad\qquad\;\;\,{\rm Spin}\;
{\rm singlet }~P~{\rm states}\\
&&{\cal A}_{S=1,L=1} = {\cal E}_{\alpha\beta}
{{\rd}\over{\rd q_\beta}} \tr\left.\left[{\cal C}\, 
{\mit \Pi}_1^\alpha 
{\cal A}\right]\right|_{q=0}\quad\qquad{\rm Spin}\; 
{\rm triplet }~P~{\rm states}
\ea    
${\cal A}$ being the standard QCD amplitude for production (or decay) of the 
$\QQ$ pair, amputated of the heavy quark
spinors.    
The amplitudes ${\cal A}_{S,L}$ will then have to be squared, summed over the
final degrees of freedom and averaged over the initial ones.
The selection of the appropriate total angular momentum quantum number is done
by performing the proper polarization sum. We define:
\be                                                  
\Pi_{\alpha\beta} \equiv -g_{\alpha\beta} + {{P_\alpha P_\beta}\over{M^2}}
\; ,
\ee 
where $M=2m$.
The sum over polarizations for a $^3S_1$ state, which is still a vector even in
$D=4-2\ep$ dimensions, is then given by:
\be
\sum_{J_z} \ep_\alpha\ep_{\alpha'}^* = \Pi_{\alpha\alpha'}
\ee
In the case of $^3P_J$ states, the three multiplets corresponding to $J=0,1$
and 2 correspond to a scalar, an antisymmetric tensor and a symmetric
traceless tensor, respectively. We shall denote their polarization tensors by 
${\cal E}_{\alpha\beta}^{(J)}$. The sum over polarizations is then given by:
\ba                           
&&{\cal E}_{\alpha\beta}^{(0)}{\cal E}_{\alpha'\beta'}^{(0)*} =
{1\over{D-1}} \Pi_{\alpha\beta}\Pi_{\alpha'\beta'}
\label{polsum0}
\\
&&\sum_{J_z} {\cal E}_{\alpha\beta}^{(1)}{\cal E}_{\alpha'\beta'}^{(1)*} =
{1\over{2}} [\Pi_{\alpha\alpha'}\Pi_{\beta\beta'} - 
             \Pi_{\alpha\beta'}\Pi_{\alpha'\beta}]
\label{polsum1}
\\
&&\sum_{J_z} {\cal E}_{\alpha\beta}^{(2)}{\cal E}_{\alpha'\beta'}^{(2)*} =
{1\over{2}} [\Pi_{\alpha\alpha'}\Pi_{\beta\beta'} + 
             \Pi_{\alpha\beta'}\Pi_{\alpha'\beta}] - 
{1\over{D-1}} \Pi_{\alpha\beta}\Pi_{\alpha'\beta'}
\label{polsum2}
\ea
for the $\threePzero$, $\threePone$ and $\threePtwo$ states respectively.
Total contraction of the polarization tensors gives the number of polarization
degrees of freedom in $D$ dimensions. Therefore
\be
N_{^3S_1} = \sum_{J_z} \ep_\alpha\ep_{\alpha}^* = \Pi_{\alpha\alpha} = D-1 = 3 -2 \ep
\ee      
for the $\threeSone$ state and
\be
N_J = \sum_{J_z} {\cal E}_{\alpha\beta}^{(J)}{\cal E}_{\alpha\beta}^{(J)*}
\ee
for the $\threePJ$  states, with
\be
N_0 = 1, \; \;  N_1 = {{(D-1)(D-2)}\over 2}=(3-2\ep)(1-\ep), \; \;
N_2 = {{(D+1)(D-2)}\over 2}=(5-2\ep)(1-\ep) \;.                   
\ee                                            

The application of this set of rules produces the short-distance cross
section coefficients $\hat \sigma$ or the short-distance decay widths 
$\hat \Gamma$
for the $ij\leftrightarrow\spectr^{[1,8]}$ processes:
\ba
&&d\hat\sigma(ij\to\spectr^{[1,8]}) = {1\over{2s}}
   \overline{\sum}|{\cal A}_{S,L}|^2\;d \Phi\, ,\\
&&d\hat\Gamma(\spectr^{[1,8]}\to ij) = {1\over{4m}}
   \overline{\sum}|{\cal A}_{S,L}|^2\;d \Phi \, ,
\ea                        
$s$ being the partonic centre of mass energy squared and $2m$ representing the
mass of the decaying $\QQ$ pair. To find the physical cross sections or decay
rates for the observable quarkonium state $H$ these short distance coefficients
must be properly related to the NRQCD production or annihilation matrix
elements $\langle{\cal O}_{[1,8]}^H(\spectr)\rangle$ and $\langle H|{\cal
O}_{[1,8]}(\spectr)|H\rangle$ respectively.

The above procedure contains some freedom in the choice of both the absolute
and relative normalization of the NRQCD matrix elements. We could in fact
decide to shift some overall normalization factor from the long-distance to the
short-distance matrix elements. A natural choice for the absolute normalization
of the non-perturbative matrix elements is
obtained by requiring the short-distance coefficients to coincide with those 
which appear when the expectation value of the NRQCD operators is taken between
free $\QQ$ states. The                                            
relative normalization between colour-singlet and colour-octet
operators is obtained by imposing the decomposition:     
\be       
    {\o}_{\rm tot} = {\o}_1 + {\o}_8\; . 
\ee                  
The identification of the colour-singlet and color-octet components
is obtained by using the Fierz rearrangement:
\be                     
\delta_{i'i} \delta_{j'j} = \frac{1}{\nc} \delta_{j i} \delta_{i' j'}
+ 2 T^a_{j i} T^a_{i' j'}
\ee
The identification of the spin-0 and spin-1
components is obtained by using the Fierz rearrangement:
\be        
\delta_{i'i} \delta_{j'j} = \frac{1}{2} \left[\delta_{j i} \delta_{i' j'}
+ \sigma^a_{j i} \sigma^a_{i' j'}  \right] \; ,
\ee                                       
with the normalization of the Pauli matrices fixed to the canonical one:
\be                                                                     
       \left\{ \sigma^a \, , \, \sigma^b \right\} \= 2 \delta^{ab} \; .
\ee
Combining the two results, we obtain the following decomposition:
\ba                                                              
\psi^\dagger \psi\; \chi^\dagger \chi &=&
\frac{1}{2\nc} \left( \psi^\dagger\chi\; \chi^\dagger \psi \; + \;
                      \psi^\dagger \sigma^i \chi\; \chi^\dagger \sigma^i \psi 
               \right) \nonumber \\
&+& \psi^\dagger T^a\chi \;\chi^\dagger T^a\psi \;+\;
   \psi^\dagger T^a\sigma^i \chi \;\chi^\dagger T^a\sigma^i \psi \; .
\ea                                                         
We find it natural to use this decomposition to identify the normalization of
the NRQCD operators, which are presented  in full detail for $S$ and $P$-wave
states in Appendix~\ref{appA}.
The resulting normalization for the colour-singlet part differs from the
usual one found in the literature. Compared to the conventions of
BBL~\cite{bbl}, our normalizations are given by\footnote{Strictly
speaking, consistency with the relativistic normalization of the $\QQ$
state imposed by eqs.~(\ref{proj_00}) and (\ref{proj_1Sz}) requires the NRQCD
matrix elements to be evaluated using a relativistic normalization for the
quarkonium state. The normalization factor $2M$, implicitly assumed in the
unitary sum over intermediate states used in eq.~(\ref{eq:3}),  rescales
however the  matrix element to the value obtained using standard
non-relativistic normalizations, up to corrections of order
$v^2$~\cite{Braaten96}.  For this reason, in the following we will consider
NRQCD matrix elements evaluated using non-relativistic normalizations, as
customary, and no additional $2M$ normalization factor is required.}:
\ba                                                             
&&{\o}_1 = \frac{{\o }_1^{\rm BBL}}{2\nc}, \\
&&{\o}_8 = {\o }_8^{\rm BBL}.
\ea        
We stress that, in addition to being more natural, 
(e.g. in the case of $S$ waves 
the non-perturbative matrix elements for colour singlet states
coincide exactly with the value of the total wave functions evaluated at the
origin),                                     
our definition is probably more adequate when trying to estimate the
order of magnitude of non-perturbative matrix elements using velocity-scaling 
rules. For example, it is known that the ratio between the 
matrix elements of the colour-octet and
colour-singlet $^3S_1$ operators should scale like $v^4$. The extraction of 
$\langle \o^{\psi,{\rm BBL}}_8(\threeSone) \rangle$ from the Tevatron
data~\cite{Cacciari95,Cho96,Beneke97}               
leads to                             
values in the range $0.65 \div 1.4 \times 10^{-2}$~GeV$^{-3}$.             
The value of                                                            
$\langle \o^{\psi,{\rm BBL}}_1(\threeSone) \rangle$ is given by 
1.3~GeV$^{-3}$. The ratio between the two, of the order of $0.5 \div 1 \times
10^{-2}$, is quite smaller than the estimated              
value of $v^4 \sim 0.06$. Using our normalization of the NRQCD operators, the
ratio becomes $3 \div 6 \times
10^{-2}$, which is consistent with the velocity scaling rule.
This behaviour is confirmed by similar
results obtained in the case of $\psi^{\prime}$ and $P$-wave states.

Having done this, the cross sections and the decay rates read:
\ba                             
&&\sigma(ij\to\spectr^{[1,8]}\to H) = \hat \sigma(ij\to\spectr^{[1,8]}) 
{{\langle \o_{[1,8]}^H(\spectr)\rangle}
\over{N_{col} N_{pol}}},\\[10pt]
&&\Gamma(H\to\spectr^{[1,8]}\to ij) = \hat \Gamma(\spectr^{[1,8]}\to ij)
\langle H|\o_{[1,8]}(\spectr)|H\rangle.
\ea
$N_{col}$ and $N_{pol}$ refer to the number of colours and polarization states
of the $\QQ[\spectr]$ pair produced. They are given by 1 for singlet
states or $N_c^2-1$ for octet states, and by the $D$-dimensional $N_J$'s defined
above. Dividing by these colour and polarization degrees of freedom in the cross
sections is necessary as we had summed over them in the evaluation of the short
distance coefficient $\hat\sigma$. Analogously, these degrees of freedom are 
summed over in the                      
decay matrix element, but averaged as part of the definition of the short
distance width $\hat\Gamma$, when mediating over the $\QQ$ initial 
state.

The matrix elements appearing in the equations above are of course meant to be
the bare $D$-dimensional ones whenever $D$-dimensional cross sections or decay
rates are to be calculated. Making use of their correct mass dimension,
$3-2\ep$ and $5-2\ep$ for $S$ and $P$ wave states respectively (see
Section~\ref{sec:opren}), gives the right dimensionality to $D$-dimensional
cross sections and widths, i.e. $2-D = -2+2\ep$ and 1, respectively.

\subsection{Digression on the use of $\gamma_5$ in dimensional regularization}
Once the right counting of degrees of freedom is taken into account via
the polarization sums eqs.~(\ref{polsum0},\ref{polsum1},\ref{polsum2}) ,
further potential inconsistencies
which could, a priori, spoil the possibility of using projectors 
on $D$-dimensional amplitudes, are related to the 
the definition of the matrix $\gamma_5$ in arbitrary
dimensions.           
Several prescriptions have been introduced and 
shown to be consistent in specific problems. The most common ones are
the naive dimensional regularization (NDR)~\cite{axialcurrent} and the 
t'Hooft-Veltman dimensional regularization (HVDR) ~\cite{tHooftVeltman}. 

Within the
first prescription,  $\gamma_5^{N}$ is defined by the property that it
anti-commutes with all $\gamma^\mu$ in $D$ dimensions
\be
\{\gamma_5^{N}, \gamma^\mu \} = 0 \, , \qquad \mu=0,1,\cdots, D-1,
\label{ngamma5}
\ee
and it satisfies
\be
(\gamma_5^{N})^2=1 \, .
\ee
while, in contrast to (\ref{ngamma5}), in the t'Hooft-Veltman prescription
it holds
\be
\gamma_5^{HV} \equiv i \gamma^0 \gamma^1 \gamma^2 \gamma^3 \, ,
\label{hvgamma5}
\ee
and 
\ba
\{\gamma_5^{HV}, \gamma^\mu \} &=& 0 \, , \;\;\; \mu=0,1,2,3 \, , \\
\left[ \gamma_5^{HV}, \gamma^\mu  \right] &=& 
 0 \, , \;\;\;  \mu=4, \cdots, D-1 \, .
\ea
Since it is defined explicitly by construction in eq.~(\ref{hvgamma5}), 
$\gamma_5^{HV}$ is unique and well-defined, while to this respect, 
$\gamma_5^{N}$,  is  an ambiguous object.
For example, while the HVDR scheme allows to recover the standard
Adler-Bell-Jackiw anomaly, the NDR does not without additional 
{\it ad hoc} prescriptions.

We do not expect any ambiguity to arise in our case, since in QCD parity is
conserved and since no $\gamma_5$ insertions are present inside the loop
amplitudes we will be evaluating.
Extra care should instead be applied when dealing with, for example,
corrections to 
the  decay  $ b \to {\cal Q}_{c\bar c} \, s$.  
To verify the independence from the scheme,  we did the calculation  
of virtual diagrams contributing to NLO decay widths and
cross sections of $^1S_0^{[1,8]}$ states
in both  in the HVDR and NDR schemes. The 
results were found to coincide exactly, diagram by diagram.  
             
Finally we would like to compare our technique with the ones
available in the literature.
Braaten and Chen~\cite{Braaten96,Braaten97}  
determine the 
short-distance coefficients via the matching technique,  
generalizing to $D-1$ spatial dimensions their threshold expansion method. 
This technique is briefly
presented in Appendix \ref{appc} and has been used in this paper as a 
check for the virtual contributions in the NLO calculations.
Within this approach a quantity that is closely related to the
cross section for the production of a $\QQ$ pair with total
momentum $P$ is calculated using perturbation theory in full QCD and
expanded in powers of the relative 3-momentum ${\bf q}$ of the $\QQ$
pair and in a few  non relativistic matrix elements. To 
exploit this procedure one is forced to choose a representation for
$\gamma$-matrices. Braaten and Chen write
\begin{equation}
\gamma^0  \;=\; 
\left( \begin{array}{cc} 
        1 &  0  \\ 
        0 & -1 
        \end{array} \right) \;, 
\qquad
\gamma^i \;=\;  
\left( \begin{array}{cc} 
             0    & \sigma^i \\ 
        -\sigma^i &    0        
        \end{array} \right) \;, \quad i=1,\dots D-1 \;.
\label{gammas}                                   
\end{equation}
Once this choice has been made, one easily realizes
that no room is left for selecting a $\gamma_5$ different
from  
\be
\gamma_5^{BC} \;=\; 
\left( \begin{array}{cc} 
        0 & 1  \\ 
        1 & 0 
        \end{array} \right) \;,
\label{bcgamma5}               
\ee
which, as a simple calculation shows, satisfies eq. (\ref{ngamma5}).
This observation shows that in  
the $D$-dimensional threshold expansion method  
an implicit choice for dealing with $\gamma_5$ is made and it
exactly corresponds to $\gamma_5^{BC}=\gamma_5^{N}$.

\section{General description of the calculation of higher-order corrections}
\label{sec:general}
In this section we briefly describe the strategy of the calculation of 
higher-order corrections to decay widths and to total 
production cross sections. The
reader who is not interested in the details of the calculation, can just read
this section to get an idea of the general framework, and can skip the
following sections where all the required calculations are carried out. 
                                                                        
A consistent calculation of higher-order corrections entails the evaluation of
the real and virtual emission diagrams, carried out in $D$ dimensions. The UV
divergences present in the virtual diagrams are removed by the standard
renormalization. The IR divergences appearing after the
integration over the phase space of the emitted parton are cancelled by
similar divergences present in the virtual corrections, or by higher order
corrections to the long-distance matrix elements~\cite{bbl}.
Collinear divergences, finally, are either cancelled by similar divergences
in the virtual corrections or, in the case of production, by factorization
into the NLO parton densities. 
The evaluation of the real emission matrix elements in $D$ dimensions 
is usually particularly complex, and is presumably the main reason that has
prevented the so far the calculation of NLO corrections to the production of
$P$-wave states. In this paper we follow an approach already employed
in~\cite{Mele91}, whereby the structure of soft and collinear singularities in
$D$ dimensions is extracted by using universal factorization properties of the
amplitudes. Thanks to these factorization properties, that will be discussed in
detail in the following section, 
the residues of all IR and collinear poles in $D$ dimensions
can be obtained without an explicit calculation of the full
$D$-dimensional real matrix elements. They only require, in general,
knowledge of the $D$-dimensional Born-level amplitudes, a much simpler task.
The isolation of these residues allows to
carry out the complete cancellations of the relative poles in $D$ dimensions,
leaving residual finite expressions which can then be evaluated exactly
directly in $D=4$ dimensions. In this way one can avoid the calculation of the
full $D$-dimensional real-emission matrix elements. 
Furthermore, the four-dimensional real matrix elements that will be required
have been known in the literature for quite some time~\cite{csm,Cho96}.

\section{Soft emission behaviour}
\label{sec:soft}
We discuss in this section the factorization properties of the real-emission
amplitudes in the soft-emission limit. The factorization formulae 
presented here will be used in the following sections to isolate       
the IR poles and cancel them against the singularities of the virtual
processes.
\begin{figure}[t]
\begin{center}
\epsfig{file=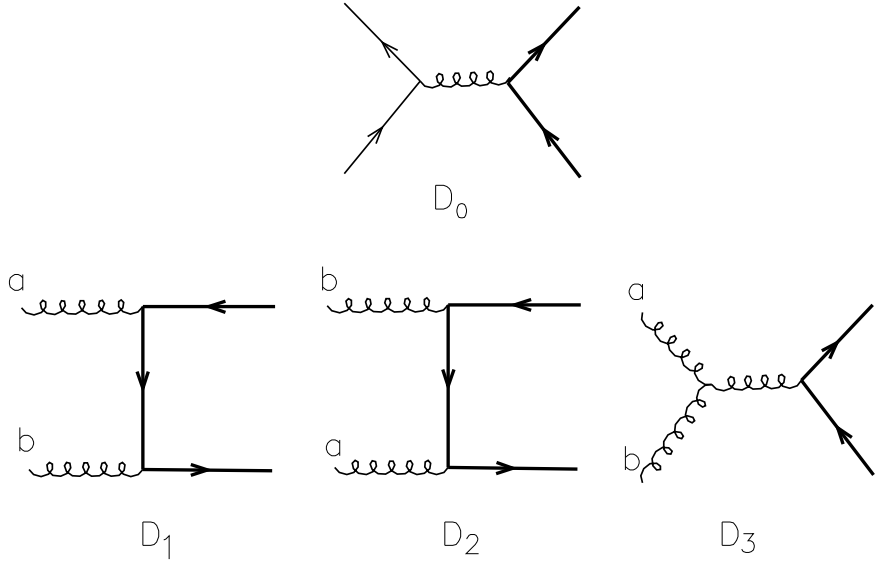,width=12cm,clip=}
\parbox{12cm}{
\caption{\label{fig:ggborn} 
\small Diagrams for the $\qq$ and $gg$ Born amplitudes.}}
\end{center}                                 
\end{figure}

\subsection{Soft factorization in $3g$ amplitudes}
We start by considering the case of decays into gluons. At the Born level, the
relevant diagrams are shown in fig.~\ref{fig:ggborn}. The decay amplitude
(before projection on a specific quarkonium state) can
be written as follows:                             
\be
  A_{\rm Born} \= (ab)_{ij} \; (D_1+D_3) \; + \; (ba)_{ij} \; (D_2-D_3) \; ,
\ee           
where we introduced the short-hand notation:
\be                     
      (a\dots b)_{ij} \= \left( T_a \dots T_b \right)_{ij} \; ,
\ee   
where the $T^a$ matrices are normalized according to
$\tr(T^a T^b)= \delta^{ab}/2$ (more details on our conventions for the colour
algebra can be found in Appendix~\ref{appA}).
The terms $D_1$, $D_2$ and $D_3$ correspond to the three diagrams appearing in
fig.~\ref{fig:ggborn} with the colour coefficients removed. Using this
notation, the amplitude for emission of a soft gluon with momentum $k$ and
colour label $c$ can be written as follows~\cite{Bassetto83}:
\ba                                                                   
    A_{\rm soft} &=&  
   g(cab)_{ij} \left[ \eik{Q} - \eik{a} \right]  (D_1+D_3) +
   g(acb)_{ij} \left[ \eik{a} - \eik{b} \right]  (D_1+D_3) 
  \nn \\  &+&
   g(abc)_{ij} \left[ \eik{b} - \eik{\overline{Q}} \right]  (D_1+D_3) +
   g(cba)_{ij} \left[ \eik{Q} - \eik{b} \right]  (D_2-D_3) 
  \nn \\  &+&
   g(bca)_{ij} \left[ \eik{b} - \eik{a} \right]  (D_2-D_3) +
   g(bac)_{ij} \left[ \eik{a} - \eik{\overline{Q}} \right]  (D_2-D_3) \; ,
\ea                                                                      
where $Q$ and $\overline{Q}$ are the momenta of the heavy quarks, and $a$, $b$
indicate the momenta and colour labels of the final state gluons.

In the case of $S$-wave decay we can set $Q=\overline{Q}=P/2$. All terms
proportional to $\sp{Q}\epsilon_c~$ or to $\sp{\overline{Q}}\epsilon_c~$ 
vanish in the transverse gauge defined by the gauge vector $(k_0,-\vec{k})$. 
The soft-emission amplitude becomes:
\ba                                                         
    A_{\rm soft} &=&
         g \left[ ([a,c]b)_{ij} \, \eik{a} - (a[c,b])_{ij} \, \eik{b} \right]
         (D_1+D_3)                                      
  \nn \\  &+&
         g \left[ ([b,c]a)_{ij} \, \eik{b} - (b[c,a])_{ij} \, \eik{a} \right]  
         (D_2-D_3) \; .                                    
\ea                       
In the case of $^3P$-wave decays we also need the derivative of the decay
amplitude with respect to the relative momentum of the quark and antiquark.
In the soft-gluon limit, we obtain:
\ba                                                         
    \left(\frac{dA_{\rm soft}}{dq_{\alpha}} \right)_{q=0} &=&
         g \left[ ([a,c]b)_{ij} \, \eik{a} - (a[c,b])_{ij}\, \eik{b} \right]
            \frac{d(D_1+D_3)}{dq_{\alpha}}                     
  \nn \\  &+&
         g \left[ ([b,c]a)_{ij} \, \eik{b} - (b[c,a])_{ij} \, \eik{a} \right]  
            \frac{d(D_2-D_3)}{dq_{\alpha}} 
  \nn \\  &+& 
   2g\,\frac{\epsilon_c^{\alpha}}{ \sp{P}{k} } \left\{
   \left[ (abc)_{ij} + (cab)_{ij} \right] (D_1+D_3)+                  
   \left[ (bac)_{ij} + (cba)_{ij} \right] (D_2-D_3) \right\}
\ea                                                                      
It is easy to prove with an explicit calculation
that the two terms proportional to $\epsilon_c^{\alpha}$
give vanishing contributions, in the soft-gluon limit,
when projected on all $^3P_J$ states.
Therefore the structure of the soft limit of the derivative of the
amplitude is similar to that of the amplitude itself.      

We now proceed to prove explicitly the factorization of the soft-gluon
emission probability. We will indicate with $\overline{D}_i$ ($i=1,2,3$)
the projection of the diagrams $D_i$ on a given quarkonium state, and 
with $\overline{A}_{\rm Born}$ and $\overline{A}_{\rm soft}$ the projected
amplitudes for the Born and soft-emission processes, respectively.

The colour coefficients for the decay of a colour-singlet state can be obtained
by using the projector $\delta_{ij}\delta_{kl}/\nc$. The colour-singlet
amplitudes therefore become:
\ba                                                         
    \overline{\aoneb} &=&
               \frac{1}{2\nc}\delta_{ab}\delta_{ij}
               (\overline{D}_1+\overline{D}_2)  \; ,  \\
    \overline{\aones} &=&                   
          \frac{i g f^{acb}}{2\nc}\delta_{ij} \left[ \eik{a} -\eik{b} \right]
               (\overline{D}_1+\overline{D}_2)  \; ,                       
\ea                        
and simple colour algebra gives the final result for the soft-emission:
\be  \label{eq:csfact}
    \sum_{\rm col, pol} \; \vert\overline{\aones}\vert^2 \=
    \frac{2\sp{a}{b} }{(\sp{a}{k} ) (\sp{b}{k} )} \ca g^2
    \sum_{\rm col, pol} \; \vert\overline{\aoneb}\vert^2 
\ee                                              

The colour-octet case can be obtained from the following relation:
\be                                                               
    \sum_{\rm col, pol} \; \vert\overline{\aeight}\vert^2 \=
    \sum_{\rm col, pol} \; \vert\overline{A}\vert^2  \; - \;
    \sum_{\rm col, pol} \; \vert\overline{\aone}\vert^2
\ee                                                      
The first term on the right-hand side can be easily obtained with standard
colour algebra, and the final result is:
\be  \label{eq:octfact}
    \sum_{\rm col, pol} \; \vert\overline{\aeights}\vert^2 \=
     \ca g^2 \left[\frac{2\sp{a}{b} }{(\sp{a}{k} )(\sp{b}{k} )} -
      \frac{2\sp{a}{b} }{(\sp{P}{k} )^2 } \right]\sum_{\rm col, pol} \; \vert\overline{\aeightb}\vert^2 
\ee                                              

\subsection{Soft factorization in light-quark processes}
\label{sec:lightquark}
An approach similar to that used in the previous subsection can be applied
to the case of soft-gluon emission in processes where the $\QQ$ pair $\q$ 
decays into light-quark pairs ($\q \to \qq $).
The Born-level amplitude (which is non-vanishing only in the colour-octet case)
is given by:
\be      
  A_{\rm Born} \= \frac{1}{2} \left[ \delta_{ik}\delta_{jl} - \frac{1}{\nc}
              \delta_{ij}\delta_{kl} \right] D_0 \; ,     
\ee                             
where $(i,j)$ and $(k,l)$ are the color indices of heavy and light quark pairs,
respectively, and where $D_0$ is shown in fig.~\ref{fig:ggborn}, 
with the colour factors
removed and before projection onto the quarkonium quantum numbers. 
The amplitude for emission of a soft gluon with momentum $k$ and
colour label $c$ can be written as follows:
\ba                                                         
    A_{\rm soft} \= g D_0 \!\!\!\!\!\!\! && \left\{ 
   \frac{T^{c}_{ik}}{2}\delta_{jl}  
        \left[\eik{Q} - \eik{q} \right] \; + \;
   \frac{T^{c}_{jl}}{2}\delta_{ik}
        \left[\eik{\overline{q}} - \eik{\overline{Q} } \right]  \right.
  \nn \\  && - \left.
   \frac{T^{c}_{ij}}{2\nc} \delta_{kl}  
        \left[\eik{Q} - \eik{\overline{Q} } \right]
\; + \; 
   \frac{T^{c}_{kl}}{2\nc} \delta_{ij}  
   \left[\eik{q} - \eik{\overline{q} } \right]
   \right\} \; .                                                       
\ea        
The colour-singlet and colour-octet projections are then given by:
\ba                                                              
    \asins &=&  \frac{g D_0}{2\nc}
   T^{c}_{ij} \delta_{lk}  \left[\eik{Q} - \eik{\overline{Q} } \right]
   \\                                                      
    \aeights &=&      A_{\rm soft} -    \asins 
   \; .                          
\ea        

We study first the case of $S$-wave decays.
As before, we can set to zero all
terms proportional to $\sp{Q}\epsilon_c~$ and $\sp{\overline{Q}}\epsilon_c~$
by choosing a transverse gauge.
The colour-singlet amplitude obviously vanishes, and the colour-octet one
reduces itself to:                     
\be                                                         
    A_{\rm soft} \= \frac{g D_0}{2} \left\{
   \left[ T^{c}_{ik}\delta_{jl} -    \frac{1}{\nc} T^{c}_{kl}\delta_{ij} \right]
   \eik{q} + 
   \left[ T^{c}_{lj}\delta_{ik} -    \frac{1}{\nc} T^{c}_{kl}\delta_{ij} \right]
   \eik{\overline{q}} \right\} \; .
\ee                                    
An explicit calculation gives the following factorization formula, which is
non-vanishing only in the $^3S_1$ case since the $^1S_0$ Born amplitude
vanishes:
\be  \label{eq:coqSfact}                                          
    \sum_{\rm col, pol} \; \vert\overline{\aeights}\vert^2 \=
    g^2\; \sum_{\rm col, pol}\vert\overline{\aeightb}\vert^2 
    \left[ \frac{2\sp{q}{\bar q} }{(\sp{q}{k} )(\sp{\bar q}{k} )} \cf
         - \frac{2M^2}{(\sp{P}{k} )^2} \frac{\ca}{2} \right]\, ,
\ee                                                         
where $M=2 m $.

In the case of $P$ waves we need to consider as well the derivative with
respect to the relative momentum of the heavy quarks:
\be                                                         
    \left(\frac{dA_{\rm soft}}{dq_{\alpha}} \right)_{q=0}\=
            g D_0 \left(
      \frac{\epsilon_c^{\alpha}}{\sp{P}{k} }
     - \frac{k^{\alpha} }{(Pk)^2} (P \epsilon_c) \right)
   \left[ T^{c}_{ik}\delta_{jl} +T^{c}_{lj}\delta_{ik}  
  -    \frac{2}{\nc} T^{c}_{ij}\delta_{kl} \right]  \;.
\ee
The terms proportional to $dD_0/dq_{\alpha}$ vanish because $D_0$ corresponds to
an $S$-wave process. Denoting by $\cal{C}$ the overall colour coefficient in
the previous equation,
choosing a transverse gauge where $\epsilon_c \cdot P=0$
and using the projection operators for $^3P_J$-waves given in 
section~\ref{sec:projectors}, 
we obtain the following result for the quarkonium-decay amplitude:
\be  \label{eq:54}
     \overline{A_{\rm soft}}(^3P_J) 
      \= g\frac{1}{Pk} {\cal C} \;
      \overline{A_{\rm Born}}     
      (^3S_1(\epsilon_{\rm eff})) \;.
\ee                                                                
The amplitude was written in terms of the amplitude for the production of a
$^3S_1$ colour-octet 
state, with an {\em effective} polarization $\epsilon_{\rm eff}$
given in terms of the polarizations of the 
soft gluon and of the $^3P_J$ state as follows:
\be
   \epsilon^{J}_{{\rm eff}, \beta} \=
   \epsilon_c^{\alpha} {\cal E}_{\alpha\beta}^{J}(P)  \; .
\ee                                                        
The correlations between the polarization of the gluon and of the ${\cal Q}$
state induce a dependence of the soft-emission amplitude on the relative    
direction of the gluon and of the light quarks. 
Nevertheless, if we average over the soft gluon $D-1$ spatial 
directions we get:
\be
 \int \frac{d\Omega^{D-1}_k}{\Omega^{D-1}}
 \sum_{\rm pol} \epsilon_c^{\alpha} \epsilon_c^{\beta *}= 
  \frac{D-2}{D-1} \Pi_{\alpha\beta}(P) \; ,
\ee                                        
and one can easily compute the sum over {\em effective} polarizations   
\be
\int \frac{d\Omega^{D-1}_k}{\Omega^{D-1}}
    \sum_{\epsilon_c}  \epsilon^{J}_{{\rm eff}, \alpha} 
                       \epsilon^{J *}_{{\rm eff}, \beta}
        = -  N_J \frac{D-2}{D-1} \Pi_{\alpha\beta}(P) \; .
\ee                                                       
Here $N_J$ is the number of degrees of freedom of the $^3P_J$ state in $D$
dimensions. 
With the above expressions at hand, it is straightforward to 
square the amplitude in eq.~(\ref{eq:54}) and obtain
\be
\label{eq:coqPfact}           
\int d\Omega^{D-1}_k
     \overline{\sum}\vert A_{\rm soft}(^3P_J) \vert^2 =\Omega^{D-1}
     \frac{(D-2)^2}{(D-1)^2} 
\overline{\sum_{\rm colours}} \vert {\cal C} \vert^2 
\frac{4g^2 }{ (\sp{P}{k} )^2 }
\, N_J \, \overline{\sum} \vert A_{\rm Born}(^3S_1^{[8]})\vert^2 \,  \, ,   
\ee

The above formula embodies the well known singular behaviour of the $P$-wave
decay into light quarks~\cite{BarbieriIR}, 
to be absorbed into the colour-octet $^3S_1$ NRQCD 
matrix element, as shown in ref.~\cite{bbl} and 
discussed in Section~\ref{sec:opren}. 
This formula also shows that this phenomenon is present not
only for colour-singlet $P$ wave states, but for colour octet ones as well.
The colour factors (averaged over initial states) are given by:
\ba
\overline{\sum_{\rm colours}} \vert {{\cal C}^{[1]}} \vert^2  
     &=& \cf  \; , \\
\overline{\sum_{\rm colours}} \vert {{\cal C}^{[8]}} \vert^2  
     &=& \Bf  \; , 
\ea                                                  
for the colour-singlet and colour-octet cases, respectively. $\cf$ and $\Bf$
are defined in Appendix \ref{appA}.

\section{Decay Processes at NLO}        
\label{sec:decay}
The previous study of the soft behaviour of the real-emission amplitudes
will now be used to calculate the full set of NLO corrections to decay and
production of quarkonium states. 
The general factorization formulae that we proved in the previous section will
be useful to efficiently extract the singular behaviour of the
real-emission amplitudes, using the general formalism of
refs.~\cite{Mele91}. In conjunction with the universal
behaviour of amplitudes in the collinear regions, the knowledge of the residues
of the soft singularities will enable us to extract all poles of infrared and
collinear origin, cancel them against the poles in the virtual amplitudes and
in the parton densities, and be left with finite terms which only depend on the 
real-emission matrix elements in 4 dimensions. This technique 
saves us from the complex evaluation of the full matrix elements for real
emission in $D$ dimensions. In this section we confine ourselves to the case of
decay rates. We will discuss the production cross-sections in Section
\ref{sec:production}.

\subsection{Real emission corrections in $D$ dimensions}
\subsubsection{Kinematics and factorization of soft and collinear singularities}
Let us consider the three-body decay processes $\q^{[1,8]} \to k_1 + k_2 + k_3$, 
where $\q^{[1,8]}\equiv\QQ[\spectr^{[1,8]}]$. 
They are described in terms of the invariants                                        
\be
       x_i = \frac{2P k_i}{M^2} \, , \quad\quad  \sum x_i=2 \, ,
\ee                                                             
where $P$ is the momentum of the decaying $\QQ$ pair of mass $M= 2 m $.
The three-body phase space in $D=4-2\eps$ dimensions is given by:
\ba                 
      d\Phi_{(3)} &=& \frac{M^2}{2(4\pi)^3} \;
       \left(\frac{4\pi}{M^2}\right)^{2\eps}
      \; \frac{1}{\Gamma(2-2\eps)} \;
      \prod_{i=1}^{3}\,(1-x_i)^{-\eps} \,dx_i
      \delta(2-\sum x_i) \nn \\                 
      &=& \Phi_{(2)} \frac{N}{K}
      \prod_{i=1}^{3}\,(1-x_i)^{-\eps} \,dx_i \delta(2-\sum x_i) \; ,
\ea                                           
where $\Phi_{(2)}$ is the total two-body phase space  in $D$ dimensions:
\be                                                                     
      \Phi_{(2)} \= \frac{1}{8\pi }
       \left(\frac{4\pi}{M^2}\right)^{\eps}
      \; \frac{\Gamma(1-\eps)}{\Gamma(2-2\eps)}\label{phitwo} \; ,
\ee   
and $N$ and $K$ are defined as
\ba         
    N &=& \frac{M^2}{(4\pi)^2} \; \left( \frac{4\pi}{M^2} \right) ^{\eps}
      \; \Gamma(1+\eps) \; , \\                                             
    K &=& \Gamma(1+\eps)\Gamma(1-\eps) \sim 1+\eps^2\frac{\pi^2}{6} \;.
\ea
For future reference, it is useful to define as well the following function:
\be \label{eq:feps}                                  
    \feps{M^2} \; = \; 
    \left( \frac{4\pi\mu^2}{M^2} \right) ^{\eps}
      \; \Gamma(1+\eps) \; ,     
\ee
where $\mu$ is the renormalization scale. 

Collinear and soft singularities in the matrix element squared 
\be
{\cal{M}}(x_1,x_2,x_3) \= \overline{\sum}\vert A(x_1,x_2,x_3)
     \vert ^2                                              
\ee
appear as single poles in either of the terms $1-x_i$. As a consequence, the
function 
\be
     f(x_1,x_2,x_3) \= \prod_{i=1}^{3}\,(1-x_i) {\cal{M}}(x_1,x_2,x_3)
\ee                 
will be finite throughout the phase space.
In terms of the function $f$, the differential decay width can be written as:
\be                                            
     d\Gamma \= \C
   \prod_{i=1}^{3}\,(1-x_i)^{-1-\eps} \,dx_i \, \delta(2-\sum x_i) \,
                 f(x_1,x_2,x_3) \; .                                 
\ee                                 
It is useful to introduce the variables $(x,y,z)$,
defined by:                                                     
\ba
  x_1 &=& 1 - x  y \\
  x_2 &=& 1 - x (1-y) \\
  x_3 &=& z
\ea
With these variables
\ba
&&   \prod_{i=1}^{3}\,(1-x_i) \=
   x^2\, (1-x) \, y \, (1-y) \; , \\
&&   \prod_{i=1}^{3}\, dx_i \, \delta(2-\sum x_i) \=
   x dx \, dy  \;,             
\ea               
with $z=x$.
The total decay width can then be written as:
\be                                    \label{eq:gamma}
  \Gamma \= \C \, \frac{1}{{\cal{S}}} \int_0^1 d\, x \int_0^1 d\, y \;
         x^{-1-2\eps} \, (1-x)^{-1-\eps} \, \left[ y(1-y) \right]^{-1-\eps}
         f(x,y) \,                       
\ee               
where $\cal{S}$ is a symmetry factor, equal to $3!$ in the case of 3-gluon
decays, and equal to 1 in the case of $g\qq$ decays. The function
\be
f(x,y)\equiv f(x_1=1-xy,x_2=1-x+xy,x_3=x)
\ee          
is finite for all values of $x$ and $y$ within the integration domain.
Therefore all singularities of the total decay rate can be easily extracted by
isolating  the $\eps\to 0$ poles from the factors in eq.(\ref{eq:gamma})
explicitly depending on $x$ and $y$, as discussed in the next subsection.
                                                                         
\subsubsection{Colour-singlet three-gluon decays}
For the sake of definiteness, we present now the detailed calculations
relative to the decay of a colour singlet state into three gluons. 
The  colour octet can be worked out
in a similar manner, and will be given with fewer details in the
following.

To exploit the total symmetry of the three gluons
in the final state we restrict the integration to the domain
$ x_3< x_1, x_2 $. In terms of the variables $x$ and $y$, this
corresponds to                
\be 
0 < x < \frac{2}{3} \quad , \quad 2-\frac{1}{x} < y < \frac{1}{x}-1
\;.
\ee                      
This choice corresponds to integrating over one third of the entire
phase space, multiplying the result by the 
multiplicity factor 3.            
To proceed, we make use of the following expansions valid for small $\eps$:
\be                                                                        
     x^{-1-2\eps} \= -\frac{4^\eps}{2\eps} \delta(x) +
        \left( \frac{1}{x} \right)_{1/2} \, - \,
    2\eps \left( \frac{\log\, x }{x}\right)_{1/2} \, + \, {\cal{O}}(\eps^2) \, .
\ee                                               
where we introduced the ``1/2'' distributions defined as follows:
\be
   \int_{0}^{1/2} \, dx \, \left[d(x)\right]_{1/2} t(x) \=
   \int_{0}^{1/2} \, dx \; d(x) \;\left[t(x)-t(0)\right] \; .
\ee                                                   
We then obtain:
\be      
    x^{-1-2\eps} \, (1-x)^{-\eps} = -\frac{4^\eps}{2\eps} 
     \delta(x)  +                                         
\left(\frac{1}{x}\right)_{1/2} -\eps \left\{ \frac{\log(1-x)}{x} + 
       2\left[\frac{\log\, x}{x}\right]_{1/2}
          \right\} \, + \, {\cal O}(\eps^2)  
\ee

The total decay width can then be written as the sum
of three terms:
\be
      \Gamma \= \Gamma_{x=0} + \Gamma_{0<x<1/2} + \Gamma_{1/2<x<2/3}\; , 
\ee                                                                
where
\ba
  \Gamma_{x=0} &=& -\frac{4^\eps}{2\eps} 
         \C \, \frac{3}{{\cal{S}}} \int_0^1 d\, y \;
         \, \left[ y(1-y) \right]^{-1-\eps}
         f(0,y)\\[10pt]
  \Gamma_{0<x<1/2} &=& \C \, \frac{3}{{\cal{S}}} 
         \int_0^{1/2} d\, x \int_0^1 d\, y \;
         \, \left[ y(1-y) \right]^{-1-\eps}
         \frac{f(x,y)}{(1-x)}   \nonumber \\
         && \times \, \left[
       \left(\frac{1}{x}\right)_{1/2} -\eps \left\{ \frac{\log(1-x)}{x} + 
       2\left[\frac{\log\, x}{x}\right]_{1/2} \right\} \right]\, \\[10pt]
  \Gamma_{1/2<x<2/3} &=& \C \, \frac{3}{{\cal{S}}} 
       \int_{1/2}^{2/3} d\, x \int_{2-1/x}^{1/x-1} d\, y \;
         \frac{1}{x(1-x)y(1-y)} 
         f(x,y) \, .
\ea               
Notice that the last term is finite, since no pole is present within the
integration domain. It 
can therefore be computed directly in 4-dimensions.

We shall now proceed to the evaluation of each of these terms, starting from
the first one.                    
The $x\to 0$ limit is the soft limit for $k_3$. In this limit,
eq.(\ref{eq:csfact}) gives:
\be                                       
{\cal{M}}(x,y) \; \stackrel{x\to 0}{\longrightarrow} \;
 \frac{16\pi \ca \asb}{M^2} \, \frac{1}{x^2 y (1-y)} \, {\cal{M}}_{\rm Born} \;,
\ee                  
where $\asb=\as\mu^{2\eps}$ is the dimensional coupling constant in $D$
dimensions and $\mborn$ is the Born amplitude squared and
averaged in $D$ dimensions for the $\q \to gg$ processes. As a result,               
\ba                     
  \Gamma_{x=0} & \equiv& -\frac{4^\eps}{2\eps} \Cggg2 \int\, dy \, 
  \left[ y (1-y)
  \right]^{-1-\eps} f(x=0,y) \\                     
  &=&  \, \Cggg2 \, \frac{16\pi \ca \asb}{M^2}     \,
  \mborn \, \left(\frac{1}{\eps^2}  \;+\frac{2}{\eps} \log\, 2  +\,
  2 \log^2 \, 2 -\,\frac{\pi^2}{3} \right) \; .       
\ea                
In the cases in which $\mborn=0$, such as for example $\q =\threePone$,
$\mborn=0$ and the left-hand side of the above equation, together with
$\Gamma_{x=0}$, vanish. 

In order to evaluate $\Gamma_{0<x<1/2}$, we use the
following relation:                                            
\be
  \left[y(1-y)\right]^{-1-\eps} \= -\frac{1}{\eps} \left[ \delta(y)+\delta(1-y)
  \right] + \left[ \left( \frac{1}{y}\right)_+ + \left( \frac{1}{1-y}\right)_+
  \right] \, + \, {\cal{O}}(\eps)
\ee                              
With obvious notation, we can therefore decompose $\Gamma_{0<x<1/2}$ as
follows:                                                   
\be
  \Gamma_{0<x<1/2} \= \Gamma_{y=0} + \Gamma_{y=1} + \Gamma_{\rm{finite}} \; .
\ee                                                                         
The limits $y\to 0$ and $y\to 1$ correspond to the collinear limits
$k_2\parallel k_3$ and $k_1\parallel k_3$, respectively. 
In these limits we can apply  the universal factorization of collinear
singularities
\be          
      f(x,y=0) \= f(x,y=1) \=   
      \frac{8 \pi \asb}{M^2} \; \pgg{x} \; \mborn \, x(1-x) \; ,
\ee                                                         
where $ \pgg{x} $  is the $D$-dimensional 
Altarelli-Parisi splitting kernel:
\be                                                               
  \pgg{x} \= 2\ca \left[ \frac{x}{1-x} + \frac{1-x}{x} + x(1-x) \right] \; ,
\ee                                                           
and obtain the following result:
\ba
  \Gamma_{y=0} \= \Gamma_{y=1} &=& \frac{16\pi\asb\ca}{M^2} 
         \Cggg2 \mborn  \nn\\
      &&\times \left[ \frac{11}{12\eps} - \frac{1}{\eps} \log \,2+
             \frac{45}{16} -\frac{\pi^2}{4}+ \frac{11}{12} \log \, 2 -
              \log^2 \, 2 \, \right] \; .
\ea                                        
We can now collect all the results obtained so far in the following expression
\be
  \Gamma(\q \to 3g) \= \Gamma^{[1]}_{\rm{univ}} + 
                          \Gamma_{\rm{finite}} \; ,
\ee                                              
where 
\ba
  \Gamma^{[1]}_{\rm{univ}} &\equiv & \left[ \Gamma_{x=0}+
                              \Gamma_{y=0}+\Gamma_{y=1}\right]   \nn \\
  &=&
      \frac{\ca\as}{\pi} \, \gborn \,
      \feps{M^2} \,
      \left( \frac{1}{\eps^2} + \frac{11}{6\eps} +\frac{45}{8}    
    -\frac{5}{6}\pi^2+\frac{11}{6} \log\, 2 \right) 
\ea                                                 
is a universal term, whose dependence on the quarkonium type is only implicit
in the factor $\gborn$, and where                                            
\be
    \Gamma_{\rm{finite}} \= \Cggg2 
    \int_0^{\frac{1}{2}} \, dx \, \int_0^1 \, dy \, f(x,y)
  \left[ \left( \frac{1}{y}\right)_+ + \left( \frac{1}{1-y}\right)_+\right]
         \left( \frac{1}{x}\right)_{\frac{1}{2}} + \; \Gamma_{1/2<x<2/3}
\ee                                    
is a finite expression, explicitly dependent on the quarkonium state through
the specific form of the function $f(x,y)$. In particular, this is the only
term which survives in cases where $\mborn=0$, consistently with the form taken
by eq.(\ref{eq:gamma}) in the limit $\eps\to 0$.
Since $\Gamma_{\rm{finite}}$ is free of soft and collinear singularities, it
can be evaluated directly in four dimensions. In particular, this implies that
the matrix elements for the real process $\q \to ggg$ can be calculated in
$D=4$ dimensions, with a significant reduction in the complexity of the
calculation. The evaluation of the $\Gamma_{\rm{finite}}$ contributions is
a tedious but straightforward calculation, which makes use of the 4-dimensional
matrix elements given in ref.~\cite{csm} for the colour-singlet states.
The explicit evaluation leads to the following result 
(valid for $\gborn\neq 0$):
\be                                                                            
  \Gamma(\q ^{[1]} \to ggg) \=
      \frac{\ca\as}{\pi} \Gamma_{\rm Born}(\q^{[1]} \to gg)
      \feps{M^2}           
     \left( \frac{1}{\eps^2} + \frac{11}{6\eps} +
     A^{[1]}_{\q}\right) \; ,                   
\ee              
with
\ba
  \label{eq:a1finite}
  &&   A^{[1]}_{\oneSzero} \= \frac{181}{18} -\frac{23}{24}\pi^2 \; , \\
  &&   A^{[1]}_{\threePzero} \= \frac{400}{81} -\frac{61}{144}\pi^2 \; , \\
  &&   A^{[1]}_{\threePtwo} \= \frac{2113}{216} -\frac{401}{384}\pi^2 \; . 
\ea                                                            
The processes $\threeSone\to ggg$ and $\threePone\to ggg$  are completely 
finite. In absence of real or virtual contributions from two-gluon final
states, they provide the LO contribution to the gluonic decay rates:
\ba                                                     
 \Gamma^H (\threeSone^{[1]} \to ggg) &=&
      4\,\ascube\, C_2(F)
       \left(-1+ \frac{\pi^2}{9} \right)
      {{\langle H|\o_1(\threeSone)|H\rangle}\over{m^2}} \\
  \label{eq:anfinite}
 \Gamma^H (\threePone^{[1]} \to ggg) &=&
      \ascube\,  
    \ca\,\cf \left( \frac{587}{27} -\frac{317}{144}\pi^2 \right)
    {{\langle H|\o_1(\threePone)|H\rangle}\over{m^4}}. 
\ea                                                            
These results agree with previous evaluations (see e.g. 
Schuler~\cite{Schuler94}).

We would like to point out an interesting fact related to the size of the
non-universal finite corrections appearing in eqs.~(\ref{eq:a1finite}) through
(\ref{eq:anfinite}):
although the size of the contributions from the rational number and from the
term proportional to $\pi^2$ are large in each case, 
there is always a cancellation 
between them which reduces their sum to about 1/10 of their individual value.
The same accurate cancellations take place in the processes that will be   
considered in the following subsections.   
In all cases the finite coefficient $A^{[1,8]}_{\q}$ is a number of order 1.

\subsubsection{Colour-octet three-gluon decays}                          
The case of colour-octet decays can be analysed in a similar way, the only
difference being in the soft limit of the amplitude squared, given by
eq.(\ref{eq:octfact}):
\be                                       
{\cal{M}}(x,y) \; \stackrel{x\to 0}{\longrightarrow} \;
 \frac{16\pi \ca \asb}{M^2} \, 
   \frac{1-y (1-y)}{x^2 y (1-y)} \, {\cal{M}}_{\rm Born} \;,
\ee             
where $\mborn$ is the Born amplitude squared and
averaged in $D$ dimensions for the $\q^{[8]}\to gg$ processes.
Following the same steps as above, we get:
\be
  \Gamma(\q^{[8]}\to 3g) \= \Gamma^{[8]}_{\rm{univ}} + 
                          \Gamma_{\rm{finite}} \; ,
\ee                                              
where 
\be
  \Gamma^{[8]}_{\rm{univ}} 
  \=
      \frac{\ca\as}{\pi} \, \gborn \,\feps{M^2}
      \left( \frac{1}{\eps^2} + \frac{7}{3\eps} +\frac{53}{8}    
    -\frac{5}{6}\pi^2+\frac{17}{6} \log\, 2 \right)\, ,
\ee
is the universal term for colour octet decays. Once the finite part is
calculated, making use of the 4-dimensional amplitudes derived in
ref.~\cite{Cho96}, we obtain  the following results:
\be                                                                            
  \Gamma(\q ^{[8]} \to ggg) \=
      \frac{\ca\as}{\pi} \gborn(\q^{[8]} \to gg)
     \feps{M^2}
     \left( \frac{1}{\eps^2} + \frac{7}{3\eps} +
     A^{[8]}_{\q }\right) \; ,                   
\ee             
with
\ba
  &&   A^{[8]}_{\oneSzero} \= \frac{104}{9} -\pi^2 \; , \\
  &&   A^{[8]}_{\threePzero} \= \frac{875}{162} -\frac{10}{27}\pi^2 \; , \\
  &&   A^{[8]}_{\threePtwo} \= \frac{4679}{432} -\frac{73}{72}\pi^2 \; . 
\ea                                                            

The processes $\threePone^{[8]}\to ggg$ and $\threeSone^{[8]}\to ggg$  are 
completely finite because of the vanishing of two-gluon amplitudes. They read:
\ba
&&\Gamma^H(\threePone^{[8]} \to ggg) =
      \ascube\;\ca\,\Bf    
     \left( \frac{1369}{54} -\frac{23}{9}\pi^2 \right)
     {{\langle H|\o_8(\threePone)|H\rangle}\over{m^4}}\, ,\\[10pt]
&&\Gamma^H(\threeSone^{[8]} \to ggg) =
      5\;\ascube    
     \left(- \frac{73}{4} +\frac{67}{36}\pi^2 \right)
     {{\langle H|\o_8(\threeSone)|H\rangle}\over{m^2}}\, .
\ea

\subsubsection{Colour singlet $\qq g$ decays}                          
The $\oneSzero^{[1]}\to\qq g$ decay rate can be calculated along the same lines
as above, starting from eq.~(\ref{eq:gamma}). No soft-gluon pole is present for
$x\to 0$, and the collinear singularities are extracted by using the
Altarelli-Parisi factorization. The result is given by:
\ba
\Gamma^H(\oneSzero^{[1]}\to\qq g) = 
\nf\Gamma^H_{\rm Born}(\oneSzero^{[1]}\to gg)                 
\frac{\as}{\pi}\frac{\feps{M^2}}{K}\tf
\left[ -\frac{2}{3\epsilon} -\frac{16}{9}\right] \, .              
\ea
The process $\threeSone^{[1]}\to\qq g$ is equal to zero.
The results for $P$-wave colour-singlet states are given by:
\ba                                         
&&\Gamma^H(\chijs\to q\overline q g)= 
     \nf \Gamma^H_{\rm Born}(\chijs\to g g)
\frac{\as}{\pi}\frac{\feps{M^2}}{K}\tf
\left(-\frac{2}{3\epsilon}\right) \nn\\             
&& -\frac{4}{9} \nf \cf \ascube 
\frac{\mu^{2\ep}\langle H |\o_1(^3P_J)|H\rangle}{m^4}     
\left[\frac{1}{\epsilon} \frac{\feps{M^2}}{K}
\left(\frac{4 \pi \mu^2}{M^2}\right)^\epsilon
\frac{\Gamma(1-\epsilon)}{\Gamma(2-2\epsilon )} + a_J\right]\, 
\label{eq:qqg1}
\ea
where 
\ba
a_0 = \frac{28}{3}\;\; , \;\;a_1 = \frac{7}{3}\;\; , \;\;a_2 = \frac{53}{15}
\label{aa}
\ea
and $f_{\epsilon}$ is defined in Appendix ~\ref{appA}.
                   
In the above  equations the $1/\ep$ poles proportional to $\tf$ 
are due to the gluon splitting into collinear quarks.
They will be cancelled by the virtual corrections to the
two-gluon decay process. 
The singular $J$-independent pieces, proportional to $\cf$, come from the
soft-gluon pole isolated in eq.~(\ref{eq:coqPfact}), and
will be cancelled by the                        
addition of the $\threeSone^{[8]}\to\qq$ contribution to the
decay width (see Section \ref{sec:opren}).

\subsubsection{Colour octet  $\qq g$ decays}
The process $\oneSzero^{[8]}\to \qq g$ is completely analogous to the colour 
singlet one:
\ba
\Gamma^H(\oneSzero^{[8]}\to \qq g) = \nf
\Gamma^H_{\rm Born}(\oneSzero^{[8]}\to gg) 
\frac{\as}{\pi}
\frac{\feps{M^2}}{K}\tf
\left[ -\frac{2}{3\epsilon} -\frac{16}{9}\right]\, .
\ea

The process $\threeSone^{[8]}\to \qq g$ shows collinear and IR singularities 
that are expected to cancel against the virtual ones: 
\ba
\Gamma^H(\threeSone^{[8]}\to q\overline q g) 
&=&\Gamma^H_{\rm Born}(\threeSone^{[8]}
\to q\overline q)\frac{\as}{\pi}\feps{M^2} 
\left\{\cf \left(\frac{1}{\ep^2}+\frac{3}{2\ep}\right) +\ca\frac{1}{2\ep} 
\right.\nn\\ &&+\left.\cf\left(\frac{19}{4}
-\frac{2}{3}\pi^2\right) +\frac{11}{6}\ca\right\}\, .
\label{qqgh}
\ea    
   
The colour-octet $P$-wave decays exhibit the same singularity structure as the
colour-singlet case, and are given by:
\ba                                         
&&\Gamma^H(\chijh\to q\overline q g)= 
     \nf \Gamma^H_{\rm Born}(\chijh\to g g) \frac{\as}{\pi}
\frac{\feps{M^2}}{K} \tf        
\left(-\frac{2}{3\epsilon}\right) \nn\\&& -\frac{4}{9}\nf \Bf \ascube 
\frac{\mu^{2\ep} \langle H |\o_8(^3P_J)|H\rangle}{m^4}
\left[\frac{1}{\epsilon} 
\frac{\feps{M^2}}{K}
\left(\frac{4 \pi\mu^2}{M^2}\right)^\epsilon
\frac{\Gamma(1-\epsilon)}{\Gamma(2-2\epsilon )}  
+ a_J\right]
\label{eq:qqg8}
\ea
and the quantities $a_J$ are defined in (\ref{aa}). $\Bf$ is defined in
Appendix \ref{appA}.

\subsection{Virtual corrections}          
We present in this section the results of the calculation of the 1-loop
diagrams necessary for the evaluation of                            
the virtual corrections to the production and decay matrix elements.
The final results for the decay rates, obtained by combining 
the results of this section with those on real
emission presented in the previous sections, are collected in Appendix
\ref{appNLO}.

\begin{figure}[t]
\begin{center}
\epsfig{file=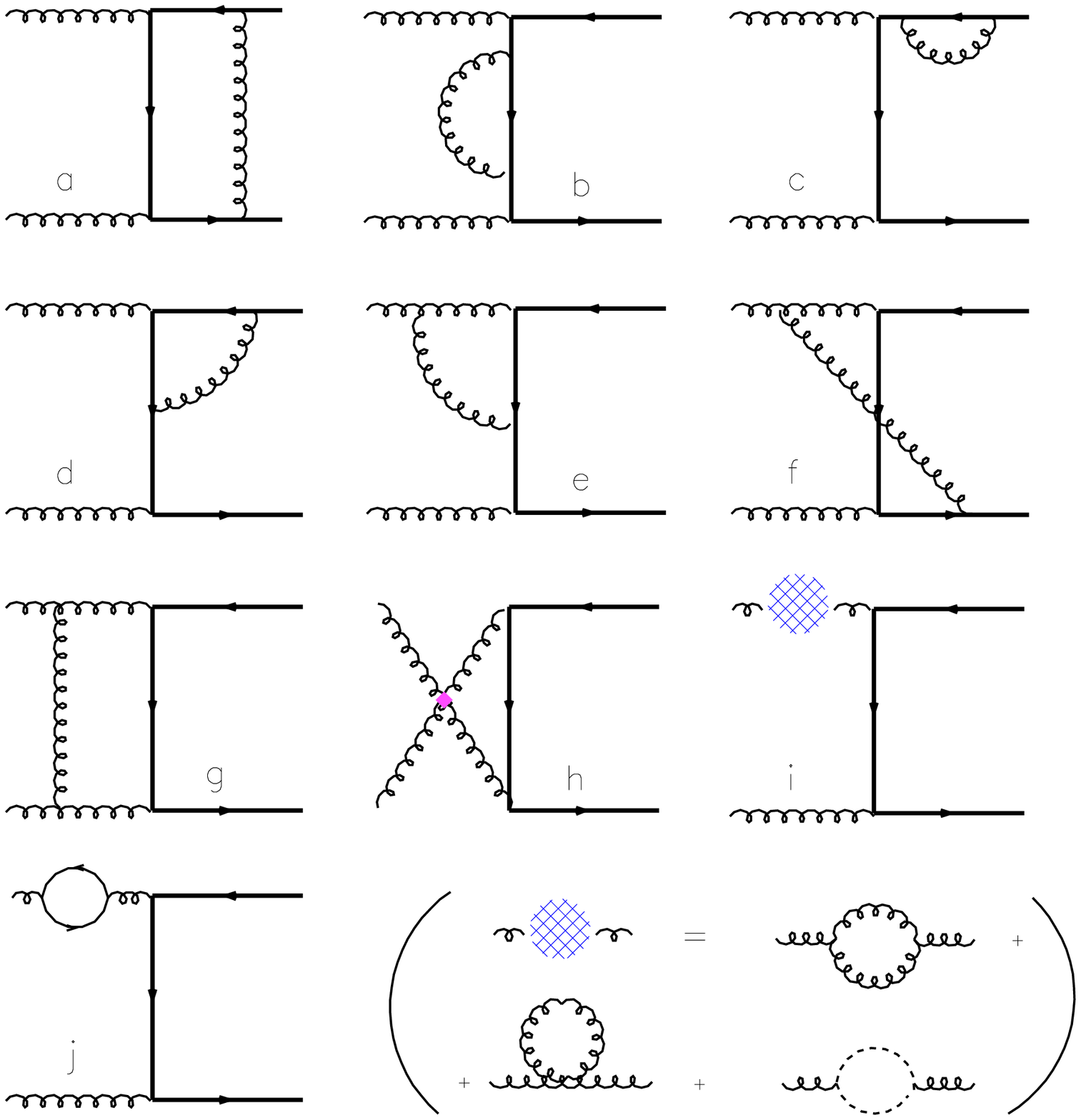,
%             bbllx=10pt,bblly=60pt,bburx=550pt,bbury=770pt,
%              bbllx=30pt,bblly=160pt,bburx=540pt,bbury=660pt,
            width=12cm,clip=}
\ccaption{}{\label{fig:gg} Feynman diagrams contributing to the one-loop 
corrections to the processes $\QQ[^3P_0^{[1,8]}] \to gg$,  
$\QQ[^3P_2^{[1,8]}] \to gg$ and $\QQ[^1S_0^{[1,8]}] \to gg$.
}                                                       
\end{center}
\end{figure}

The explicit calculation of the virtual contribution $\Gamma_{V}$ 
was done using dimensional regularization, both for the UV and the 
IR divergences. 
We have introduced the parameter $v$ defined in the Appendix~\ref{appA} in 
order to regularize the Coulomb singularity. The quantity $v$ represents 
the velocity of the heavy quarks in the quarkonium rest frame, and is kept 
small but finite. The Coulomb singularities appear as poles in the relative 
velocity, i.e. as $1/2v$. 
The relevant Feynman diagrams
are shown  in figures ~\ref{fig:gg} and ~\ref{fig:qq}, 
and the results are given diagram by diagram
in tables~\ref{tab:virtual1S0}, \ref{tab:virtual3P0}, 
\ref{tab:virtual3P2} and \ref{tab:psi8}. 
In these tables we report the contribution of each diagram $k$,
indicating separately the colour factors \fk\                              
relative to the colour singlet and colour octet cases. 

The expressions \dk\ appearing in the tables are defined by the following
equation:
\be
  \Gamma_V \= \gborn
         \frac{\as}{\pi}           
      \feps{M^2} \sum_k \dk\fk \; ,
\ee                      
where the sum extends over the set of diagrams.

The virtual corrections are both infrared and ultraviolet divergent. While the
infrared poles are cancelled when adding the real corrections, the ultraviolet
ones are instead disposed of via coupling constant and heavy quark mass
renormalization. The latter has already been explicitly performed in the 
on-shell scheme, which amounts to replacing the bare mass by
\be
m^{\rm bare} = m\left[1-\frac{3\as}{4\pi}\cf\left(\frac{1}{\eu} - \gamma_E +
\log4\pi + \frac{4}{3}\right)\right]\, .
\ee
The coupling constant renormalization has instead been left undone, to give in
principle anyone the possibility to perform it in a scheme of his own choice.
We will do it in the $\MSB$ scheme through the replacement of the bare coupling
constant by
\be
\asb= \mu^{2\ep}\as\left[1-\frac{\as}{2\pi}b_0
\left(\frac{1}{\eu} - \gamma_E +
\log4\pi\right)\right]\, .
\ee

We point out that we carried out the evaluation of the virtual corrections
using two independent techniques. In addition to using the $D$-dimensional
projection operators constructed in sec.~\ref{sec:projectors}, we also used the
threshold-expansion technique introduced in ref.~\cite{Braaten96,Braaten97},
and reviewed here in  Appendix ~\ref{appc}.
The results obtained in the two cases coincide exactly, diagram by diagram,
providing an important consistency check of our calculation
showing the equivalence of our $D$-dimensional projections to the threshold
expansion method.

\subsubsection{Colour-singlet $gg$ virtual decays} 
The singularity structure of the virtual corrections is dictated by the
renormalization properties of the theory, by the universal form of the Coulomb
limit, and by requirement that soft and collinear singularities cancel against
the real corrections evaluated above. The form of the virtual corrections to
the decay rate is therefore the following:
\ba                         
\label{eq:virtual}
  \Gamma_V^H(\q^{[1]} \to gg) &=& \gbh(\q^{[1]} \to gg)
         \frac{\as}{\pi} \feps{M^2} 
\nn \\                                                               
&&\times                                
    \left\{ \frac{b_0}{\epsuv} + \cf\frac{\pi^2}{2v} -
     \ca \left( \frac{1}{\epsilon^2} + \frac{11}{6\epsilon} \right)
      +                                                        
     \frac{2}{3\epsilon} \nf \tf + B^{[1]}_{\q}\right\}\; ,
\ea                                                      
where we explicitly labelled the $\eps$'s to indicate their origin,
and where all of the state dependence is included in the finite factor 
$B_{\q}$. The heavy quark-antiquark relative velocity $2v$, $b_0$ and 
$f_{\epsilon}(M^2)$ are
defined in Appendix~\ref{appA}. $\nf$ is the number of flavours lighter than
the heavy, bound one, as a consequence of the heavy quark never appearing in
the virtual loops. 
       
{\renewcommand{\arraystretch}{1.8}
\begin{table}[t]
%\scriptsize
\begin{center} 
\begin{tabular}{|l|l|l|l|} \hline
Diag. &${\cal D}_k$ &$f_k^{[1]}$&$f_k^{[8]}$\\ \hline\hline

 a& $ \frac{\pi^2}{2v} +\frac{1}{\ep} -2 +2\log 2   $&$\cf$
&$\caf$\\ \hline
 b& $  -\frac{1}{2 \eu}-1 + 3\log 2 $&\cf &\cf \\ \hline
 c&  $  -\frac{1}{2 \eu}-\frac{1}{\ep}-2-3\log 2 $ &\cf&\cf\\ \hline
 d & $  \frac{1}{ \eu} -2 \log 2 +\frac{\pi^2}{4} $ &\caf&\caf \\ \hline
 e & $ -\frac{3}{\eu}+\frac{1}{\ep^2}+\frac{1}{\ep}- 4+2 \log 2 -\frac{\pi^2}{6}
    $ &$-\frac{1}{2}$\ca &$-\frac{1}{2}$\ca\\ \hline               
 f & $  \frac{1}{\ep^2}+\frac{1}{\ep}
 +2-4\log 2-\frac{5}{12}\pi^2 $&$\frac{1}{2}\ca$ &$0$\\\hline
 g & $  -\frac{1}{\ep^2}-\frac{1}{\ep} -2 +2 \log 2 +\frac{2}{3}\pi^2  $&\ca
    &$\frac{1}{2}\ca$\\  \hline
 h & $ 0 $&\ca &\ca\\  \hline
 i & $ \frac{5}{6\eu}-\frac{5}{6\ep}  $ &\ca &$\ca$\\\hline
 j & $ \left( -\frac{2}{3\eu}+\frac{2}{3\ep}\right) \nf $ &\tf &\tf\\ \hline
\end{tabular}
\ccaption{}{\label{tab:virtual1S0} 
First column: diagram labels according to the conventions of fig.~\ref{fig:gg}.
Second column: diagram-by-diagram contribution to the 
virtual corrections   
to the matrix element squared for the
processes $\QQ[^1S_0^{[1,8]}] \to gg$ and  with colour factors set to
1.
Third column: colour factors relative to the $gg$ colour singlet decay.
Fourth column: colour factors relative to the $gg$ colour octet decay.
}
\end{center}                          
\end{table}

{\renewcommand{\arraystretch}{1.8}
\begin{table}[t]
%\scriptsize
\begin{center} 
\begin{tabular}{|l|l|l|l|} \hline
Diag. &${\cal D}_k$ &$f_k^{[1]}$&$f_k^{[8]}$\\ \hline\hline

 a & $ \frac{\pi^2}{2v} +\frac{1}{\ep} -\frac{4}{9} +\frac{32}{9}\log 2
    +\frac{\pi^2}{12}  $&$\cf$&$\caf$\\ \hline
 b & $  -\frac{1}{2 \eu}-\frac{13}{9} +\frac{5}{9}\log 2 $&\cf &\cf \\ \hline
 c & $  -\frac{1}{2 \eu}-\frac{1}{\ep}-2-3\log 2 $ &\cf&\cf\\ \hline
 d & $  \frac{1}{ \eu} +\frac{14}{9}-\frac{10}{9}\log 2 +\frac{\pi^2}{6} $
    &\caf&\caf \\ \hline
 e & $ -\frac{3}{\eu}+\frac{2}{3\ep^2}+\frac{20}{9\ep}- \frac{11}{3}+\frac{2}{3}
     \log 2 -\frac{\pi^2}{9} $ &$-\frac{1}{2}$\ca &$-\frac{1}{2}$\ca\\ \hline 
 f & $  \frac{4}{3 \ep^2}-\frac{2}{9\ep}             
    +1-\frac{20}{3}\log 2-\frac{5}{9}\pi^2 $&$\frac{1}{2}\ca$ &$0$\\\hline
 g & $  -\frac{4}{3\ep^2}+\frac{2}{9\ep} -\frac{14}{27} +\frac{14}{27}\log 2
    +\frac{13}{18}\pi^2  $&\ca &$\frac{1}{2}\ca$\\  \hline
 h & $  -\frac{19}{27}+\frac{70}{27}\log 2  $&\ca &\ca\\  \hline
 i & $ \frac{5}{6\eu}-\frac{5}{6\ep}  $ &\ca &$\ca$\\\hline
 j & $ \left( -\frac{2}{3\eu}+\frac{2}{3\ep}\right) \nf $ &\tf &\tf\\ \hline
\end{tabular}
\ccaption{}{\label{tab:virtual3P0} 
Same as table~\ref{tab:virtual1S0}, for  $^3P_0^{[1,8]}$ states.}
\end{center}                                                       
\end{table}

{\renewcommand{\arraystretch}{1.8}
\begin{table}[t]
%\scriptsize
\begin{center} 
\begin{tabular}{|l|l|l|l|} \hline
Diag. &${\cal D}_k$ &$f_k^{[1]}$&$f_k^{[8]}$\\ \hline\hline

 a & $ \frac{\pi^2}{2v} +\frac{1}{\ep} -\frac{5}{3} +\frac{7}{3}\log 2
    +\frac{\pi^2}{8}  $&$\cf$&$\caf$\\ \hline
 b & $  -\frac{1}{2 \eu}-\frac{1}{6} +\frac{11}{6}\log 2 $&\cf &\cf \\ \hline
 c & $  -\frac{1}{2 \eu}-\frac{1}{\ep}-2-3\log 2 $ &\cf&\cf\\ \hline
 d & $  \frac{1}{ \eu} -\frac{1}{6}-\frac{7}{6}\log 2 -\frac{\pi^2}{8} $
    &\caf&\caf \\ \hline
 e & $-\frac{3}{\eu}+\frac{3}{8\ep^2}+\frac{17}{16\ep}-
   \frac{59}{96}-\frac{\pi^2}{16} $ &$-\frac{1}{2}$\ca &$-\frac{1}{2}$\ca\\ \hline 
 f & $  \frac{13}{8 \ep^2}+\frac{15}{16\ep}
    +\frac{107}{96}-\frac{11}{2}\log 2-\frac{67}{48}\pi^2 $&$\frac{1}{2}\ca$
    &$0$\\\hline
 g & $  -\frac{13}{8\ep^2}-\frac{15}{16\ep} -\frac{17}{288} +\frac{89}{18}\log
    2 +\frac{37}{48}\pi^2  $&\ca &$\frac{1}{2}\ca$\\ \hline
 h & $  -\frac{5}{9}-\frac{10}{9}\log 2  $&\ca &\ca\\  \hline
 i & $ \frac{5}{6\eu}-\frac{5}{6\ep}  $ &\ca &$\ca$\\ \hline
 j & $  \left(-\frac{2}{3\eu}+\frac{2}{3\ep}\right) \nf $ &\tf &\tf\\ \hline
\end{tabular}
\ccaption{}{\label{tab:virtual3P2} 
Same as table~\ref{tab:virtual1S0}, for  $^3P_2^{[1,8]}$ states.}
\end{center}                                    
\end{table}

Summing the contribution of all diagrams, we obtain the following results for
the colour singlet coefficients $B_{\q}^{[1]}$ :
\ba                                    
     && B^{[1]}_{\oneSzero} \= \; 
               \cf\left(-5+\frac{\pi^2}{4}\right)
               +\ca\left(1+\frac{5}{12}\pi^2\right) \, ,\\
     && B^{[1]}_{\threePzero} \=      
               \cf \left( -\frac{7}{3}+\frac{\pi^2}{4} \right) +
               \ca \left( \frac{1}{3}+\frac{5}{12}\pi^2 \right) \; ,
               \\                                               
     && B^{[1]}_{\threePtwo} \=  -4\cf +
                          \ca \left( \frac{1}{3}+\frac{5}{3}\log 2 +
                          \frac{\pi^2}{6} \right) \; .
\ea

We recall that the virtual
corrections to processes which are absent at the Born level vanish.

\subsubsection{Colour-octet virtual $gg$ decays}
Summing the contributions in tables~\ref{tab:virtual1S0}, \ref{tab:virtual3P0}
and \ref{tab:virtual3P2}, we get the following result for the colour-octet $gg$
decays:
\ba      
         \Gamma_V^H(\q^{[8]} \to gg) &=& \gbh(\q^{[8]} \to gg)
         \frac{\as}{\pi} \feps{M^2} 
\nn \\                                                               
\label{eq:virtual8}                              
    && \times \left\{  \frac{b_0}{\epsuv} + (\caf)\frac{\pi^2}{2v} -
     \ca \left( \frac{1}{\epsilon^2} + \frac{7}{3\epsilon} \right)+
     \frac{2}{3\epsilon} \nf \tf + B^{[8]}_{\q} \right\}\; ,
\ea    
where the state-dependent finite parts are given by:
\ba                             
     && B^{[8]}_{\oneSzero} \= \; \cf\left(-5+\frac{\pi^2}{4}\right)+\ca\left(2+\frac{7}{24}\pi^2\right)\, ,  \\
     && B^{[8]}_{\threePzero}          
             \=  \cf \left( -\frac{7}{3}+\frac{\pi^2}{4} \right) +
 \ca \left( \frac{17}{54}+\frac{35}{27}\log 2 + \frac{7}{24}\pi^2 \right) \; ,\\
     && B^{[8]}_{\threePtwo} \=  -4\cf +
                          \ca \left( \frac{23}{36}+\frac{7}{9}\log 2 +
                          \frac{5}{12}{\pi^2} \right) \; .
\ea

\begin{figure}[t]
\begin{center}
\epsfig{file=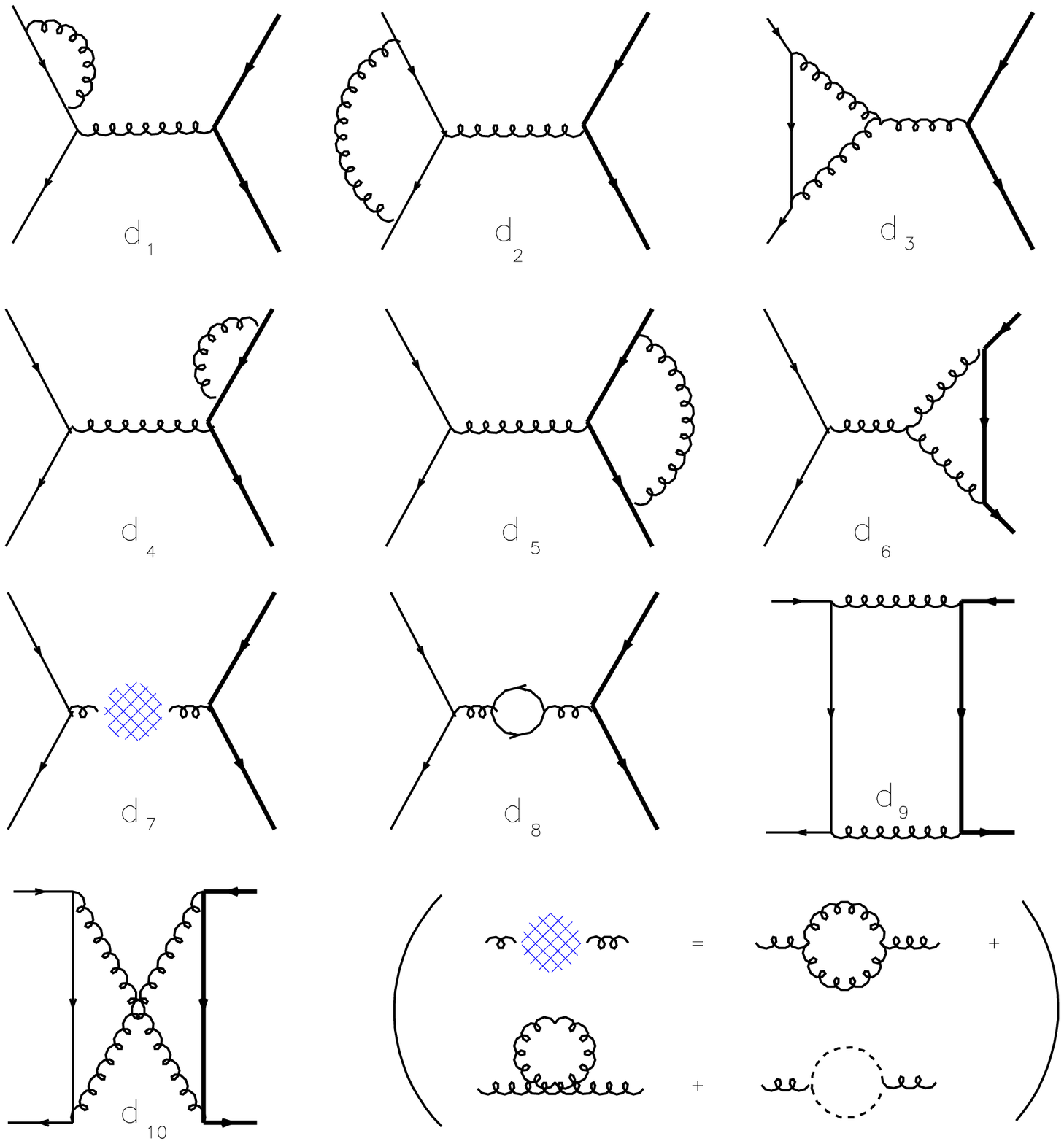,
%             bbllx=10pt,bblly=60pt,bburx=550pt,bbury=770pt,
%              bbllx=30pt,bblly=160pt,bburx=540pt,bbury=660pt,
            width=12cm,clip=}
\ccaption{}{\label{fig:qq} Feynman diagrams contributing to the one-loop 
corrections to the process $\qq\to\QQ[^3S_1^{[8]}]$.}
\end{center}                                        
\end{figure}

{\renewcommand{\arraystretch}{1.8}
\begin{table}[t]
%\scriptsize
\begin{center} 
\begin{tabular}{|l|l|l|} \hline
Diag. &${\cal D}_k$ &$f_k$ \\ \hline\hline

 $d_1$& $  -\frac{1}{2 \eu} + \frac{1}{2 \ep} $&
 $\cf$\\ \hline
 $d_2$& $  \frac{1}{2
 \eu}-\frac{1}{{\ep}^{2}}-\frac{2}{\ep}-4+\frac{2}{3}{\pi}^2 $ &\caf \\ \hline
 $d_3$&  $  \frac{3}{2 \eu}-\frac{2}{\ep}-1 $ &$\frac{1}{2}$\ca \\ \hline
 $d_4$ & $ -\frac{1}{2 \eu}-\frac{1}{\ep}-2-3\log
 2 $ &\cf \\ \hline
 $d_5$ & $ \frac{\pi^2}{2v}+\frac{1}{2
\eu}+\frac{1}{\ep}- 2 + 3 \log 2  $ &\caf\\ \hline 
 $d_6$ & $  \frac{3}{2 \eu}
 +\frac{8}{3}+\frac{13}{3}\log 2 $&$\frac{1}{2}$\ca \\  \hline
 $d_7$ & $ \frac{5}{6}\frac{1}{\eu} +
\frac{31}{18}  $ &\ca\\  \hline
 $d_8$ & $ \left( -\frac{2}{3 \eu} -\frac{10}{9}\right)  \nf $&\tf\\  \hline
 $d_9$ & $ \frac{1}{{\ep}^{2}} 
 - \frac{{\pi}^2}{6}  $ & $2\cf - \ca$\\ \hline
 $d_{10}$ & $  - \frac{1}{ {\ep}^{2} } + \frac{{\pi}^2}{6}  $ &$ 2\cf -
 \frac{1}{2}\ca$\\ \hline
\end{tabular}
\ccaption{}{\label{tab:psi8}
Diagram-by-diagram contribution to the virtual corrections   
to the matrix element squared for the  
processes $\QQ[^3S_1^{[8]}] \to q \bar q$.
Diagram labels according to the conventions of fig.~\ref{fig:qq}.}
\end{center}                                                     
\end{table}

\subsubsection{Colour-octet $\qq$ virtual decays}
The only relevant process in this channel is relative to the $\psih$ configuration, and the  
diagrams are shown in fig.~\ref{fig:qq}. The
contributions of the various diagrams are given in table~\ref{tab:psi8}, and
their sum is
\ba         
\label{eq:virtualqq}
\Gamma_V^H(\threeSone^{[8]} \to \qq) &=& \gbh(\threeSone^{[8]} \to \qq)
         \frac{\as}{\pi} \feps{M^2} 
\nn\\                               
&&\times \!\! \left\{\left(\caf\right) \frac{\pi^2}{2v} +
   \frac{\b0}{\eu} + \cf\left(-\frac{1}{\ep^2}-\frac{3}{2\ep}\right)
   -\ca\frac{1}{2\ep}+ A[\psih]\right\}\label{virtpsih} , 
\ea                                                          
with
\be
B[\psih]= \cf\left(-8+\frac{2}{3}\pi^2\right)+\ca\left(\frac{50}{9} +
\frac{2}{3}\log 2 -\frac{\pi^2}{4}\right) -\frac{10}{9}\nf\tf \, .
\label{apsih} 
\ee

\section{Cancellation of IR singularities within NRQCD}
\label{sec:opren}
Throughout this paper, we  witness the cancellation or removal of both
infrared/collinear or ultraviolet singularities by the usual QCD mechanisms:
soft/virtual cancellation, collinear factorization, mass and coupling 
constant
renormalization. Two exceptions exist: the $1/v$ Coulomb singularity 
in the virtual corrections, see eqs.~(\ref{eq:virtual}, \ref{eq:virtual8},
\ref{eq:virtualqq}), and the $1/\ep$ infrared singularity which 
appears in the $^3P_J^{[1,8]} \to \qq g$ 
real correction processes in eqs.~(\ref{eq:qqg1}, \ref{eq:qqg8})
(or in $\qq \to ^3P_J^{[1,8]} g$, as discussed in the next section) 
cannot be eliminated via these standard mechanisms.
Their removal is strictly related with the NRQCD factorization
approach to quarkonium production and decay. It is within
this approach that one finds a rigorous solution to this problem, previously
dealt with in an empiric way, by absorbing the Coulomb singularity into the
Bethe-Salpeter wave function and cutting off the infrared singularity 
with the binding energy of the quarkonium.
         
Within NRQCD, one can determine the short distance coefficients of the 
various
operators by performing a {\sl matching} between cross sections calculated in
perturbative QCD and perturbative NRQCD. Since, by definition, the two
theories are equivalent in the long distance regime, all singularities of
infrared origin appear equally in both calculations, and hence cancel in the
matching.

Our approach  does not perform an explicit matching, but makes use
of projection techniques and then complements the short distance cross 
sections and decay widths with the
non-perturbative NRQCD matrix element describing the transition to the
observable quarkonium state. Rather than seeing the Coulomb/infrared
singularities disappear in the matching, we will therefore cancel them by
taking into account radiative corrections to the NRQCD matrix elements.

\begin{figure}[t]
\begin{center}
\epsfig{file=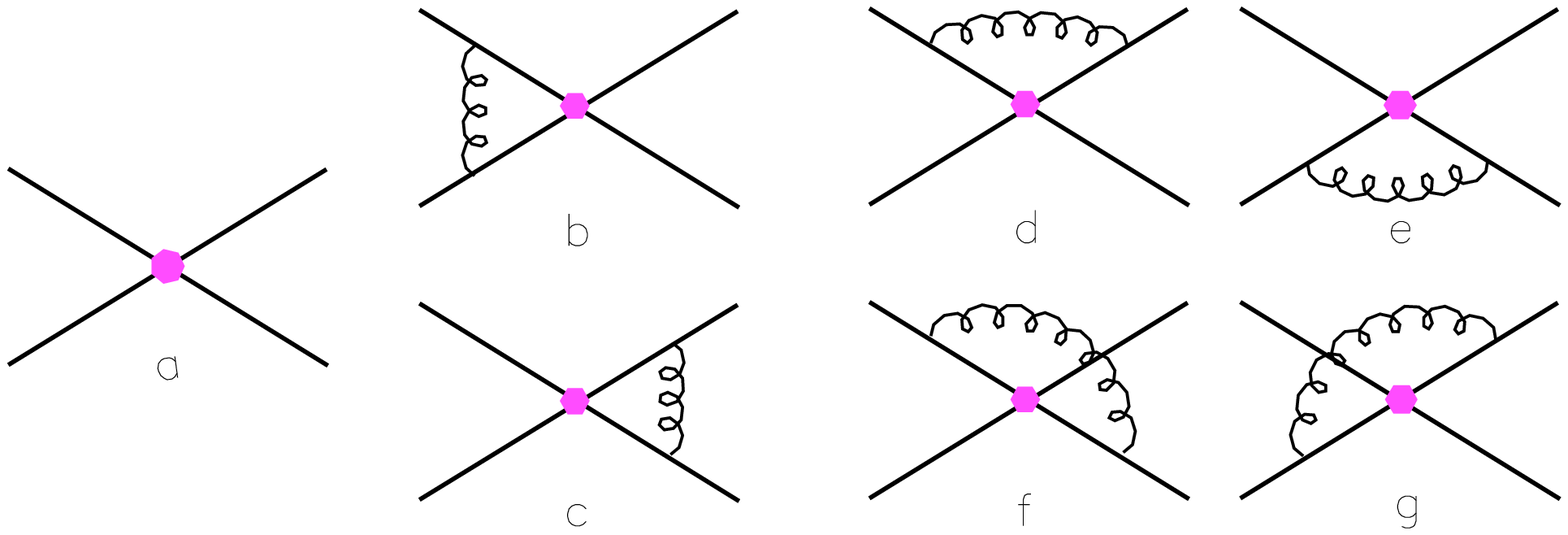,width=14cm,clip=}
\parbox{12cm}{
\caption{\label{fig:nrqcd} \small Diagrams for NLO corrections
to the NRQCD matrix elements.}}
\end{center}
\end{figure}

Let us first describe with some details the cancellation of the Coulomb
singularity. We discuss for definiteness 
the case of decays, but the same applies to production cross sections.
The Coulomb-singular part of the virtual correction to a              
two-partons decay of a $\q \equiv \QQ [\spectr]$ state reads:
\be
\hat\Gamma_V(\q\to ij) = \hat\Gamma_{\rm Born} \frac{\as}{\pi}
      \feps{M^2}
      \left({\cal C} {\pi^2\over{2v}} + \cdots\right)\,.
\label{nr:gamma}
\ee
The coefficient ${\cal C}$ is a colour factor which, for $\q$
in a colour-singlet or octet state, takes the value
\ba                          
{\cal C}^{[1]} &=& \cf \, ,\nonumber\\
{\cal C}^{[8]} &=& \caf \, .
\label{c18}
\ea

The above expression (\ref{nr:gamma}), multiplied by the NRQCD matrix element
$\langle H|{\o}(\q)|H\rangle$, provides the final answer for the
virtual corrections to the decay width of the physical quarkonium state $H$.
However, to get rid of the Coulomb singularity, we must first evaluate in 
NRQCD the radiative corrections to this matrix element. This was done for the
first time in ref.~\cite{bbl}, and we reproduce their argument for
illustration.

$\langle H|{\o }(\q)|H\rangle$ is depicted in 
fig.~\ref{fig:nrqcd}(a) in lowest order,
and the higher order corrections important for the Coulomb singularities are
shown in figs.~\ref{fig:nrqcd}(b,c). Let us consider, for instance,
diagram~\ref{fig:nrqcd}(b).

Working in Coulomb gauge, the gluon propagator is given by the sum of a
transverse part,
\be
G^{\mu\nu}_{trans}(k^2) = - {i\over k^2}\left(g^{\mu\nu} +
                      {{k^\mu k^\nu}\over{{\bf k}^2}}\right)
                      (1-\delta_{\mu 0})(1-\delta_{\nu 0}),
\ee
and of a Coulomb part,
\be
G^{\mu\nu}_{Coul}(k^2) = {i\over{{\bf k}^2}}\delta_{\mu 0}\delta_{\nu 0}.
\ee
It is this second part responsible for the singularity we are looking for.

Inserting this term only and making use of NRQCD Feynman rules we get for
diagram~\ref{fig:nrqcd}(b)
\be
I_b = -i g_s^2 \int {{d^4k}\over{(2\pi)^4}} {1\over{{\bf k}^2}}
{1\over{\left[p_0 + k_0 - {{({\bf p} + {\bf k})^2}\over{2m}} + 
i\ep\right]}}\;
{1\over{\left[p_0 - k_0 - {{({\bf p} + {\bf k})^2}\over{2m}} + 
i\ep\right]}}\, ,
\ee
with the on-shell condition $p_0 = {\bf p}^2/2m$, $p$ being the heavy quark
momentum and $k$ the gluon one. The integral over $k_0$ can
be performed by contour integration, closing the contour in the lower half
of the complex $k_0$ plane and picking up the pole at 
$k_0 = - p_0 + ({\bf p} + {\bf k})^2/2m - i\ep$. The result is
\be
I_b = g_s^2 m \int {{d^3k}\over{(2\pi)^3}} {1\over{{\bf k}^2}}
{1\over{{\bf k}^2 + 2{\bf p}\cdot{\bf k} - i\ep}}\, .
\ee
Being divergent, this integral must be performed within some regularization
technique. Using dimensional regularization, and defining ${\bf v} = {\bf 
p}/m$,
$v = |{\bf v}|$, we find
\be
I_b = {{\pi\as}\over{4v}}(1 + {\rm infrared~singular~imaginary~part})\, .
\ee

From the diagram~\ref{fig:nrqcd}(c)  
we find $I_c$, similar to $I_b$ but with opposite
imaginary part. Summing the two contributions, the overall Coulomb
correction to the matrix element is therefore:
\be
\langle H|{\o }(\q)|H\rangle = 
\langle H|{\o }(\q)|H\rangle_{\rm Born} 
\left(1 + {\cal C} {{\pi\as}\over{2v}}\right).
\ee
The colour factor ${\cal C}$ evaluates to
the same defined above in (\ref{c18}), according to the operator we are 
correcting (the central blob in figure~\ref{fig:nrqcd}) being a
colour singlet or a colour octet one.

For the decay width we have then
\ba
\Gamma^H [\q \to ij] &=& \left(\hat\Gamma_{\rm Born} + \hat\Gamma_V(\q\to 
ij) + 
\cdots\right) 
\langle H|{\o }(\q)|H\rangle_{\rm Born} \nonumber\\
&=& \left[\hat\Gamma_{\rm Born} + \hat\Gamma_{\rm Born} 
\frac{\as}{\pi}          
      \left({\cal C} {\pi^2\over{2v}} + \cdots\right)\right]
\langle H|{\o }(\q)|H\rangle
\left(1 - {\cal C} {{\pi\as}\over{2v}}\right) \, .
\ea
Upon inspection, we can see that the singular $1/v$ terms cancel. Since no 
finite                                                    
parts are introduced in this process, we can just drop the singular Coulomb
term whenever it appears in the virtual contributions to cross sections and
decay widths.

Next we consider the problem of the infrared singularity which appears for
instance in the $^3P_J^{[1,8]} \to \qq g$ real correction process. We have
mentioned in Section \ref{sec:lightquark} how the residue of this 
singularity 
is proportional
to the matrix element squared for the process $^3S_1^{[8]} \to \qq$, making
it possible to absorb the singularity in the colour-octet wave function
which accompanies this process.
Again, NRQCD puts this cancellation on rigorous grounds. Radiative 
corrections
to the colour octet matrix element $\langle H| \o_8 (^3S_1)|H\rangle$ 
display a
singularity which matches the one appearing in the $P$-state decay 
calculation.
When summing $P$-wave and $^3S_1$ octet processes, the singularity cancels
exactly. Some finite parts are however introduced by this process, which 
makes it necessary to carefully evaluate the diagrams involved.
         
Let us then consider the radiative corrections depicted in diagrams
\ref{fig:nrqcd}(d,e,f,g). It is this time the transverse part of the gluon
propagator to be relevant, and the loop integral reads
\be
I_d = ig_s^2 
     \int {{d^4k}\over{(2\pi)^4}} {{\bfp\cdot\bfpp - (\bfp\cdot\bfk)
(\bfpp\cdot\bfk)/\bfk^2}\over{m^2(k_0^2 - \bfk^2 + i\ep)}}
{1\over{\left[p_0 - k_0 - {{({\bf p} - {\bf k})^2}\over{2m}} + i\ep\right]}}
{1\over{\left[p_0' - k_0 - {{({\bf p}' - {\bf k})^2}\over{2m}} + 
i\ep\right]}}\, .
\ee
Contour-integrating over $k_0$ around the $k_0 = \vert\bfk\vert - i\ep$ pole 
we find
\be
I_d = g_s^2 
      \int {{d^3k}\over{(2\pi)^3}} {{\bfp\cdot\bfpp - (\bfp\cdot\bfk)
(\bfpp\cdot\bfk)/\bfk^2}\over{2|\bfk|m^2}}
{1\over{\left[-|\bfk| - {{{\bf k}^2}\over{2m}} + 
{{\bfp\cdot\bfk}\over{m}}+ i\ep\right]}}
{1\over{\left[-|\bfk| - {{{\bf k}^2}\over{2m}} + 
{{\bfpp\cdot\bfk}\over{m}}+ i\ep\right]}}
\, .
\ee
This integral is infrared divergent but ultraviolet finite. However, it  has 
been argued \cite{Braaten97, Manohar97, Beneke97a} that 
the correct way to  perform 
it is to first expand \cite{bbl} the integrand in powers
of $\bfp/m$, $\bfpp/m$ and $\bfk/m$, since it is for small values of these
momenta that NRQCD is valid (though alternative approaches exist, see for 
instance ref.~\cite{rg97}). Such an expansion produces
\be                       
I_d = {{g_s^2 }\over{2m^2}}
      \int {{d^3k}\over{(2\pi)^3}} {{\bfp\cdot\bfpp - (\bfp\cdot\bfk)
(\bfpp\cdot\bfk)/\bfk^2}\over{|\bfk|^3}}\, ,
\ee
which is now {\sl both} infrared and ultraviolet divergent. We can 
perform it by using dimensional regularization and obtain
\be           
I_d = {{4\pi\asb}\over{2m^2}} {{\bfp\cdot\bfpp}\over{3\pi^2}} {1\over 2}
\left({1\over\eu} - {1\over\ei}\right)\, ,
\ee
where $\ei$ and $\eu$ are poles of infrared and ultraviolet origin
respectively. It is to be noted that this ultraviolet divergence arises  within
NRQCD, and has nothing to do with the ultraviolet divergences which  appear in
the virtual corrections within full QCD, to be subsequently removed by
renormalization of the coupling constant.
                                         
The other three diagrams give identical result but with different colour
factors when including the colour structure of the $\o_8(^3S_1)$ 
operator we are correcting. Using eqs.~(\ref{Tabba},\ref{Tabab}) we 
get for the sum of the four diagrams   
\be
I = {{4\asb}\over{3\pi m^2}}\left({1\over\eu} - {1\over\ei}\right)
\o_8(^3S_1) \left[ \cf \frac{1\otimes 1}{2 \nc} +  \Bf 
T^a\otimes T^a\right] \bfp\cdot\bfpp \, .
\ee

The diagrams depicted in fig. \ref{fig:nrqcd}  formally describe 
annihilation 
matrix elements $\langle H|\o(\q)|H\rangle$, but the result is identical 
for the production ones $\langle \o^H(\q)\rangle$. Recalling 
the definition of $\o_{[1,8]}(^3P_J)$ (see Appendix~\ref{appA}) we can write
\ba
\langle H|\o_8(^3S_1)|H\rangle &=& \langle H|\o_8(^3S_1)|H\rangle_{\rm 
Born} + 
{{4\asb}\over{3\pi m^2}}\left({1\over\eu} - {1\over\ei}\right) \nn\\
&&\times
\left[\cf \sum_{J=0}^2 \langle H|\o_{1}(^3P_J)|H\rangle +
\Bf \sum_{J=0}^2 \langle H|\o_{8}(^3P_J)|H\rangle\right] 
\, .
\label{3s1corr}
\ea
The presence of the ultraviolet singularity indicates that the 
$\langle H|\o_8(^3S_1)|H\rangle$
operator needs renormalization. The suitable counterterm in the $\MSB$ 
scheme 
is such that the relation between the bare ($D$-dimensional) and the 
renormalized matrix element reads
\ba
\langle H|\o_8(^3S_1)|H\rangle &=& \mul^{-2\ep}\left\{\langle
H|\o_8(^3S_1)|H\rangle^{(\mul)} + 
{{4\as}\over{3\pi m^2}}\left({1\over\eu} +\ln 4\pi -
\gamma_E\right)\right.\nonumber\\
&&\times\left.\left[\cf \sum_{J=0}^2 
\langle H|\o_{1}(^3P_J)|H\rangle +
\Bf \sum_{J=0}^2 \langle H|\o_{8}(^3P_J)|H\rangle
\right]\right\},
\label{nrqcdct}
\ea
$\mul$ being the NRQCD  renormalization scale. 
Since the renormalized matrix elements
on the right hand side have mass dimension three while the bare 
$D$-dimensional
one on the left has dimension $D-1$ (a factor of $2(D-1)$ coming from the 
four
spinor fields of the operator plus a factor of $-(D-1)$ from the 
nonrelativistic
normalization of the $|H\rangle$ states, $\langle H({\bf p})|H({\bf 
p'})\rangle 
= (2\pi)^{D-1} \delta^{D-1}({\bf p} - {\bf p'})$), the $\mul^{-2\ep}$ 
compensates for the difference.

Using eqs.(\ref{3s1corr},\ref{nrqcdct}) (no renormalization is  necessary 
at this order for the $\langle H|\o_{[1,8]}(^3P_J)|H\rangle$ matrix elements) 
we find
\ba
\langle H|\o_8(^3S_1)|H\rangle_{\rm Born} &=& 
\mul^{-2\ep}\langle H|\o_8(^3S_1)|H\rangle^{(\mul)} +
\left({1\over\ei} +\ln 4\pi -\gamma_E\right)
\left(\frac{\mu}{\mul}\right)^{2\ep}
{{4\as}\over{3\pi m^2}} \nn\\
\!\!\!\!\!&&\times\left[\cf \sum_{J=0}^2 \langle 
H|\o_{1}(^3P_J)|H\rangle +
\Bf \sum_{J=0}^2 \langle 
H|\o_{8}(^3P_J)|H\rangle\right]\,.
\label{3s1full}
\ea

This equation can now be used to cancel the infrared pole in the $\threePJ\to
\qq g$ processes given in eqs.~(\ref{eq:qqg1}, \ref{eq:qqg8}): summing to 
these
decays the colour-octet one $\threeSone^{[8]}\to \qq$ and substituting 
the bare
matrix element with the expression given above one finds (restricting 
ourselves to the infrared singular parts):                         
\ba          
&&\sum_J\Gamma^H(\chijs\to\qq g) + \Gamma^H(\threeSone^{[8]}\to\qq) =\nn\\
&=& \sum_J\left\{- \frac{4}{9} \nf \cf     \ascube
\frac{\mu^{2\ep}\langle H |\o_1(^3P_J)|H\rangle}{m^4}             
\left[\frac{1}{\ei} 
\frac{\feps{M^2}}{K}
\left(\frac{4 \pi\mu^2}{M^2}\right)^\epsilon
\frac{\Gamma(1-\epsilon)}{\Gamma(2-2\epsilon )} + a_J\right] + \mbox{ 
IR-finite}\right\}\nn\\
&&+ {{\assq \pi}\over{m^2}}
\left({{4\pi\mu^2}\over{4m^2}}\right)^{\epsilon}
{{\Gamma(1-\ep)}\over{\Gamma(2-2\ep)}} {{(1-\ep)}\over{(3-2\ep)}}\nn\\
&&\times  
\left({1\over\ei} +\ln 4\pi -\gamma_E\right)
\left(\frac{\mu}{\mul}\right)^{2\epsilon}
{{4\as}\over{3\pi m^2}}
\nf \cf \mu^{2\epsilon}\sum_J\langle H|\o_{1}(^3P_J)|H\rangle 
+ \mbox{ 
IR-finite}\nn\\                                                       
&=& \sum_J\left\{ - \frac{4}{9} \cf \ascube
\frac{\langle H |\o_1(^3P_J)|H\rangle}{m^4}\left[a_J + \frac{1}{3} +
\log\frac{\mul^2}{4m^2}\right] + \cdots\right\}
\ea

\section{Production processes}
\label{sec:production}

\subsection{Kinematics and factorization of soft and collinear singularities}
The technique used to extract the soft and collinear singularities from the
real emission processes is similar to the one employed in the study of decays.
The kinematics of the process $k_1+k_2 \to P + k_3$, where $P$ is the momentum
of the heavy quark pair (~identified by $\q^{[1,8]}$~) and $k_i$ are the momenta of the massless
partons, can be described in terms of the standard Mandelstam variables
$s$, $t$ and $u$, defined by
\ba
     s &=& (k_1+k_2)^2 \; , \\      
     t &=& (k_2-k_3)^2 \;\equiv\; -\frac{s}{2}(1-x)(1-y) \; , \\
     u &=& (k_1-k_3)^2 \;\equiv\; -\frac{s}{2}(1-x)(1+y) \; .   
\ea                                                          
Here we also have introduced the Lorentz invariant dimensionless variables
$x=4m^2/s$ and $y$ ($-1<y<1$), defined by the above equations.
In the center-of-mass frame of the partonic collisions, the variable $y$
becomes the cosine of the scattering angle $\theta$. In terms of $x$ and $y$
the total partonic cross section can be written in $D$ dimensions as follows:
\be                                                               
   \sigma \= \frac{1}{2s} \int d\Phi_{(2)}(x,y) \; \m(x,y) \; ,
\ee                                                        
where  $\m=\overline{\sum}\vert A\vert^2$ is the spin- and colour-averaged matrix
element squared in $D$ dimensions and $d\Phi_{(2)}(x,y)$ is the $D$-dimensional
two-body phase space:                                           
\be
  d\Phi_{(2)}(x,y) \= \frac{4^{\eps}}{K} \, 
      \left(\frac{4\pi}{s}\right)^{\eps} \, \Gamma(1+\eps) \,
      \frac{1}{16\pi} \, (1-x)^{1-2\eps} \, (1-y^2)^{-\eps} \, dy \; .
\ee                                                      
The soft and collinear singularities are associated to the vanishing of $t$ or
$u$, which appear at most as single poles in the expression of \m. One can
therefore introduce the finite, rescaled amplitude squared \mbar:
\be
     \m \= \frac{1}{ut} \, \mbar \= \frac{4}{s^2(1-x)^2(1-y^2)} \, \mbar \; .
\ee                                                                
In terms of \mbar, the partonic cross section reads as follows:
\ba                
  &&   \sigma(x) \=
     \frac{4C}{s^2} \, (1-x)^{-1-2\eps} \, \int_{-1}^{1} \, dy \,
     (1-y^2)^{-1-\eps} \, \mbar(x,y)\, dy \; , \\
  && C \= \frac{4^{\eps}}{K} \,  
      \left(\frac{4\pi}{s}\right)^{\eps} \, \Gamma(1+\eps) \,
      \frac{1}{32\pi s}  \; .               
\ea                          
As in the case of decays, soft and collinear singularities are now all
exhibited by the universal poles which develop as $x\to 1$ and $y^2\to 1$. The
residues of these poles can be derived without an explicit calculation of the
matrix elements, as they only depend on the  universal structure of collinear
and soft singularities. We will carry out an explicit evaluation of these
residues in the case of colour-singlet production in the $gg\to\q^{[1]}g$
process in the next subsection. The other cases are similar, and will be
discussed with  fewer details in the following.
                               
\subsubsection{$gg\to g\q^{[1]}$ processes}

\begin{figure}[t]
\begin{center}
\epsfig{file=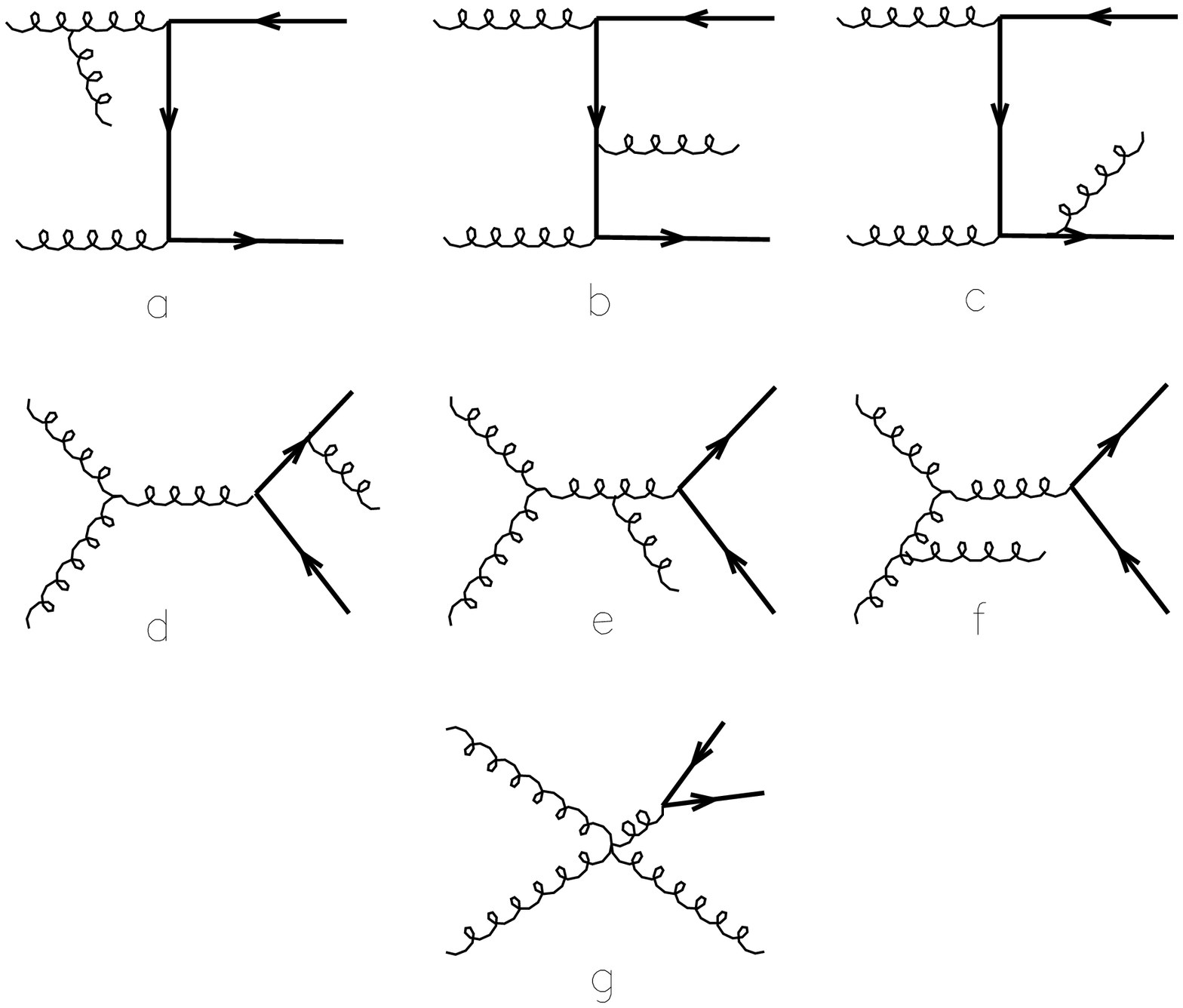,width=12cm,clip=}
\parbox{12cm}{       
\caption{\label{figgg-g} \small Diagrams for the real corrections to the $gg$
channels. Permutations of outgoing gluons and/or reversal of fermion lines are
always implied.}}
\end{center}
\end{figure}

We start by considering the soft limit, $x\to 1$. The
following distributional identity holds for small $\eps$:
\be                                       
     (1-x)^{-1-2\eps} \= -\frac{\beta^{-4\eps}}{2\eps} \delta(1-x) +
        \left( \frac{1}{1-x} \right)_\rho \, - \,
    2\eps \left( \frac{\log(1-x)}{1-x}\right)_\rho \, + \, {\cal{O}}(\eps^2) \, 
\ee                                                                           
where $\rho=M^2/S_{had}$,
$\beta=\sqrt{1-\rho}$, and the $\rho$-distributions are defined by:
\be                                                                       
   \int_{\rho}^{1} \, dx \; \left[d(x)\right]_{\rho} t(x) \=
   \int_{\rho}^{1} \, dx \; d(x) \; \left[t(x)-t(1)\right] \; .
\ee                                                   
We can therefore write, with obvious notation,
\be
   \sigma(x) \= \sigma_{x=1} + \sigma_{x\neq 1} \; .
\ee                                                
The first terms on the right-hand side is given by the following expression:
\be  \label{eq:sigx1}
  \sigma_{x=1} \=    
     - \frac{4C}{M^4} \, \frac{\beta^{-4\eps}}{2\eps} \, \delta(1-x) \,
      \int_{-1}^{1} \, dy \,               
     (1-y^2)^{-1-\eps} \, \mbar(x=1,y)\; , \\
\ee                                   
The $x\to 1$ limit of \mbar\ can be easily derived from eq.(\ref{eq:csfact}):
\ba                                 
  \m(x,y) & \stackrel{x\to 1}{\longrightarrow} & 
       g^2 \, \ca \, \frac{4s}{ut} \, \mborn \,  \\ 
  \mbar(x,y) & \stackrel{x\to 1}{\longrightarrow} & 
       4s \, g^2 \, \ca \, \mborn \; ,
\ea                                                                    
where \mborn\ is the $D$-dimensional Born amplitude squared for the $gg\to \q$
process, which is independent of $y$. The integration over $y$ of
eq.(\ref{eq:sigx1}) is elementary, and leads to the following result:
\be
   \sigma_{x=1} \=                          
      \frac{\feps{s}}{\eps^2} \,
      H \, \ca \, \frac{\as}{\pi} \beta^{-4\eps} \, \sborn \;,
\ee                                  
where $H$ is defined by   
\be              
   H \= \frac{\Gamma(1-\eps)}{\Gamma(1+\eps)\Gamma(1-2\eps)} \=
      1 - \frac{\pi^2}{3} \eps^2 + {\cal O}(\eps^3) \; ,
\ee                                                 
and \sborn\ is the $D$-dimensional Born cross section:
\be       \label{eq:sigmaB}
  \sborn \= \frac{\pi}{M^4} \; \mborn \;
\delta(1-x) \; \equiv \; \sborno \delta(1-x) \; .
\ee

The collinear singularities remaining in $\sigma_{x\neq 1}$ can be factored out
by using the following distributional identity:            
\ba
   (1-y^2)^{-1-\eps} &=& -\left[ \delta(1-y)+\delta(1+y)\right]
  \frac{4^{-\eps}}{2\eps} \nn \\
  && + \frac{1}{2}\left[\left(\frac{1}{1-y}\right)_+ + 
    \left(\frac{1}{1+y}\right)_+ \right] + {\cal O}(\eps) \; ,
\ea                                                           
where the distributions on the right-hand side are defined by:
\be                            
   \int_{-1}^{1} \, dy \, \left(\frac{1}{1\pm y}\right)_+ t(y) \=
   \int_{-1}^{1} \, dy \, \frac{1}{1\pm y} \left[t(y)-t(\mp 1)\right] \; .
\ee                                                                       
The contribution $\sigma_{x\neq 1}$ can then be split into three terms:
\be              
   \sigma_{x\neq 1} \= \sigma_{y=1}+\sigma_{y=-1}+\sigma_{\rm finite} \; .
\ee                                                                      
The term $\sigma_{\rm finite}$ has no residual divergences, and is given by the
following expression:                 
\be  \label{eq:finite}
  \sigma_{\rm finite} \= \frac{2C}{s^2} \left( \frac{1}{1-x}\right)_{\rho}
  \, \int_{-1}^{1} \, dy \, \left[\left(\frac{1}{1-y}\right)_+ +       
    \left(\frac{1}{1+y}\right)_+ \right] \, \mbar(x,y) \; .
\ee                                         
This piece explicitly depends on the nature of the quarkonium state produced.
For processes whose Born contribution vanishes, this is the only non zero term.

The remaining  pieces, $\sigma_{y=\pm 1}$, are given by:
\be           
  \sigma_{y=\pm 1} \=
  - \frac{4C}{s^2} \, \frac{4^{-\eps}}{2\eps} \,
  \left[ \left( \frac{1}{1-x} \right)_\rho        
       - 2\eps \left( \frac{\log(1-x)}{1-x}\right)_\rho \right] 
     \, \mbar(x,y=\pm 1) \; .
\ee     
The limits for $y\to \pm 1$ of $\mbar(x,y)$ are universal, thanks to the
factorization of collinear singularities:                  
\be
  \mbar(x,y) \stackrel{y\to\pm 1}{\longrightarrow} 8\pi \, s \, 
   \asb P_{gg}(x)
  \frac{1-x}{x} \mborn \; ,
\ee                        
where $P_{gg}(x)$ is the standard $D=4$ 
Altarelli-Parisi splitting kernel, defined
in Appendix~\ref{appA}.
Using these relations we get:
\be   \label{eq:sigcol} 
  \sigma_{y=\pm 1} \=
  - \frac{1}{\epsbar} \, \left(\frac{\mu^2}{s}\right)^{\eps} \,
    \frac{\as}{2\pi} \, \pgg{x}  \, (1-x)x \,  \sborno 
    \, \left[ \left( \frac{1}{1-x} \right)_\rho        
       - 2\eps \left( \frac{\log(1-x)}{1-x}\right)_\rho \right] \; ,
\ee                                                                 
where 
\be
   \frac{1}{\epsbar} \= \frac{1}{\eps} -\gamma_{\rm E} + \log(4\pi) \; ,
\ee
The collinear poles take the form dictated by the factorization theorem.
According to this the parton cross section can be written as:
\ba  \label{eq:isfact}
     {\rm d}\sigma_{ij}(p_1,p_2) &=& \sum_{k,l}
     {\rm d}\hat\sigma_{kl}(x_1 p_1,x_2 p_2) 
     \Gamma_{ki}(x_1) \Gamma_{lj}(x_2) {\rm d}x_1 {\rm d}x_2 \;,
\\                    
     \Gamma_{ij}(x)  &=&
      \delta_{ij} \delta(1-x) \;-\; \frac{1}{\epsbar} \,\frac{\as}{2\pi} 
      \left(\frac{\mu^2}{\mufsq}\right)^{\epsilon}
      {\cal P}_{ij}(x) \; + \; K_{ij}(x) \; ,
\ea                                   
where ${\rm d}\hat\sigma$ is free of collinear singularities as $\epsilon\to
0$. Here we allowed the factorization scale $\muf$ to differ from the
renormalization scale $\mu$.
The functions ${\cal P}_{ij}(x)$ are the $D=4$  Altarelli-Parisi splitting
kernels, collected in Appendix~\ref{appA}, and the factors $K_{ij}$ are
arbitrary functions, defining the factorization scheme. In this paper we adopt
the \MSB\ factorization, in which $K_{ij}(x)=0$ for all $i,j$. For the
definition of $K_{ij}(x)$ in the DIS scheme, see for example
ref.~\cite{Kuehn93}. The collinear
factors $\Gamma(x)$ are usually reabsorbed into the hadronic parton densities,
and the physical cross section is then expressed as:
\be                                                 
     {\rm d}\sigma_{H_1 H_2}(p_1,p_2) \= \sum_{k,l}
     {\rm d}\hat\sigma_{kl}(x_1 p_1,x_2 p_2,\muf)  
     F_{kH_1}(x_1,\muf) F_{lH_2}(x_2,\muf) {\rm d}x_1 {\rm d}x_2 \;,
\ee         
where $F_{kH}(x,\muf)$ is the density of the parton $k$ in the hadron $H$,
evaluated at the factorization scale $\muf$.
Expanding eq.~(\ref{eq:isfact}) order by order in $\as$, we extract the
following counter-terms $\sigma^{(c)}_{y=\pm 1}$:
 defined by                                       
\be \label{eq:sigct}
  \sigma^{(c)}_{y=\pm 1} \= \frac{1}{\epsbar} \,\frac{\as}{2\pi} \left(
  \frac{\mu^2}{\mufsq}\right)^{\eps} \, \cpgg{x}\, x \sborno \; ,
\ee                                                  
with
\be
{\cal P}_{gg}(x) =
2\ca\left[\frac{x}{(1-x)_{\rho}}+\frac{1-x}{x}+x(1-x)\right]+\left(\b0+4
\ca\log\beta\right)\delta(1-x)\, .
\ee

Putting together all pieces, we come to the final result for the real emission
cross section:
\ba
\sigma^H[g\,g\to \q^{[1]} \,  g] &=&\frac{\as}{\pi}
\sigma_0^H[gg\to \q^{[1]}]\nn\\        
&&\times \left\{\feps{s}  \left[\ca\left(\frac{1}{\ep^2}+\frac{11}{6\ep}  
   -\frac{\pi^2}{3}
   +8\log^2\beta\right) -\frac{2}{3\ep}\nf\tf \right]\delta(1-x)\right.\nn\\ 
&&+ \left.\left[x {\cal P}_{gg}(x)\log\frac{s}{\qf} + 2
   x(1-x) P_{gg}(x)\left(\frac{\log(1-x)}{1-x}\right)_{\rho}\right.\right.\nn\\
&&+ \left.\left.\left(\frac{1}{1-x}\right)_{\rho} f_{gg}[\q^{[1]}](x)
  \right]\right\}, \quad \, \quad \nn \\[10pt] 
&& [\q^{[1]}=\etas,\chizs,\chits]  \; ,
\ea
where $\sigma_0^H[gg\to \q^{[1]}]$ is the $D$-dimensional, Born-level partonic
cross section for the production of the quarkonium state $H$ via the
$\q^{[1]}$ intermediate state, after removal of the $\delta(1-x)$ term (see
eq.~(\ref{eq:sigmaB})).
The finite functions $f_{gg}(x)$, obtained from the explicit evaluation of
eq.~(\ref{eq:finite}), are collected in Appendix~\ref{appNLO}.
                                                
The result for the $^1S_0^{[1]}$ state agrees with previous
calculations~\cite{Kuehn93,Schuler94}. The results for the $P$ waves are new. 
In the case of $\threeSone$ and $\threePone$  production the $\oacube$ process
is the leading order one. Integration over the emitted gluon phase space is
completely finite, and the results have been known for long time in the
literature (see e.g. ref.~\cite{Schuler94}). For completeness, we collect them
in Appendix~\ref{appNLO}.

\subsubsection{$qg\to q\q^{[1]}$ processes}                               
The Born level processes $qg\to \q^{[1]}$ identically vanish. As a result no IR
divergence is present at $\oacube$ and virtual corrections are not present. The
only singularities appearing at this order come from the emission of the 
final-state quark collinear to the initial-state one.
The behaviour of the amplitude in the $y\to 1$ collinear limit is again
controlled  by the Altarelli-Parisi splitting functions:
\be                                                     
  \mbar(x,y) \stackrel{y\to 1}{\longrightarrow} 8\pi \, s \, \asb 
  P_{gq}(x) \frac{1-x}{x} \mborn \; .
\ee                                  
In analogy to the $gg$ case, one introduces the following counter-term in the 
\MSB\ scheme:                                             
\be \label{eq:sigctqq}
  \sigma^{(c)}_{y=1} \= \frac{1}{\epsbar} \,\frac{\as}{2\pi} \left(
  \frac{\mu^2}{\mufsq }\right)^{\eps} \, \cpgq{x}\, x\sborno \; ,
\ee                                                         
where $\cpgq{x}$~is defined in Appendix~\ref{appA}.
                                       
Following a procedure analogous to the one detailed in the case of $gg$
production, we find the following result:  
\ba
\sigma^H[g\,q\to \q^{[1]} \,q] &=& \frac{\as}{\pi}
\sigma_0^H[gg\to \q^{[1]}]\nn\\
&&\times \left\{\left[\frac{x}{2}
   {\cal P}_{gq}(x)\log\frac{s(1-x)^2}{\qf} + \cf\frac{x^2}{2}\right]+
   f_{gq}[\q^{[1]}](x)\right\}\, , \nn\\[10pt]
&& [\q^{[1]}=\etas,\chizs,\chits] \label{eq:qgP1}\; ,
\ea                                              
where the functions $f_{gq}(x)$ are collected in
Appendix~\ref{appNLO}, together with the result for 
$\threePone^{[1]}$
production, for which no collinear singularity is present to start with.
Our result for $^1S_0^{[1]}$ production agrees with those presented
in refs.~\cite{Kuehn93,Schuler94}. In the case of $\threePzero^{[1]}$ and 
$\threePtwo^{[1]}$
we differ from ref.~\cite{Schuler94} by the factor $\cf x^2/2$ in
eq.(\ref{eq:qgP1}). This piece arises from the \MSB\ factorization
prescription, which dictates use of the $D=4$ Altarelli-Parisi kernel in the
collinear counter-term, eq.~(\ref{eq:sigctqq}). 

%\dschio[g\,q\to \chios\,q] &=& \frac{\pi^2\ascube}{{(2
%   m)}^7}
%   f_{gq}[\chios](x)\langle 0 | {\cal O}^H_1(\threePone)| 0 \rangle
%\ea
% 
%\dschij[g\,q\to \psih\,q] &=& \frac{\as}{\pi}\szchij[\psih]\times\nn\\
%&&\left\{\left[\frac{x}{2}
%   P_{qg}(x)\log\frac{s(1-x)^2}{\qf} + \tf x^2(1-x)\right] +
%   f_{gq}[\psih](x)\right\}
%\ea

\subsubsection{$\qq\to g\q^{[1]}$ processes}
In this case no collinear divergences are present upon integration over the
final-parton phase space. No virtual corrections are present either.
Infrared divergences however appear, associated to the 
presence of an intermediate $^3S_1^{[8]}$ state:
\ba
\sigma^H[\qq\to \chijs \,g] &=& -\frac{\cf}{D_F^2}
\frac{256\pi^2\ascube\mu^{4\ep}}{9 {(2
   m)}^7}\beta^{-4\ep}\feps{s}  \left(\frac{1}{\ep} +
   \frac{4}{3}\right)\delta(1-x)
   \langle \o^H_1(\threePJ)\rangle \nn\\
&&+ \frac{\pi^2\ascube\mu^{4\ep}}{{(2m)}^7}
   \left(\frac{1}{1-x}\right)_{\rho} f_{\tiny\qq}[\chijs](x)
   \langle \o^H_1(\threePJ)\rangle\qquad [J=0,1,2]\, ,
\label{qqg1p}
\ea
where $\df$ is defined in Appendix~\ref{appA}.
The infrared divergences is absorbed by the inclusion of the $\qq
\to {}^3S_1^{[8]}$ process, as discussed in Section~\ref{sec:opren}.
The resulting finite expressions and the functions $f_{qq}(x)$, as well as the
trivial result for the $\etas$ production, 
are collected in Appendix~\ref{appNLO}.    
%\ba                                                                  
%\dschij[\qq\to\psih \, g] &=& \frac{\as}{\pi}\szchij[\psih]\nn\\
%&&\left\{\feps{s}  \left[\cf\left(\frac{1}{\ep^2}+\frac{3}{2\ep}
%   -\frac{\pi^2}{3} +8\log^2\beta\right)+\right.\right.\nn\\
%&&\left.\left. 
%\ca\left(\frac{1}{2\ep}+1-2\log\beta\right)\right]\delta(1-x)+\right.\nn\\
%&&\left[x {\cal P}_{qq}\log\frac{s}{\qf} +\cf
%   x(1-x) + 2 x(1-x) P_{qq}(x)\left(\frac{\log(1-x)}{1-x}\right)_{\rho}\right]
%   +\nn\\
%&&\left.\left(\frac{1}{1-x}\right)_{\rho}
%   f_{\tiny\qq}[\psih](x)\right\}
%\ea

\subsubsection{$gg\to g\q^{[8]}$ processes}
The diagrams of fig.~\ref{figgg-g} 
have been evaluated in 4-dimensions for colour-octet production
in ref.~\cite{Cho96}.  Since the Born cross sections   
$gg \to \q^{[8]}$ do not all vanish we expect both infrared and collinear
singularities to appear.
After subtraction of the collinear poles in the $\overline{MS}$ scheme the
partonic cross sections read
\ba      
\dsh[g\,g\to\psih\,g] &=& \frac{\ascube \pi^2}{{(2 m)}^5}
f_{gg}[\psih](x)
\langle \o^H_8(^3S_1)\rangle\\
&&\nn\\[20pt]
\dsh[g\,g\to\chioh\,g] &=& \frac{\ascube \pi^2}{{(2 m)}^7}
f_{gg}[\chioh](x)
\langle \o^H_8(^3P_1) \rangle\\
&&\nn\\[20pt]
\dsh[g\,g\to \q^{[8]}\,g] &=& \frac{\as}{\pi}\szh[gg\to \q^{[8]}] \nn\\
&&\times \left\{\feps{s}   \left[\ca\left(\frac{1}{\ep^2}+
\frac{7}{3\ep} +1-\frac{\pi^2}{3} -2\log\beta+8\log^2\beta\right)-
 \frac{2}{3\ep}\nf\tf
\right]\delta(1-x)  \right.\nn\\
&&+ \left.\left[x {\cal
P}_{gg}(x)\log\frac{s}{\qf} + 2 x(1-x)
P_{gg}(x)\left(\frac{\log(1-x)}{1-x}\right)_{\rho}\right.\right. \nn\\
&&+\left.\left.\left(\frac{1}{1-x}\right)_{\rho} f_{gg}[\q^{[8]}](x)
\right]\right\}\, ,\nn\\[10pt]                                      
&&[\q^{[8]}=\etah,\chizh,\chith]\, ,
\ea                                 
with the functions $f_{gg}[\q^{[8]}](x)$ collected in Appendix~\ref{appNLO}.
                                                                            
\subsubsection{$qg\to q\q^{[8]}$ processes}
These processes proceed via the diagrams of fig. \ref{figgq-q}. Diagrams b) and
c) only contribute to $\psih$ production.

\begin{figure}[t]
\begin{center}
\epsfig{file=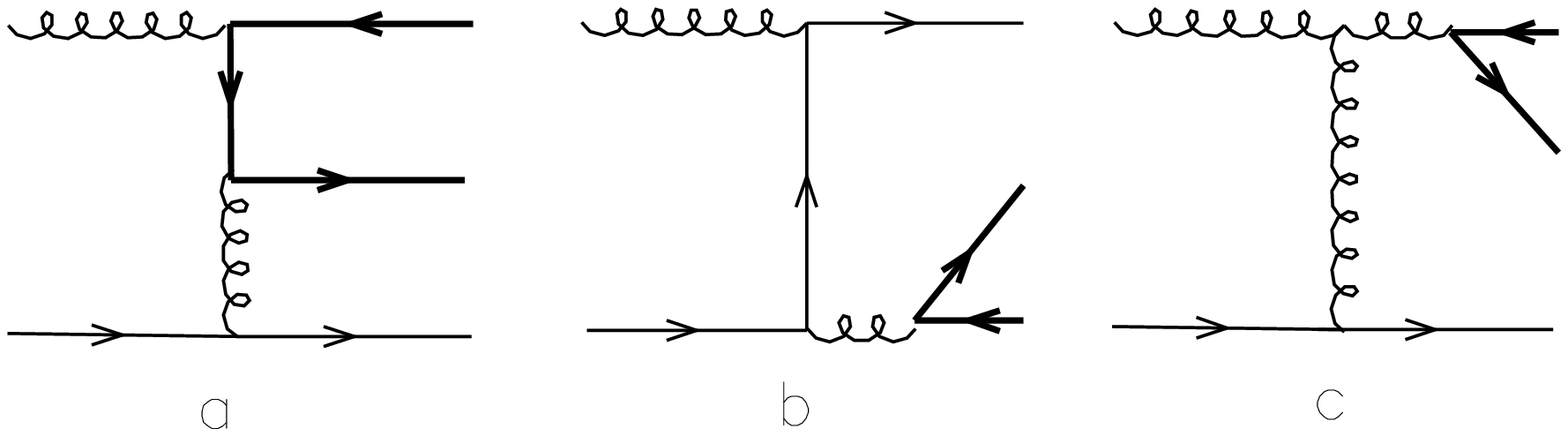,width=12cm,clip=}
\parbox{12cm}{
\caption{\label{figgq-q} \small Diagrams for  the $gq$
channels. Reversal of fermion lines is always implied.
}
}
\end{center}
\end{figure}

Since no ${\cal O}(\assq)$ cross sections exist for these channels the results
are expected to be infrared finite, only collinear singularities being allowed
as long as $gg \to \q^{[8]}$ is non vanishing. Notice the special case of
$\psih$ production, whose collinear singularity is rather subtracted by the
$\qq \to \psih $ Born term.

The cross sections read:
\ba 
\dsh[g\,q\to \psih\,q]&=& \frac{\as}{\pi}\szh[\qq\to\psih] \nn\\
&& \times\left\{\left[\frac{x}{2}
{\cal P}_{qg}(x)\log\frac{s(1-x)^2}{\qf} + \tf
x^2(1-x)\right]+ f_{gq}[\psih](x)\right\}\\ 
&&\nn\\
&&\nn\\
\dsh[g\,q\to \chioh\,q] &=& \frac{\pi^2\ascube}{{(2 m)}^7}
f_{gq}[\chioh](x)
\langle \o^H_8(^3P_1)\rangle\\
&&\nn\\
&&\nn\\
\dsh[g\,q\to \q^{[8]}\,q] &=& \frac{\as}{\pi}\szh[gg\to\q^{[8]}] \nn\\
&&\times\left\{\left[\frac{x}{2}
{\cal P}_{gq}(x)\log\frac{s(1-x)^2}{\qf} + \cf\frac{x^2}{2}\right]
+  f_{gq}[\q^{[8]}](x)\right\}\\[10pt]
&&[\q^{[8]}=\etah,\chizh,\chith]\nn
\ea

\subsubsection{$\qq\to g\q^{[8]}$ processes}
\begin{figure}[t]
\begin{center}
\epsfig{file=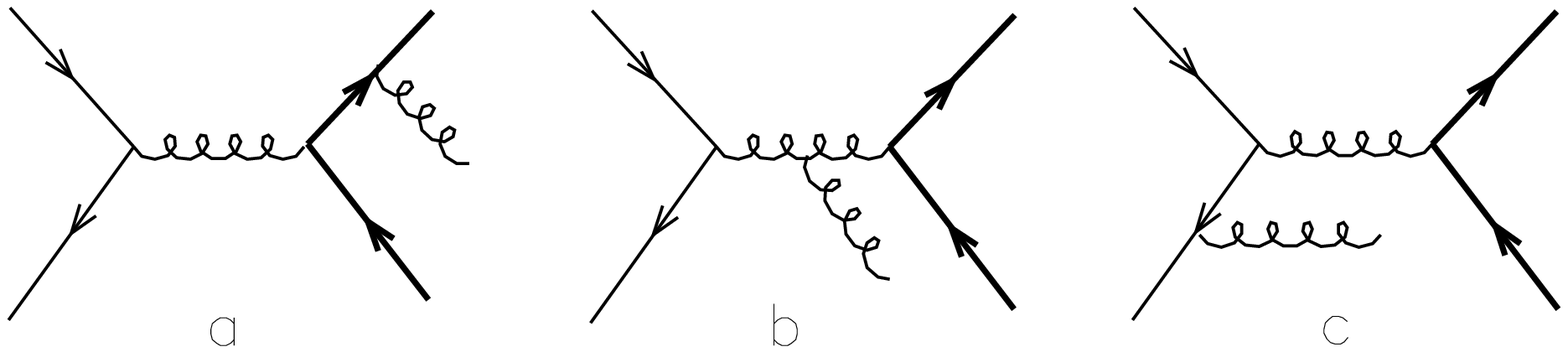,width=12cm,clip=}
\parbox{12cm}{
\caption{\label{figqq-g} \small Diagrams for the real corrections to the $\qq$
channels. Permutations of outgoing gluons and/or reversal of fermion lines are
always implied.
}
}
\end{center}
\end{figure}
The diagrams for these channels are depicted in fig. \ref{figqq-g}. 
After subtraction of the collinear singularities the cross sections read
%\ba
%\dsh[\qq\to\etah \, g] &=& \frac{\ascube \pi^2}{(2m)^5}
%f_{\tiny\qq}[\etah](x)
%\langle \o^H_8(^1S_0) \rangle\\
%&&\nn\\[20pt]
%\ea
\ba
\dsh[\qq\to\psih \, g] &=&\frac{\as\mu^{2\ep}}{\pi}\szh[\qq\to\psih]\times
  \left\{\feps{s}  \left[\cf\left(\frac{1}{\ep^2}+\frac{3}{2\ep}
   -\frac{\pi^2}{3} +8\log^2\beta\right)\right.\right.\nn\\
&&+ \left.\left.
\ca\left(\frac{1}{2\ep}+1-2\log\beta\right)\right]\delta(1-x)\right.\nn\\
&&+\left[x {\cal P}_{qq}(x)\log\frac{s}{\qf} +\cf
   x(1-x) + 2 x(1-x) P_{qq}(x)\left(\frac{\log(1-x)}{1-x}\right)_{\rho}\right]
   \nn\\
&&+\left.\left(\frac{1}{1-x}\right)_{\rho}
   f_{\tiny\qq}[\psih](x)\right\}\\
&&\nn\\[20pt]
\dsh[\qq\to\chijh \,g] &=& -\frac{\Bf}{D_F^2}
\frac{256\pi^2\ascube\mu^{6\ep}}{9 {(2m)}^7}
  \beta^{-4\ep}\feps{s}  \left(\frac{1}{\ep} +
  \frac{4}{3}\right)\delta(1-x)
\langle \o^H_8(^3P_J) \rangle \nn\\
&&+\frac{\pi^2\ascube}{{(2
   m)}^7}
\left(\frac{1}{1-x}\right)_{\rho} f_{\tiny\qq}[\chijh](x)
\langle \o^H_8(^3P_J)\rangle\qquad [J=0,1,2]\, ,
\label{qqg8p}
\ea
and $\Bf$ and $\df$ are defined in Appendix~\ref{appA}.

A few comments are in order.
  
The production of $\psih$ shows singularities of both collinear and
infrared type. The latter will be cancelled by the addition of the virtual
corrections to $\qq\to\psih$.

The $\dsh[\qq\to\chijh \,g]$ cross sections only show a singularity
which is neither collinear nor infrared in the usual sense, i.e. to be
cancelled by the addition of virtual corrections. The Born cross sections
$\qq\to\chijh$ are indeed zero, therefore not allowing for such cancellations
to take place. This singularity is originated by the gluon emission from the
heavy quark line, and it can only be eliminated, in the spirit of the
factorization approach, by its re-absorption into the matrix element 
$\langle\o^H_8(^3S_1) \rangle $
when adding to this cross section the degenerate one (in the infrared endpoint)
$d\sigma^{H}[\qq\to \psih]$ (see Section~\ref{sec:opren}).

The process $q\overline q\rightarrow\oneSzero^{[8]}g$ is completely finite and it is shown in the Appendix~\ref{appA} together with the functions $f_{\tiny\qq}[\q^{[8]}]$.

\subsection{Inclusion of virtual corrections}

In order to obtain the full ${\cal O}(\ascube)$ correction 
we need to calculate the one-loop QCD
corrections to the four non vanishing Born processes $\qq\to\QQ[\psih]$, 
$gg\to\QQ[^1S_0^{[1,8]}]$, $gg\to\QQ[^3P_0^{[1,8]}]$ and 
$gg\to\QQ[^3P_2^{[1,8]}]$. The relevant
diagrams for the $\qq$ and $gg$ channels respectively are shown in figs.
\ref{fig:qq} and \ref{fig:gg}.
 The diagrams are depicted in fig. \ref{fig:qq} for the 
$\qq$ initiated process and in fig. \ref{fig:gg} for the $gg$ one. The full
virtual  correction is constructed from the various parts listed in the tables
by:
\be
\sigma_V^H[ij\to\q] = \frac{\as\mu^{2\ep}}{\pi}\sigma_0^H[ij\to\q]
\,  f_{\epsilon}(s) \sum_k {\cal D}_k f_k\,\delta(1-x). 
\ee
Explicit expressions for $\sum_k {\cal D}_k f_k$ can be read out of
eqs~(\ref{eq:virtual},~\ref{eq:virtual8},~\ref{eq:virtualqq}). The sums of real
and virtual corrections take simple forms. In the case of gluon-initiated colour-singlet production one obtains:                
\ba
\dsh[g\,g\to \q^{[1]} \, X] &=& \szh[gg\to\q] \left(\delta(1-x) 
+ \frac{\as}{\pi}                                
\left\{ \;  \left[  B_{\q^{[1]}} - \ca\frac{\pi^2}{3} \right.\right.\right.
       \nn \\
  &+& \left. \left. \left. 8\ca \,\log^2\beta
     + 2\, {b_0}\,\log {{\mur }\over {\muf}} +
     8\ca\log\beta \log\frac{2m}{\muf} \right]\; \delta(1-x)\right.\right.
\nn\\ 
&+&\left[x {\overline P}_{gg}(x)\log\frac{4m^2}{x\qf} + 2     
   x(1-x) P_{gg}(x)\left(\frac{\log(1-x)}{1-x}\right)_{\rho}  \right.\nn\\
&+& \left. \left. \left.\left(\frac{1}{1-x}\right)_{\rho} f_{gg}[\q^{[1]}](x)
   \right]\right\}\right)\, ,\qquad  [\q^{[1]}=\etas,\chizs,\chits]
\ea                                                                
where the coefficients $B_{\q^{[1]}}$ are the finite parts of the virtual
corrections, defined by eq.~(\ref{eq:virtual}).
A similar expression holds for the  $ g\,g\to \q^{[8]}\, X$ and   $q\bar q\to \psih\, X$ NLO cross sections.
The final results for the finite sums of real plus virtual corrections are
collected in Appendix~\ref{appNLO}.

\section{Conclusions}

We have performed  a calculation of next-to-leading-order QCD
corrections to total hadronic cross sections and to light-hadron decay rates of 
heavy quarkonium states. Both colour singlet and colour octet contributions  
are included.

We have extended the technique  of covariant projections to $D$ dimensions,
to use it within dimensional regularization. Where common results exist, they
have been found in agreement with the ones given by the
threshold-expansion technique recently introduced by Braaten and Chen.

All the singularities which develop during the calculation are found to cancel
either via the usual QCD mechanisms (soft/virtual cancellation, collinear
factorization, mass and coupling renormalization) or via NRQCD cancellation
mechanisms, by taking into account radiative corrections to the NRQCD matrix
elements. The final formulas are therefore self-contained and do not need
ad-hoc prescriptions to produce finite results.

Results for  quarkonia    
photoproduction and decay into one photon plus light hadrons can be easily
extracted from these calculations: explicit results will be presented in a
forthcoming publication.

A phenomenological study of quarkonia NLO total production  cross sections and
decay rates, including all the  processes calculated in the present work, will
be the subject of a separate publication.

\vspace{1cm}
{\bf Acknowledgments.} We thank Peter Cho for providing us with his code
for the evaluation of the 4-dimensional, $\oacube$
colour-octet matrix elements, and
Paolo Nason for sharing with us his 
results on massive scalar integrals.
M.C. and A.P. thank the CERN Theory
Division for the hospitality on various occasions and for supporting their
stay there while  part of this work was performed.

%%%%%%%%%%%%%%%%%%%%%%%%%%%%%%%%%%%%%%%%%%%%%%%%%%%%%%%%%%%%%%%%%%%%%%%
%
%
%   Appendici
%
%
%%%%%%%%%%%%%%%%%%%%%%%%%%%%%%%%%%%%%%%%%%%%%%%%%%%%%%%%%%%%%%%%%%%%%%%

\appendix
\section{Symbols and notations}
\label{appA}
This Appendix collects the meaning of various symbols which are used throughout
the paper.
\\[0.2cm]
\underline{Kinematical factors}:
\be                            
M = 2m \; , \quad
v={\sqrt{1-\frac{M^2}{s}}}\; , \quad
\rho= \frac{M^2}{S_{had}}\; , \quad
\beta=\left(1-\rho\right)^{\frac{1}{2}}\; ,
\ee
where $s$ is the partonic center of mass energy squared and $S_{had}$ is the  
hadronic one. $v$ is the velocity of the bound (anti)quark in 
the quarkonium rest frame, $2v$ being then the relative velocity of the quark
and the antiquark. We also define:
\be                                          
\feps{Q^2} = \left(\frac{4 \pi\mu^2}{Q^2}\right)^{\epsilon}\Gamma(1+\ep) 
= 1+\ep\left(-\gamma_E +\log(4\pi) + \log{\mu^2 \over Q^2}\right) +
  {\cal O}(\epsilon^2) \,,
\ee   
and we denote a perturbative $\QQ$ state with generic spin and angular momentum
quantum numbers and in a colour-singlet or colour-octet state
by the symbol                                         
\be
{\cal Q}^{[1,8]}  \equiv \QQ[\spectr^{[1,8]}] \; .
\ee                                               
\\[0.2cm]
\underline{Altarelli-Parisi splitting functions}. Several functions related to
the AP splitting kernels enter in our calculations. We collect here our
definitions:
\ba                                             
&& P_{qq}(x) = \cf\left[\frac{1+x^2}{1-x}-\ep(1-x)\right]  \\
&&{\overline P}_{qq}(x) = \cf\frac{1+x^2}{(1-x)_{\rho}}\\
&&{\cal P}_{qq}(x) = {\overline P}_{qq} +\cf \left(\frac{3}{2}+4
\log\beta\right)\delta(1-x) \\
&&P_{qg}(x) = \tf\left[x^2+(1-x)^2-2\ep\ x(1-x)\right]\\
&&{\cal P}_{qg}(x) = \tf\left[x^2+(1-x)^2\right]\\
&&P_{gq}(x)=\cf\left[\frac{1+(1-x)^2}{x}-\ep\ x\right]\\
&&{\cal P}_{gq}(x)=\cf\left[\frac{1+(1-x)^2}{x}\right]\\
&&P_{gg}(x) = 2\ca\left[\frac{x}{1-x}+\frac{1-x}{x}+x(1-x)\right]\\
&&{\overline P}_{gg}(x) = 
2\ca\left[\frac{x}{(1-x)_{\rho}}+\frac{1-x}{x}+x(1-x)\right] \\
&&{\cal P}_{gg}(x) = {\overline P}_{gg}(x) + \left(\b0+4
\ca\log\beta\right)\delta(1-x)
\ea
where 
\be           
\b0 = \frac{11}{6}\ca- \frac{2}{3} \tf\nf
\ee
with $\nf$ number of flavours {\sl{lighter}} than the bound one.  
The $P_{ij}$ are the $D$-dimensional splitting
functions which appear in the factorization of collinear singularities from
real emission,while the functions ${\cal P}_{ij}$ are the four-dimensional AP
kernels, which enter in the \MSB\ collinear counter-terms.
The $\rho$-distributions are defined by:                  
\be                                                                       
   \int_{\rho}^{1} \, dx \, \left[d(x)\right]_{\rho} t(x) \=
   \int_{\rho}^{1} \, dx \; d(x) \; \left[t(x)-t(1)\right] \; .
\ee                                                   
\\[0.2cm]
\underline{Colour Algebra}
\ba
[T^a, T^b] &=& i f^{abc} T^c \\
\{T^a, T^b\} &=& d^{abc} T^c + \frac{\delta^{ab}}{\nc}
\ea
\be
\begin{array}{ll}
 \tr(T^a T^b) = \tf \delta^{ab} &\tf = \frac{1}{2}\\
\sum_a (T^a T^a)_{ij} = \cf \delta_{ij}
&\cf = \frac{N_c^2-1}{2\nc} = \frac{4}{3}\\
\sum_{bc} f^{abc} f^{ebc} = \ca \delta^{ae} &\ca = \nc = 3 \\
\sum_{bc} d^{abc} d^{ebc} = 4 \Bf \delta^{ae} 
&\Bf = \frac{N_c^2-4}{4 \nc} = \frac{5}{12} \\
\sum_{abc} d^{abc} (T^a T^b T^c)_{ij} = C_2(F) \delta_{ij}
&C_2(F) = \frac{(N_c^2-4) (N_c^2-1)}{4 N_c^2} = \frac{10}{9} \\
\end{array}
\ee
\ba
\df &=& \sum_i \delta_{ii} = \nc = 3 \\
\da &=& \sum_a \delta^{aa} = N_c^2 -1 =8 
\ea
The following formulas were  found to be useful:
\ba
\sum_a T^a_{ij} T^a_{kl}   &=& \frac{1}{2}
\left( \delta_{il} \delta_{jk} -          
\frac{1}{\nc} \delta_{ij} \delta_{kl} \right)\\ 
\tr (T^a T^b T^c) &=& \frac{1}{4} ( d^{abc} + i f^{abc} ) \\
\tr (T^a \{T^b ,T^c\})&=& \frac{1}{2} d^{abc}\\
\cf \df &=& \da / 2\\
C_2(F) \df &=& \Bf \da 
\ea
\\[0.2cm]
\underline{NRQCD operators}.
Although we never make explicit use of them, we collect here for ease of
reference the definitions of the NRQCD operators relative to the states
considered in this work (see also
\cite{bbl} for details and a comparison with theirs). 
In the case of $S$ states we have: 
\ba                               
&&{\o}_1(\oneSzero) = \psi^\dagger\frac{1}{\sqrt{2}}\frac{1}{\sqrt{\nc}}\chi
              \chi^\dagger\frac{1}{\sqrt{2}}\frac{1}{\sqrt{\nc}}\psi\\
&&{\o}_1(\threeSone) = 
 \psi^\dagger\frac{\mbox{\boldmath $\sigma$}}{\sqrt{2}}\frac{1}{\sqrt{\nc}}\chi
\cdot \chi^\dagger\frac{\mbox{\boldmath $\sigma$}}{\sqrt{2}}\frac{1}{\sqrt{\nc}}\psi\\
&&{\o}_8(\oneSzero) = \psi^\dagger\frac{1}{\sqrt{2}}\sqrt{2}T^a\chi
                      \chi^\dagger\frac{1}{\sqrt{2}}\sqrt{2}T^a\psi\\
&&{\o}_8(\threeSone) = \psi^\dagger\frac{\mbox{\boldmath $\sigma$}}{\sqrt{2}}\sqrt{2}T^a\chi
            \cdot  \chi^\dagger\frac{\mbox{\boldmath $\sigma$}}{\sqrt{2}}\sqrt{2}T^a\psi
\ea
and for $P$ states:
\ba                
&&{\o}_1(^1P_1) = \psi^\dagger(-{i\stackrel{\leftrightarrow}{\bf D}\over 2})
           \frac{1}{\sqrt{2}}\frac{1}{\sqrt{\nc}}\chi \cdot
              \chi^\dagger(-{i\stackrel{\leftrightarrow}{\bf D}\over 2})
              \frac{1}{\sqrt{2}}\frac{1}{\sqrt{\nc}}\psi\\
&&{\o}_1(^3P_0) = \frac{1}{3}
           \psi^\dagger(-{i\stackrel{\leftrightarrow}{\bf D}\over 2}\cdot
            \frac{\mbox{\boldmath $\sigma$}}{\sqrt{2}}   )
           \frac{1}{\sqrt{\nc}}\chi
              \chi^\dagger(-{i\stackrel{\leftrightarrow}{\bf D}\over 2}\cdot
              \frac{\mbox{\boldmath $\sigma$}}{\sqrt{2}})
              \frac{1}{\sqrt{\nc}}\psi\\
&&{\o}_1(^3P_1) = \frac{1}{2}
           \psi^\dagger(-{i\stackrel{\leftrightarrow}{\bf D}\over 2}\times
            \frac{\mbox{\boldmath $\sigma$}}{\sqrt{2}}   )
           \frac{1}{\sqrt{\nc}}\chi\cdot
              \chi^\dagger(-{i\stackrel{\leftrightarrow}{\bf D}\over 2}\times
              \frac{\mbox{\boldmath $\sigma$}}{\sqrt{2}})
              \frac{1}{\sqrt{\nc}}\psi\\
&&{\o}_1(^3P_2) = 
           \psi^\dagger(-{i\stackrel{\leftrightarrow}{D^{(i}}\over 2}
            \frac{\sigma^{j)}}{\sqrt{2}}   )
           \frac{1}{\sqrt{\nc}}\chi
              \chi^\dagger(-{i\stackrel{\leftrightarrow}{D^{(i}}\over 2}
              \frac{\sigma^{j)}}{\sqrt{2}})
              \frac{1}{\sqrt{\nc}}\psi\\
&&{\o}_8(^1P_1) = \psi^\dagger(-{i\stackrel{\leftrightarrow}{\bf D}\over 2})
           \frac{1}{\sqrt{2}}\sqrt{2}T^a\chi \cdot
              \chi^\dagger(-{i\stackrel{\leftrightarrow}{\bf D}\over 2})
              \frac{1}{\sqrt{2}}\sqrt{2}T^a\psi\\
&&{\o}_8(^3P_0) = \frac{1}{3}
           \psi^\dagger(-{i\stackrel{\leftrightarrow}{\bf D}\over 2}\cdot
            \frac{\mbox{\boldmath $\sigma$}}{\sqrt{2}}   )
           \sqrt{2}T^a\chi
              \chi^\dagger(-{i\stackrel{\leftrightarrow}{\bf D}\over 2}\cdot
              \frac{\mbox{\boldmath $\sigma$}}{\sqrt{2}})
              \sqrt{2}T^a\psi\\
&&{\o}_8(^3P_1) = \frac{1}{2}
           \psi^\dagger(-{i\stackrel{\leftrightarrow}{\bf D}\over 2}\times
            \frac{\mbox{\boldmath $\sigma$}}{\sqrt{2}}   )
           \sqrt{2}T^a\chi\cdot
              \chi^\dagger(-{i\stackrel{\leftrightarrow}{\bf D}\over 2}\times
              \frac{\mbox{\boldmath $\sigma$}}{\sqrt{2}})
              \sqrt{2}T^a\psi\\
&&{\o}_8(^3P_2) = 
           \psi^\dagger(-{i\stackrel{\leftrightarrow}{D^{(i}}\over 2}
            \frac{\sigma^{j)}}{\sqrt{2}}   )
           \sqrt{2}T^a\chi
              \chi^\dagger(-{i\stackrel{\leftrightarrow}{D^{(i}}\over 2}
              \frac{\sigma^{j)}}{\sqrt{2}})
              \sqrt{2}T^a\psi
\ea
Writing all previous operators as products of fermion bilinears,
$\o(n)=\o_2(n)\o_2(n)$,
we also recall the definition of the various NRQCD
matrix-elements appearing in the production cross sections:
\be
\langle {\cal O}^H_{[1,8]}(n)\rangle \= \sum_X 
  \langle 0\vert {\cal O}_2^{(n)\dagger}\vert H\, X\rangle
\langle H\, X\vert {\cal O}_2^{(n)}\vert 0\rangle  \; .
\ee                                                      
Notice that, according to the discussion in
Section~\ref{sec:projectors},
our conventions differ slightly from those introduced in
ref.~\cite{bbl} (and labelled here as BBL):
\ba                                       
\label{eq:opnorm}
&&{\o }_1 = {1\over{2}} {1\over{\nc}}  {\o }_1^{\rm BBL}, \\
&&{\o }_8 = {1\over{2}} 2 {\cal O}_8^{\rm BBL} = {\o }_8^{\rm BBL} \; .
\ea

\section{Summary of $O(\assq)$ Results}
\label{appLO}      
The $D$-dimensional cross sections and the decay rates read
\ba
&&\sigma(ij\to\spectr^{[1,8]}\to H) = \hat \sigma(ij\to\spectr^{[1,8]}) 
{{\langle{\o }_{[1,8]}^H(\spectr)\rangle}
\over{N_{col} N_{pol}}},\\[10pt]
&&\Gamma(H\to\spectr^{[1,8]}\to ij) = \hat \Gamma(\spectr^{[1,8]}\to ij)
\langle H|{\o }_{[1,8]}(\spectr)|H\rangle\, ,
\ea
the short distance coefficients $\hat\sigma$ and $\hat\Gamma$ having been 
calculated according to the rules of Section~\ref{sec:projectors}.
$N_{col}$ and $N_{pol}$ refer to the number of colours and polarization states
of the $\QQ[^{2S+1}L_J^{[1,8]}]$ pair produced. They are given by 1 for 
singlet states or $\da = N_c^2-1$ for octet states, 
and by the $D$-dimensional $N_J$'s defined
in Section~\ref{sec:projectors}.
Recall that  the matrix elements appearing in the equations
above are meant to be the bare $D$-dimensional ones. 
Making use of their correct
mass-dimension, $3-2\ep$ and $5-2\ep$ for $S$ and $P$ wave states respectively
(see Section~\ref{sec:opren}), gives the right dimensionality
to $D$-dimensional cross sections and widths, i.e. $2-D = -2+2\ep$ 
and one.

\subsection{$O(\assq)$ decay rates}
We shall use the short-hand notation
\be
\Gamma^H(\q^{[1,8]} \to i j ) \equiv \Gamma(H  \to \q^{[1,8]} \to i j ) 
\ee                                                                     
to indicate the decay  of the physical quarkonium state
$H$ via the intermediate $\QQ$ state ${\q}^{[1,8]} = \QQ[\spectr^{[1,8]}]$.
                         
The $D$-dimensional Born level decay rates read:
\ba
&&\gbh(\oneSzero^{[1]}\to gg) =\cf \, 
{{16 \assq\mu^{4\ep}\pi^2}\over{m^2}}
\Phi_{(2)}
 (1-\ep)(1-2\ep) 
\langle H |{\o}_1(\oneSzero)|H\rangle\\&&\nn\\
&&\gbh(\threeSone^{[1]}\to gg) = 0\\
&&\gbh(\threePzero^{[1]}\to gg) =\cf \,
{{144\assq\mu^{4\ep}\pi^2}\over{m^4}}
\Phi_{(2)}
{{1-\ep}\over{3-2\ep}} 
\langle H |{\o}_1(\threePzero)|H\rangle \\&&\nn\\
&&\gbh(\threePone^{[1]}\to gg) =0\\
&&\gbh(\threePtwo^{[1]}\to gg) = \cf \, \,
{{32 \assq\mu^{4\ep}\pi^2}\over{m^4}}
\Phi_{(2)}
{{6-13\ep+4\ep^2}\over{(3-2\ep)(5-2\ep)}}
\langle H |{\o}_1(\threePtwo)|H\rangle 
\\&&\nn\\ 
&&\gbh(\threeSone^{[8]}\to \qq) = \nf {{8\assq\mu^{4\ep} \pi^2}\over{m^2}}
\Phi_{(2)}
{{1-\ep}\over{3-2\ep}}  
\langle H |{\o}_8(\threeSone)|H\rangle \\
&&\gbh(\oneSzero^{[8]}\to gg) =\Bf  {{16 \assq\mu^{4\ep}\pi^2}\over{m^2}}
\Phi_{(2)}
 (1-\ep)(1-2\ep) 
\langle H |{\o}_8(\oneSzero)|H\rangle\\&&\nn\\
&&\gbh(\threePzero^{[8]}\to gg) =\Bf
{{144\assq\mu^{4\ep}\pi^2}\over{m^4}}
\Phi_{(2)}
{{1-\ep}\over{3-2\ep}} 
\langle H |{\o}_8(\threePzero)|H\rangle \\
&&\gbh(\threePone^{[8]}\to gg) =0\\
&&\gbh(\threePtwo^{[8]}\to gg) = \Bf
{{32\assq\mu^{4\ep}\pi^2}\over{m^4}}
\Phi_{(2)}
{{6-13\ep+4\ep^2}\over{(3-2\ep)(5-2\ep)}}
\langle H |{\o}_8(\threePtwo)|H\rangle
\ea
where $\Phi_{(2)}$ is the two-body phase space given in the equation (\ref{phitwo}).

\subsection{$O(\assq)$ cross sections}
We shall use the short-hand notation
\be
\sigma^H(i j  \to \q^{[1,8]} ) \equiv \sigma(i j  \to {\q}^{[1,8]} \to H) 
\ee                                        
to indicate the production process  of the physical quarkonium state
$H$ via the intermediate $\QQ$ state ${\q}^{[1,8]} = \QQ[\spectr^{[1,8]}]$.
                         
The $D$-dimensional Born cross sections read:
\ba 
&&\sbh(gg \to \oneSzero^{[1]})= \frac{\cf}{D_A^2}
{{2 \assq\mu^{4\ep}\pi^3}\over{m^5}}
{{1-2\ep}\over{1-\ep}} \delta(1-x) \langle {\o}^H_1(\oneSzero)\rangle\\
&&\sbh(gg \to \threeSone^{[1]})= 0\\
&&\sbh(gg \to \threePzero^{[1]} ) =  \frac{\cf}{D_A^2}{{18\assq\mu^{4\ep}\pi^3}\over{ m^7}}
{1\over{(1-\ep)(3-2\ep)}} \delta(1-x) \langle {\o}^H_1(\threePzero)\rangle\\
&&\sbh(gg \to \threePone^{[1]} )= 0 \\
&&\sbh(gg \to \threePtwo^{[1]} ) =  \frac{\cf}{D_A^2}{{4 \assq\mu^{4\ep}\pi^3}\over{m^7}}{{6-13\ep+4\ep^2}\over{(1-\ep)(3-2\ep)}} \delta(1-x) 
{{\langle {\o}^H_1(\threePtwo)\rangle}\over{(1-\ep)(5-2\ep)}}\\&&\nn\\
&&\sbh(\qq \to \threeSone^{[8]}) = \frac{\da}{D_F^2}{{\assq\mu^{4\ep}\pi^3}\over{2 m^5}}(1-\ep)   \delta(1-x) 
{{\langle {\o}^H_8(\threeSone)\rangle} \over {\da (3-2\ep)}}\\
&&\sbh(gg \to \oneSzero^{[8]}) = 
\frac{\Bf}{\da}{{2\assq\mu^{4\ep}\pi^3}\over{ m^5}}
{{1-2\ep}\over{1-\ep}} \delta(1-x) 
{{\langle {\o}^H_8(\oneSzero)\rangle}\over{\da}} \\&\nn\\
&&\sbh(gg \to \threePzero^{[8]}) = \frac{\Bf}{\da} {{18\assq\mu^{4\ep}\pi^3}\over{ m^7}}
{1\over{(1-\ep)(3-2\ep)}} \delta(1-x) 
{{\langle {\o}^H_8(\threePzero)\rangle}\over{\da}}\\
&&\sbh(gg \to \threePone^{[8]} )=0\\
&&\sbh(gg \to \threePtwo^{[8]}) =  \frac{\Bf}{{\da}}
 {{4 \assq\mu^{4\ep}\pi^3}\over{m^7}}
{{6-13\ep+4\ep^2}\over{(1-\ep)(3-2\ep)}} \delta(1-x) 
{{\langle {\o}^H_8(\threePtwo)\rangle}\over{\da (1-\ep)(5-2\ep)}} 
\ea

\section{Summary of  $\oacube$ Results}
\label{appNLO}                 
\subsection{Decay}
We distinguish the two different cases in which the $O(\assq)$ subprocess is  
either $\q\to\qq$ or $\q\to g g$. In the former case the inclusive $O(\ascube)$ 
inclusive annihilation rate into light hadrons ($\lh$) of the quarkonium state 
$H$ through the component $\q$ can be written as follows:      
\ba                                                                            
  \Gamma^H(\q \to \lh ) = \Gamma^H_{\rm Born}(\q\to\qq) 
   + \Gamma^H(\q \to q \bar q g)+ \Gamma^H(\q \to ggg)+
   \Gamma_V^H(\q\to\qq)
\ea   
In the latter case one has instead:  
\ba                                                                            
  \Gamma^H(\q \to \lh ) = \Gamma^H_{\rm Born}(\q\to gg) 
   +  \Gamma^H(\q \to q \bar q g) + \Gamma^H(\q \to ggg)+
   \Gamma_V^H(\q\to gg)
\ea  

Given a quarkonium state $H$, we remark that, in order to  obtain an infrared
finite result, it is necessary to sum over all the intermediate configurations
$\q$ which contribute at any given order in the velocity $v$.  Typical cases
are  given by the $\chijs, \psih$  and $\chijh, \psih$  configuration pairs
that appear at the same order in $v$ for any quarkonium state $H$. The
corresponding  NRQCD operators then mix under renormalization group equation.

The inclusive decay rate is thus given by
\be
\Gamma(H \to\lh) = \sum_\q \Gamma^H(\q\to\lh)
\ee
and we list here the most interesting $O(\ascube)$ contributions to the 
annihilation of a state $H$ into light hadrons.
                              
\ba
\Gamma^H(\oneSzero^{[1]}\to\lh) &=& \gbh(\oneSzero^{[1]}\to gg) \left\{
1+\frac{\as}{\pi}\left[ \cf\left(-5+\frac{\pi^2}{4}\right)
\right.\right.\nn\\&&+\left.\left.
\ca\left(\frac{199}{18}-\frac{13}{24}\pi^2\right) - \frac{16}{9} \nf \tf + 2
\b0 \log\frac{\mur}{2 m}\right]\right\}\\                           
\nn\\\nn\\ 
\Gamma^H(\oneSzero^{[8]}\to\lh) &=& \gbh(\oneSzero^{[8]}\to gg) \left\{
1+\frac{\as}{\pi}\left[ \cf\left(-5+\frac{\pi^2}{4}\right)
\right.\right.\nn\\&&+\left.\left.
\ca\left(\frac{122}{9}-\frac{17}{24}\pi^2\right) - \frac{16}{9} \nf \tf + 2 \b0
\log\frac{\mur}{2 m}\right]\right\}\\
\nn\\ \nn\\
\Gamma^H (\threeSone^{[1]} \to \lh) &=&
     4 \ascube  
C_2(F)\left(-1+ \frac{\pi^2}{9} \right){{\langle
H|\o_1(\threeSone)|H\rangle}\over{m^2}}\\
\nn\\\nn\\                                
\Gamma^H(\threeSone^{[8]}\to\lh) &=& \gbh(\threeSone^{[8]}\to
\qq)\left\{1+\frac{\as}{\pi}\left[-\frac{13}{4}\cf\right.\right.\nn\\
&&+\left.\left. \ca\left(\frac{133}{18}+\frac{2}{3}\log 2
-\frac{\pi^2}{4}\right)-\frac{10}{9}\nf\tf +  2 \b0 
\log\frac{\mur}{2 m}\right]\right\}\nn\\ &&+  5\,\ascube              
     \left(- \frac{73}{4} +\frac{67}{36}\pi^2 \right)
     {{\langle H|\o_8(\threeSone)|H\rangle}\over{m^2}}\\
\nn\\\nn\\
\Gamma^H(\threePzero^{[1]} \to {\rm LH} ) &=& 
 \gbh(\threePzero^{[1]} \to gg) 
 \left\{1 + \frac{\as}{\pi}\left[\cf \left(   
 -\frac{7}{3}+\frac{\pi^2}{4}       \right) \right.\right.\nn\\
  &&+\ca\left.\left.  \left( \frac{427}{81}- \frac{1}{144}\pi^2  \right) 
       + 2 b_0 \log \frac{\mur}{2 m} \right]\right\} \nn\\
  &&+ \nf{\ascube}\frac{8}{9}\cf
\left( - {29\over 6} 
- \log{\mul\over{2m}}\right)\frac{
\langle H|\o_1(\threePzero)|H\rangle}{m^4}\\
\nn\\\nn\\
\Gamma^H(\threePone^{[1]} \to {\rm LH}) &=&
      \ascube \ca\cf     
     \left( \frac{587}{27} -\frac{317}{144}\pi^2 \right)
     {{\langle H|\o_1(\threePone)|H\rangle}\over{m^4}} \nn\\
&&+\nf{\ascube}\frac{8}{9}\cf  
\left( - {4\over 3}  - \log{\mul\over{2m}}\right)
\frac{\langle H|\o_1(\threePone)|H\rangle}{m^4}\\
\nn\\\nn\\
\Gamma^H(\threePtwo^{[1]} \to {\rm LH} ) &=&
 \gbh(\threePtwo^{[1]} \to gg) 
\left\{ 1 + \frac{\as}{\pi} \Big[ -4 \,\cf  \right.\nn\\
&&+\ca \left.\left. \left( \frac{2185}{216}-\frac{337}{384}\pi^2 + \frac{5}{3} 
\log 2 \right)+ 2 b_0 \log \frac{\mur}{2 m}\right]\right\}\nn\\
&&+ \nf {\ascube}\frac{8}{9}\cf 
\left( - {29\over 15} 
- \log{\mul\over{2m}}\right)
\frac{\langle H|\o_1(\threePtwo)|H\rangle}{m^4}\\
\nn\\\nn\\
\Gamma^H(\threePzero^{[8]} \to {\rm LH} ) &=& 
\gbh(\threePzero^{[8]} \to gg) 
 \left\{1 + \frac{\as}{\pi}\left[\cf \left(   
 -\frac{7}{3}+\frac{\pi^2}{4}       \right) \right.\right.\nn\\
  &&+\ca\left.\left.  \left( \frac{463}{81}+ \frac{35}{27}\log 2- 
\frac{17}{216}\pi^2  \right) 
       + 2 b_0 \log \frac{\mur}{2 m} \right]\right\} \nn\\
  &&+ \nf{\ascube}\frac{8}{9}\Bf
\left( - {29\over 6} 
- \log{\mul\over{2m}}\right)\frac{
\langle H|\o_8(\threePzero)|H\rangle}{m^4}\\
\nn\\\nn\\
\Gamma^H(\threePone^{[8]} \to {\rm LH}) &=&
      \ascube \ca\,\Bf   
     \left( \frac{1369}{54} -\frac{23}{9}\pi^2 \right)
     {{\langle H|\o_8(\threePone)|H\rangle}\over{m^4}} \nn\\
&&+\nf{\ascube}\frac{8}{9}\Bf   
\left( - {4\over 3}  - \log{\mul\over{2m}}\right)
\frac{\langle H|\o_8(\threePone)|H\rangle}{m^4}\\
\nn\\\nn\\
\Gamma^H(\threePtwo^{[8]} \to {\rm LH} ) &=& 
\gbh(\threePtwo^{[8]} \to gg) 
\left\{ 1 + \frac{\as}{\pi} \Big[ -4 \,\cf \right.\nn\\
&&+\ca \left.\left. \left( \frac{4955}{431} + \frac{7}{9}\log 2 
-\frac{43}{72}\pi^2 
\right)+ 2 b_0 \log \frac{\mur}{2 m}\right]\right\}\nn\\
&&+ \nf {\ascube}\frac{8}{9}\Bf
\left( - {29\over 15} 
- \log{\mul\over{2m}}\right)
\frac{\langle H|\o_8(\threePtwo)|H\rangle}{m^4}
\ea
We remind the reader that every time the NRQCD factorization scale $\mul$
appears the $\Gamma^H(\threeSone^{[8]}\to\lh)$ channel  has to be added for
consistency, and the $\langle H|\o_8(\threeSone)|H\rangle^{(\mul)}$
matrix element is then understood to be the renormalized one 
(see Section~\ref{sec:opren}).

We make now some comments on the comparison of our results with previous
calculations. The calculations for $\etas$ decays have already been 
performed in
refs.~\cite{Barbieri1S0,Hagiwara}. Our results agree with theirs. The
calculation of the $\oacube$ decay rate for $\chizs$ and $\chits$ first
appeared in ref.~\cite{Barbieri3PJ},
where a massive-gluon regulator was used to treat the IR
divergencies.                                     
Ours is the first independent calculation
since then. The results differ in the $\ca$ terms, for which we find:
\ba                                                
 &&  \ca \left( \frac{427}{81}- \frac{\pi^2}{144} \right) \quad \quad J=0 \;,
  \\                                              
 &&
     \ca \left( \frac{2185}{216}-\frac{337}{384}\pi^2 + \frac{5}{3}            
           \log 2 \right) \quad \quad J=2 \;,
\ea                                         
where Barbieri et al. find instead:
\ba
 &&  \ca \left( \frac{454}{81}- \frac{\pi^2}{144} \right) \quad \quad J=0 \;,
  \\                     
 &&
     \ca \left( \frac{2239}{216}-\frac{337}{384}\pi^2 + \frac{5}{3}            
           \log 2 \right) \quad \quad J=2 \;.
\ea                                         
The differences amount to a factor $\ca/3$ and $\ca/4$ for $J=0$ and $J=2$,
respectively. We point out that the $\oas$ corrections to the
two-photon decay rates of $\chizs$ and $\chits$, which can be obtained by
setting $\ca=0$ and by 
summing diagrams $a$ through $d$ of the virtual corrections displayed in
Tables~\ref{tab:virtual3P0} and \ref{tab:virtual3P2}, coincide with the results
of ref.~\cite{Barbieri3PJ}. 
                                       
The calculation of the $\oacube$ decay rate for $\etah$ has already been
performed in ref.~\cite{Huang96}. 
The results differ in the $\ca$ terms, for which we find:
\be 
 \ca \left( \frac{122}{9}- \frac{17}{24}\pi^2  \right) \;,
\ee                                                       
while Huang et al. find:
\be
 \ca \left( \frac{479}{36}- \frac{17}{24}\pi^2  \right) \;,
\ee                                                       
which differs by a factor of $\ca/4$ from ours.
A comparison with the diagram-by-diagram breakdown of the calculation performed
in ref.~\cite{Huang96} shows agreement in the evaluation of the virtual
corrections, and the discrepancy therefore originates
from the real-emission part\footnote{After this paper was released as a
preprint, the authors of ref.~\cite{Huang96} reviewed their calculation. We
have been informed that their final result now coincides with ours.}. 
                             
\subsection{Production}
We define:  
\be
      \sigma_0^H(ij\to {\cal Q}) \delta(1-x)
      \equiv \sigma_{\rm Born}^H(ij\to {\cal Q}) 
\ee
The \oacube\ cross sections are given as a function of the variable $x=M^2/s$.
\subsubsection{Colour-singlet channels}                 
\noindent
{\bf The $gg \to \q^{[1]} X$  channels}
                      
\ba
\dsh[g\,g\to\psis\,g]&=&\frac{\ascube\pi^2}{{(2 m)}^5}
f_{gg}[\psis](x)\langle\o_1^H(\threeSone)\rangle\\
f_{gg}[\psis](x) &=& \frac{C_2(F)}{D_A^2}\frac{256x^2}{3(-1+x)^2(1+x)^3}\left[2 + x + 2 x^2 - 4 x^4  - x^5\right.\nn\\ &&\left.+ 2 x^2 (5 + 2 x + x^2)\log x\right]\\
&&\nn\\
\dsh[g\,g\to\chios\,g] &=& \frac{\ascube\pi^2}{{(2 m)}^7}
   f_{gg}[\chios](x)\langle\o_1^H(\threePone)\rangle\\
f_{gg}[\chios](x)&=&\frac{\ca\,\cf}{D_A^2}\frac{256}{9(-1+x)^4(1+x)^5}\left[
 10 - 8 x - 157 x^2 - 9 x^3 - 6 x^4 + 136 x^5  \right.\nn\\
&&+ 404 x^6+26 x^7 - 212 x^8 - 144 x^9 - 39 x^{10} - x^{11} + ( - 48 x^2 \nn\\  
&&- 48 x^3 - 480 x^4 + 84 x^5   + 204 x^6   + 336 x^7   + 312 x^8   \nn\\
&&+ 108 x^9   + 12 x^{10}) \log x\left.\right]
\ea
\ba
\dsh[g\,g\to \q^{[1]}\, X] &=& \szh[gg\to\q^{[1]}] \left(\delta(1-x) 
+ \frac{\as}{\pi}                                
\left\{ \;  \Atot [\q^{[1]}] \;   \delta(1-x)\right.\right.\nn\\ 
&&+\left[x {\overline P}_{gg}(x)\log\frac{4m^2}{x\qf} + 2
   x(1-x) P_{gg}(x)\left(\frac{\log(1-x)}{1-x}\right)_{\rho}  \right.\nn\\
&&+\left. \left. \left.\left(\frac{1}{1-x}\right)_{\rho} f_{gg}[\q^{[1]}](x)
   \right]\right\}\right)\, ,\qquad  [\q^{[1]}=\etas,\chizs,\chits]
\ea
Where:
\ba
\Atot [\etas] &=& \cf \, 
     \left( -5  + \frac{\pi^2}{4} \right)  + 
     \ca\,\left( 1 + {\pi^2 \over 12}\right)   \nn\\
     &&+ 2\, {b_0}\,\log {{\mur }\over {\muf}} +
     8\ca\log\beta \log\frac{2m}{\muf} + 8\ca \log^2\beta\\        
\Atot [\chizs] &=& \cf \,
     \left( -{7\over 3} + {{{{\pi }^2}}\over 4} \right)  + 
     \ca\,\left( {1\over 3} + {\pi^2 \over 12} \right)   \nn\\
     &&+ 2\, {b_0}\,\log {{\mur }\over {\muf}} +
     8\ca\log\beta \log\frac{2m}{\muf}+8\ca\log^2\beta\\        
\Atot [\chits ] &=& -4 \,\cf + \ca \,
     \left( {1\over 3} - {{{{\pi }^2}}\over 6} + 
     {{5}\over 3}\log 2   \right)   \nn\\
  &&+ 2\, {b_0}\,\log {{\mur }\over {\muf}} +
     8\ca\log\beta \log\frac{2m}{\muf}+ 8\ca\log^2 \beta\\    
&&\nn\\
f_{gg}[\etas](x)&=&
\frac{\ca}{6(-1 + x)(1 + x)^3} \left[
   12 + 11x^2 + 24x^3 - 21x^4 - 24x^5 \right.\nn\\
&& \left. + 9x^6 - 11x^8 + 12(-1+ 5x^2 + 2x^3 \right.\nn\\
&& \left. + x^4 + 3x^6 + 2x^7)\log x ) \right]  \\
&&\nn\\   
f_{gg}[\chizs](x)&=&\frac{\ca}{54(-1 + x)^3(1 + x)^5}\left[
 172 - 56 x - 789 x^2 + 244 x^3 + 1721 x^4  
\right.\nn\\
&&- 696 x^5- 802 x^6 + 560 x^7 -  210 x^8 + 80 x^9 + 7 x^{10} - 132 x^{11}  \nn\\
&&- 99 x^{12} + 12 (-9 + 31 x^2 - 14 x^3 - 40 x^4 + 10 x^5 + 176 x^6  \nn\\
&&- 42 x^7 - 7 x^8 - 10 x^9 + 41 x^{10} + 24 x^{11}) \log x\left.\right]\\
&&\nn\\
f_{gg}[\chits](x)&=&\frac{\ca}{36(-1 + x)^3(1 + x)^5}\left[
 106 - 32 x - 207 x^2 + 271 x^3 + 752 x^4  \right.\nn\\ 
&&- 1032 x^5- 256 x^6 + 266 x^7 - 360 x^8 + 728 x^9 + 31 x^{10} - 201 x^{11}  \nn\\
&&- 66 x^{12}+ 12( - 6 + 22 x^2   - 8 x^3   + 74 x^4   + 31 x^5   + 11 x^6    \nn\\
&&- 204 x^7   + 86 x^8   + 17 x^9   + 5 x^{10}   + 12 x^{11}) 
\log x\left.\right]
\ea

\noindent
{\bf The $gq(\bar q) \to \q^{[1]} X$  channels}                             
         
\ba
\dsh[g\,q\to {\cal Q}^{[1]}\,q] &=& \frac{\as}{\pi}\szh[gg\to\q^{[1]}]\nn\\
&&\times \left\{\left[\frac{x}{2}                                    
   P_{gq}(x)\log\frac{4m^2(1-x)^2}{x\qf} + \cf\frac{x^2}{2}\right]+
   f_{gq}[{\cal Q}](x)\right\}\, , \nn\\[10pt]
&&[\q^{[1]}=\etas,\chizs,\chits] \\
&&\nn\\                                  
\dsh[g\,q\to \psis\,q] &=& 0\\
&&\nn\\
\dsh[g\,q\to \chios\,q] &=& \frac{\ascube\pi^2}{{(2
   m)}^7}f_{gq}[\chios](x)\langle\o_1^H(\threePone)\rangle
\ea
where
\ba
&&f_{gq}[\etas](x)= \cf(-1+x)(1-\log x) \\
&&f_{gq}[\chizs](x)=\frac{\cf}{9}\left[\frac{1}{3}(-1+x)(43-14 x+4
  x^2) + (9-9x+4x^2)\log x\right]\\
&&f_{gq}[\chits](x)=\frac{\cf}{12}\left[\frac{1}{3}(-1+x)(53-16 x+20
  x^2)+(12-12x+5x^2)\log x\right]\\
&&f_{gq}[\chios](x)=\frac{\cf}{D_F D_A} \frac{128}{3}\left[\frac{1}{3}(-1+x)(-5+4x+4x^2)-x^2\log x\right] 
\ea

\noindent
{\bf The $\qq\to \q^{[1]} X$ channels}

\ba
\dsh[\qq\to\etas \, g] &=& \frac{\ascube\pi^2}{(2m)^5} f_{\tiny\qq}[\etas]\langle\o_1^H(\oneSzero)\rangle \\
&&\nn\\
\dsh[\qq\to\psis \, g] &=&0\\
&&\nn\\                           
\dsh[\qq\to \chijs\,g] 
%+ \sigma^H[\qq\to\psih] 
 &=&\frac{\ascube\pi^2}{{(2m)}^7} \left[\frac{\cf}{D_F^2}
 \frac{512}{9}\left(2\log\beta - {5\over 6} - \log{{\mul}\over{2m}}\right)
   \delta(1-x)\right.\nn\\
&&+\left.\left(\frac{1}{1-x}\right)_{\rho} f_{\tiny\qq}[\chijs](x)\right]
 \langle\o^H_1(\threePJ)\rangle \nn\\[10pt]
%&&+ \sigma^H[\qq\to\psih]\, , \nn\\[10pt]
&&   [J=0,1,2]
\ea
where
\ba
&&f_{\tiny\qq}[\etas]= \frac{\cf}{D_F^2}\frac{32}{3}x^2(1-x)\\
&&f_{\tiny\qq}[\chizs]=\frac{\cf}{D_F^2}\frac{128}{9}x^2(1-3 x)^2\\
&&f_{\tiny\qq}[\chios]=\frac{\cf}{D_F^2}\frac{256}{9}x^2(1+x)\\
&&f_{\tiny\qq}[\chits]=\frac{\cf}{D_F^2}\frac{256}{45}x^2(1+3 x + 6 x^2)
\ea
%\ba
%&&f_{\tiny\qq}[\chijs]= \frac{8}{15}f_{\tiny\qq}[\chijh]\, ,\quad(J=0,1,2)
%\ea.
Note that, alike the $\Gamma^H[\chijs\to \qq g]$ process, 
$\dsh[\qq\to \chijs\,g]$ is only infrared finite when at least the Born term of
the $\sigma^H[\qq\to\psih]$ cross section is added to it. We have performed the
cancellation and an explicit dependence on the NRQCD renormalization $\mul$
results. The NRQCD matrix
element in $\sigma^H[\qq\to\psih]$ is then understood to be the renormalized,
$\mul$ scale dependent one $\langle\o^H_8(\threeSone)\rangle^{(\mul)}$.

\subsubsection{Colour-octet channels}
\noindent                   
{\bf The $gg \to \q^{[8]} X$  channels}

\ba
\dsh[g\,g\to\psih\,g] &=& \frac{\ascube \pi^2}{{(2 m)}^5}
f_{gg}[\psih](x)\langle\o_8^H(\threeSone)\rangle\\
f_{gg}[\psih](x)&=& \frac{1}{36 (-1 + x)^2 (1 + x)^3}\left[
     108 + 153 x + 400 x^2 + 65 x^3  \right.\nn\\   
&&- \left.   356 x^4 - 189 x^5 - 152 x^6 - 29 x^7 + 
  \left( 108 x + 756 x^2     \right.\right.\nn\\ 
&&+ \left.\left. 432 x^3 +  704 x^4   + 260 x^5   + 76 x^6 \right)
\log x\right]\label{fggpsih}\\
&&\nn\\
\dsh[g\,g\to\chioh\,g] &=& \frac{\ascube \pi^2}{{(2 m)}^7}
f_{gg}[\chioh](x)\langle\o_8^H(\threePone)\rangle\\
f_{gg}[\chioh](x) &=&
 \frac{\ca\Bf}{D_A^2}\frac{256}{9(-1 + x)^4 (1 + x)^5}\left[  
        10 - 8 x - 165 x^2 - 8 x^3 - 182 x^4 + 220 x^5  \right.\nn \\
&&+ \left. 
692 x^6 + 44 x^7 - 276 x^8 - 244 x^9 - 79 x^{10} - 4 x^{11} + (- 48 x^2  - 
48 x^3    \right.\nn \\
&&- \left. 
 660 x^4 + 144 x^5   + 48 x^6   + 576 x^7   + 
 540 x^8   + 192 x^9  + 24 x^{10}) \log x\right]\\
&&\nn\\
\dsh[g\,g\to \q^{[8]}\, X] &=& \szh[gg\to\q^{[8]}]\left(\delta(1-x) 
+ \frac{\as}{\pi}
\left\{  \Atot [\q^{[8]} ] \, \delta(1-x) \right.\right.\nn\\
&&+ \left[ x {\overline P}_{gg}(x)\log\frac{4m^2}{x\qf} + 2 x(1-x)
P_{gg}(x)\left(\frac{\log(1-x)}{1-x}\right)_{\rho} \right. \nn\\
&&+ \left. \left. \left. \left(\frac{1}{1-x}\right)_{\rho} f_{gg}[\q^{[8]}](x)
   \right]\right\} \right)\, , \nn\\[10pt]
&&[\q^{[8]}=\etah,\chizh,\chith],
\ea
where:
\ba
\Atot [\etah]  &=&\cf\,\left( -5 + {{{{\pi }^2}}\over 4} \right)  + 
  \ca\,\left( 3 - {{{{\pi }^2}}\over {24}} \right)  \nn\\ 
 &&+ 2\, {b_0}\,\log {{\mur }\over {\muf}} +
     8\ca\log\beta \log\frac{2m}{\muf}  - 2\ca\log \beta  + 
     8\ca\log^2 \beta  \\    
\Atot [\chizh] &=&
\cf\,\left( -{7\over 3} + {{{{\pi }^2}}\over 4} \right)  +
  \ca\,\left( {{71}\over {54}} - {{{{\pi }^2}}\over {24}} + 
     {35\over {27}}\log 2 
      \right)  \nn\\
 &&+ 2\, {b_0}\,\log {{\mur }\over {\muf}} +
     8\ca\log\beta \log\frac{2m}{\muf}- 2\ca\log \beta  + 8\ca\log^2 \beta \\  
\Atot [\chith] &=&-4\,\cf + \ca\,\left( {{59}\over {36}} + 
       \frac{\pi^2}{12} + 
      \frac{7}{9} \log 2 
      \right)  \nn\\
 &&+ 2\, {b_0}\,\log {{\mur }\over {\muf}} +
     8\ca\log\beta \log\frac{2m}{\muf}- 2\ca\log \beta  + 8\ca\log^2 \beta \\  
&&\nn\\
f_{gg}[\etah](x)&=& \frac{\ca}{6 (-1 + x) (1 + x)^3}
   \left[12 + 23 x^2 + 30 x^3 - 21 x^4 - 24 x^5 + 9 x^6 - 6 x^7  
\right.\nn\\ 
&&- \left. 
23 x^8 + (-12 + 60 x^2  + 24 x^3  + 36 x^4  + 
     60 x^6  + 24 x^7 )\log x \right]\\
&&\nn\\
f_{gg}[\chizh](x)&=&\frac{\ca}{54 (-1 + x)^3 (1 + x)^5}
    \left[ 172 - 56 x - 785 x^2 + 254 x^3 + 1881 x^4 - 948 x^5 
\right.\nn\\ 
&&- \left. 
334 x^6 + 632 x^7 - 790 x^8 + 268 x^9 + 63 x^{10} -150 x^{11} - 207 x^{12}+
(- 108  
\right.\nn\\ 
&&+\left.
372 x^2   - 168 x^3 - 540 x^4   - 180 x^5  + 3348 x^6   - 804 x^7 + 
312 x^8   - 252 x^9    
\right.\nn\\ 
&&+\left.
840 x^{10}   + 252 x^{11}) \log x\right]\\
&&\nn\\
f_{gg}[\chith](x)&=&\frac{\ca}{36 (-1 + x)^3 (1 + x)^5}
    \left[ 106 - 32 x - 215 x^2 + 260 x^3 + 1224 x^4 - 1284 x^5 
\right.\nn\\ 
&&-\left.
400 x^6 - 76 x^7 - 424 x^8 + 1252 x^9 - 153 x^{10} - 
120 x^{11} - 138 x^{12} + (- 72      
\right.\nn\\ 
&&+\left.
264 x^2   -96 x^3 + 1080 x^4   +
36 x^5   + 1404 x^6   - 3180 x^7   + 1320 x^8   - 612 x^9    
\right.\nn\\ 
&&+\left.
516 x^{10}   + 108 x^{11}) \log x\right]\, .
\ea

\noindent
{\bf The $gq(\bar q) \to \q^{[8]} X$  channels}
                                      
\ba
\dsh[g\,q\to \psih\,q]&=& \frac{\as}{\pi}\szh[\qq\to\psih]  \nn\\
&&\times\left\{\left[\frac{x}{2}
P_{qg}(x)\log\frac{4m^2(1-x)^2}{x\qf} + \tf
x^2(1-x)\right]+ f_{gq}[\psih](x)\right\}\\ 
&&\nn\\
&&\nn\\
\dsh[g\,q\to \chioh\,q] &=& \frac{\ascube\pi^2}{{(2 m)}^7}
f_{gq}[\chioh](x)\langle\o_8^H(\threePone)\rangle\\
&&\nn\\
&&\nn\\
\dsh[g\,q\to \q^{[8]}\,q] &=& \frac{\as}{\pi}\szh[gg\to\q^{[8]}] 
\left\{ 
\left[
\frac{x}{2} P_{gq}(x)\log\frac{4m^2(1-x)^2}{x\qf} + \cf\frac{x^2}{2}
\right]             
\right.\nn\\
&&+ \left. f_{gq}[\q^{[8]} ](x)\right\},\nn\\[10pt]
&&[\q^{[8]}=\etah,\chizh,\chith]
\ea
where
\ba
f_{gq}[\psih] &=&   - \frac{\tf}{4}(1 + 3 x) (-1 + x) x \nn\\  
    &&-\frac{\ca\df}{\da}\left[\frac{1}{2}(-1+x)(2+ x+ 2x^2) - x(1+x)\log x\right] \label{fgqpsih} \\
f_{gq}[\chioh]&=& \frac{\Bf}{D_F D_A}\frac{128}{3} \left[\frac{1}{3}(-1+x)(-5+4 x+4 x^2)-x^2\log x\right]\\
f_{gq}[\etah] &=&\cf\,(-1 + x)(1 - \log x)\\
f_{gq}[\chizh] &=&\frac{\cf}{9}\left[\frac{1}{3}(-1+x)(43-14 x+4 x^2) + (9 - 9 x + 4 x^2)\log x\right]\\
f_{gq}[\chith] &=&\frac{\cf}{12}\left[\frac{1}{3}(-1+x)(53-16 x+20 x^2)+ (12 -12 x + 5 x^2)\log x\right]
\ea

\noindent
{\bf The $\qq\to \q^{[8]} X$ channels}
                             
\ba
\dsh[\qq\to\etah \, g] &=& \frac{\ascube \pi^2}{(2m)^5}
f_{\tiny\qq}[\etah](x)\langle\o_8^H(\oneSzero)\rangle\\
&&\nn\\
\dsh[\qq\to\psih \, X] &=& \szh[\qq\to\psih]
\left(\delta(1-x)+\frac{\as}{\pi}
\left\{ \left[{{-10\,\nf \,\tf }\over 9}   \right. \right. \right.\nn\\
&&+  \ca \,\left( {{59}\over 9} - {{{{\pi }^2}}\over 4} + 
     \frac{2}{3}\,\log 2 - 2\,\log \beta  \right)  \nn\\
&&+  \cf \,\left( -8 + {{{{\pi }^2}}\over 3} + 8\,\log^2 \beta  
+ (3 + 8\log\beta)\log\frac{2m}{\muf}\right)  \nn\\
    &&+\left. 2\,{b_0}\,\log {{\mur }\over {2\,m}}\,\right] \delta(1-x)\nn\\
&&+\left[x {\overline P}_{qq}(x)\log\frac{4m^2}{x\qf} +\cf
   x(1-x) + 2 x(1-x) P_{qq}(x)\left(\frac{\log(1-x)}{1-x}\right)_{\rho}\right]
   \nn\\                      
&&+\left.\left.\left(\frac{1}{1-x}\right)_{\rho}
   f_{\tiny\qq}[\psih](x)\right\}\right)
\ea
\ba
\dsh[\qq\to\chijo\,g] 
%+ \sigma^H[\qq \to\psih] 
  &=&\frac{\ascube\pi^2}{{(2m)}^7} \left[\frac{\Bf}{D_F^2}
  \frac{512}{9}\left(2\log\beta - {5\over 6} - \log{{\mul}\over{2m}}\right)
   \delta(1-x)\right.\nn\\
&&+ \left.\left(\frac{1}{1-x}\right)_{\rho} f_{\tiny\qq}[\chijo](x)\right]
 \langle\o^H_8(\threePJ)\rangle \nn\\[10pt]
%&&+ \sigma^H[\qq \to\psih]\, , \nn\\[10pt]
&&   [J=0,1,2]
\ea
with
\ba
&&f_{\tiny\qq}[\etah]=\frac{\Bf}{D_F^2}\frac{32}{3}x^2(1-x)\\
&&f_{\tiny\qq}[\psih]= - \cf x(1-x)^2 - \frac{\ca}{3}x(1+x+x^2)\\
&&f_{\tiny\qq}[\chizh]=\frac{\Bf}{D_F^2} \frac{128}{9}x^2(1-3 x)^2\\
&&f_{\tiny\qq}[\chioh]=\frac{\Bf}{D_F^2} \frac{256}{9}x^2(1+x)\\
&&f_{\tiny\qq}[\chith]=\frac{\Bf}{D_F^2} \frac{256}{45}x^2(1+3 x + 6 x^2)
\label{fqqpsihb}
\ea
In a way identical to the colour singlet case,
$\dsh[\qq\to \chijo\,g]$ is only infrared finite when at least the Born term of
the $\sigma^H[\qq\to\psih]$ cross section is added to it. We have performed the
cancellation and an explicit dependence on the NRQCD renormalization $\mul$
results. The NRQCD matrix
element in $\sigma^H[\qq\to\psih]$ is then understood to be the renormalized,
$\mul$ scale dependent one $\langle\o^H_8(\threeSone)\rangle^{(\mul)}$.

\section{Virtual contributions with the threshold expansion}
\label{appc}
In this Appendix , we show how  the threshold expansion
method~\cite{Braaten96,Braaten97}
is  used to calculate the imaginary parts  in the short-distance          
coefficients for the virtual
contributions at NLO in $\alpha_s$.
Being the $S$-wave calculations analogous but much simpler, 
we will sketch here only the derivation for $P$-waves. 
We start considering $\QQ$ scattering in an arbitrary frame in which
the pair has total momentum $ {\bf P} $ and the relative momentum of
the two quarks is small compared to the heavy quark mass $m$.
The momenta $p$ and $\bar p$ of the $Q$ and $\overline Q$ can be written as
\begin{eqnarray}
p \;=\; \mbox{$1 \over 2$} P \;+\; L {\bf q} \;,
\\
\bar p \;=\; \mbox{$1 \over 2$} P \;-\; L {\bf q} \;,
\end{eqnarray}

The components of the momenta $P$ and $L {\bf q}$ in the CM frame 
of the pair are
\begin{eqnarray}
P^\mu \bigg|_{\rm CM} &\;=\;& ( 2 E_q , \; {\bf 0} ) \;,
\\
(L {\bf q})^\mu \bigg|_{\rm CM} &\;=\;& ( 0, \; {\bf q} ) \;.
\end{eqnarray}
When boosted to an arbitrary frame in which
the pair has total  spacial momentum ${\bf P}$, these momenta are
\begin{eqnarray}
P^\mu &\;=\;&
\left( \sqrt{ 4 E_q^2 + {\bf P}^2} , \; {\bf P} \right) \;,
\\
(L {\bf q})^\mu &\;=\;& L^\mu_j \; q^j .
\end{eqnarray}
The boost matrix $L^\mu_{\ j}$, which has one Lorentz index and one Cartesian
index, has components
\label{L-boost}
\begin{eqnarray}
L^0_{\ j} &\;=\;& {1 \over 2 E_q} P^j \; , 
\\
L^i_{\ j} &\;=\;& \delta^{ij} - {P^i P^j \over {\bf P}^2}
        \;+\; {P^0 \over 2 E_q} {P^i P^j \over {\bf P}^2}  \;.
\end{eqnarray}
These expressions are of course valid only for boosts with a vanishing time
component.
We first  calculate the Born amplitude  in $D$-dimensions  for the process
$ \QQ \to g g$, i.e.  diagrams $D_1, D_2$ of fig.~\ref{fig:ggborn} 
It reads ( all the momenta are outgoing ) :
\ba
{\cal A}^{\mu \nu, ab }_{ \QQ \to g g}
&\;= \;&
 {g_s^2 \mu^{2 \epsilon} } \;   
\bar v(- \bar p) 
\Bigg[ \; {\gamma^\nu (\not \! k_2 \; +  \not \! \bar p \;+ m) \gamma^\mu
        \over 2 \bar{p} \cdot k_2} \, T^b T^a  
\nn \\
&&+ \,  
{\gamma^\mu (- \not \! p \; -  \not \! k_2 \;  + m) \gamma^\nu 
        \over 2 p \cdot k_2} \, T^a T^b \; \Bigg] u(-p) \;.
\label{amp:ccgg} 
\ea

We then proceed by two main steps.
We consider the following expressions of the spinors of the $\QQ$ pair 
in the CMS , 
\label{spinor}
\begin{eqnarray}
u({\bf q}) &=& \sqrt{E+m \over 2E}
\left( \begin{array}{c} \xi \\
        {{\bf q} \cdot \mbox{\boldmath $\sigma$} \over E+m} \xi \end{array} \right) ,
\label{uspinor}
\\
v(-{\bf q}) &=& \sqrt{E+m\over 2E}
\left( \begin{array}{c} {(-{\bf q}) \cdot \mbox{\boldmath $\sigma$} 
\over E+m}
\eta
                \\ \eta  \end{array} \right) ,
\label{vspinor}
\end{eqnarray}
where $\xi$ and $\eta$ are 2-component spinors with suppressed colour
indices. Once boosted to a frame where the pair has total
3-momentum ${\bf P}$, it is straightforward to obtain
the relevant independent quantities up to the second order in {\bf v} ,
that can be formed by sandwiching up to three Dirac matrices between $\bar v(- \bar p)$ and $u(-p)$  (${\bf v} = {\bf q}/E$ and ${\bf v}' = {\bf q}'/E$).
They read
\ba
\bar v(- \bar p) u(-p) &\; = \; & 
\; \eta^\dagger ({\bf v} \cdot \mbox{\boldmath $\sigma$}) \xi \, ,\nn
\\
\bar v(- \bar p) \gamma^\mu  u(-p) &\;=\;& L^\mu_{\ j} \;
\left( \; \eta^\dagger \sigma^j \xi \;-\; {1 \over 2} \; v^j \;
        \eta^\dagger ( {\bf v} \cdot \mbox{\boldmath $\sigma$} ) \xi \right) 
\;, \nn\\
\bar v(- \bar p) ( \gamma^\mu \gamma^\nu
           -\gamma^\nu \gamma^\mu ) u(-p) &\;=\;& {1 \over m}
( P^\mu L^\nu_{\ j} - P^\nu L^\mu_{\ j} )  \nn\\
&&\times \left( (1-v^2) \; \eta^\dagger \sigma^j \xi 
        \;+\; {1 \over 2} v^j \; 
        \eta^\dagger ({\bf v} \cdot \mbox{\boldmath $\sigma$}) \xi \right)\,,
 \nn\\
\bar v(- \bar p) ( \gamma^\mu \gamma^\nu \gamma^\lambda
           -\gamma^\lambda \gamma^\nu \gamma^\mu ) u(-p) &\;=\;&
{1 \over m} 
\left( P^\mu L^\nu_{\ i} L^\lambda_{\ j}
        + L^\mu_{\ i} L^\nu_{\ j} P^\lambda 
        + L^\mu_{\ j} P^\nu L^\lambda_{\ i} \right) \nn\\
&&\times \left( \eta^\dagger v^i \sigma^j \xi 
        \;-\; \eta^\dagger v^j \sigma^i \xi \right) .
\label{bispinors}
\ea
We note that  these formulas are basically  the ones given in Appendix A 
of Ref. \cite{Braaten97} except that  only the  terms  that are 
specific to the case of $P$-waves, are kept. 
The normalization for the spinors
is however the non-relativistic one adopted in Ref. \cite{bbl}. 
Once the amplitude in (\ref{amp:ccgg}) is simplified and rewritten using
the above amplitudes,  one is ready to perform the second step, i.e. to 
calculate:
\begin{equation}
{\rm  Im}\; {\cal M} = 
{1 \over 2}\int_{k_1} \int_{k_2} (2 \pi)^{D} \delta^{D}(P - k_1 - k_2)
\sum ( {\cal A}_{\QQ ' \to g g } )^*
       {\cal A}_{\QQ  \to g g } \;,
\label{match-gg}
\end{equation}
where $k_1$ and $k_2$ are the momenta of the gluons,
$\int_k  \;\equiv\; \int {d ^N k \over (2 \pi)^N 2 k_0} \;$
denotes the  $N=(D-1)$-dimensional integral over the phase space associated with the momentum $k$, the sum is over their
spins and colours, and the factor of ${1 \over 2}$ comes from Bose statistics.
 
The  colour structure is easily simplified by 
\begin{eqnarray}
T^a T^b \otimes T^b T^a &=& {(N_c^2-1)\over {4 N_c^2 }} \; 1 \otimes 1
        \;+\; {N_c^2-2 \over 2 N_c} \; T^a \otimes T^a ,
\label{Tabba}
\\
T^a T^b \otimes T^a T^b &=&  {(N_c^2-1)\over {4 N_c^2 }}\; 1 \otimes 1
        \;-\; {1\over N_c} \; T^a \otimes T^a ,
\label{Tabab}
\end{eqnarray}
while  the phase space integration is performed
using the formulas 
\begin{eqnarray}   
\int d\Phi_{(2)} k^i k^j & = & { \delta^{ij} \over (D-1)} \Phi (2) ; \\
\int d\Phi_{(2)} k^i k^j k^l  k^m  & = & { (\delta^{ij} \delta^{lm}+
\delta^{il}\delta^{jm}+ \delta^{im} \delta^{jl})\over {(D-1)(D+1)}}\Phi (2) ; 
\end{eqnarray} 
where
\begin{equation}
\Phi_{(2)}=\frac{1}{8\pi}\left(\frac{4\pi}{4m^2}\right)^{{4-D} \over 2 }\frac{\Gamma({{D-2}\over 2})}{\Gamma(D-2)}\, ,
\end{equation}
is the the phase space of two massless particles in $D$ dimensions.
In so doing, we eventually obtain the following expression:
\begin{eqnarray}
&&
{\rm  Im} {\cal M}= { (N_c^2-1)  
        \over  N_c^2  m^2 (D-1) (D+1) } \;
 \pi^2 \alpha_s^2  \Phi_{(2)}   \;
\nonumber \\
&&\times  \Bigg[\; \; ( 5 D + 2) \,  \,
    \xi'^\dagger {\bf v}'\cdot \mbox{\boldmath $\sigma$} \eta' 
    \eta^\dagger {\bf v} \cdot \mbox{\boldmath $\sigma$} \xi
\; + \; (2 D^2 - 3 D - 8  ) \,{\bf v}'\cdot {\bf v} 
    \xi'^\dagger  \mbox{\boldmath $\sigma$}  \eta' \eta^\dagger 
      \mbox{\boldmath $\sigma$} \xi   
\nonumber \\
&&+ (2 D^2 - 3 D - 8  ) 
    \xi'^\dagger {\bf v} \cdot \mbox {\boldmath $\sigma$}  \eta' \eta^\dagger 
      {\bf v}'\cdot\mbox{\boldmath $\sigma$} \xi 
\;\Bigg] \;.
\label{nramp:ccgg}
\end{eqnarray}

Next we isolate the individual $P$-wave contributions. 
To this aim we follow the same procedure of section II and 
Appendix A of Ref. \cite{bbl}, but generalizing it to arbitrary dimensions.
This can be accomplished by first noting that any direct product of cartesian 
vectors in  $ N $ spatial dimensions may be written as
\ba
a^ib^j & = & {{{\bf a} \cdot {\bf b}}\over D-1}\delta^{ij}
+ \left[ (a^ib^j+a^jb^i)/2-{{{\bf a} \cdot {\bf b}}\over D-1}\delta^{ij} 
\right] \; + (a^ib^j-a^jb^i)/2 \\                           
(D-1)^2 & = & \;\; 1 \qquad \oplus \; \qquad \frac{(D-2) (D+1)}{2}  \; \; 
\qquad \oplus \; \;
\frac{(D-1)(D-2)}{2}
\ea
Applying the above decomposition to the case at hand, we obtain from the NRQCD
lagrangian
the relevant coefficients for the ${}^3 P_J$ states in $D-1$ spatial dimensions
\ba                                                    
{\cal M}({}^3 P_J) &=& {1 \over m^2} \; \Bigg[
\; {f_1({}^3P_{1}) + f_1({}^3P_{2}) \over 2} \;
\frac{ {\bf v}' \cdot {\bf v} \; \sigma^i \otimes \sigma^i }{2 \nc}\;\nn\\
&&+ {f_1({}^3P_{0}) - f_1({}^3P_{2}) \over D-1} \;
\frac{ {\bf v}' \cdot \mbox{\boldmath $\sigma$} \otimes 
{\bf v} \cdot \mbox{\boldmath$\sigma$}}{2\nc}
\nn\\
&&+\; {f_1({}^3P_{2}) - f_1({}^3P_{1}) \over 2} \;
\frac{ {\bf v} \cdot \mbox{\boldmath $\sigma$} \otimes {\bf v}' 
\cdot \mbox{\boldmath$\sigma$}}{2\nc} \Bigg]
\label{M8}
\ea
Comparing this relation to  (\ref{nramp:ccgg}) we can read off 
the imaginary parts  in $ D=4-2\epsilon $ dimensions :
\ba
\label{born}
{\rm Im} {f_1({}^3P_{0} )_0}& = & 8 \pi^2 \cf 
\frac{ 9 ( 1 - \epsilon  )}{ 3 -2 \epsilon   } \Phi_{(2)}\; \alpha_s^2 
\mu^{4\epsilon} \,,   \\
{\rm Im} {f_1({}^3P_{1} )_0}& = & 0 \,,\\
{\rm Im} {f_1({}^3P_{2} )_0}& = & 16 \pi^2 \cf  
\frac{6 - 13 \epsilon + 4 \epsilon^2 }{(3 - 2 \epsilon) ( 5- 2\epsilon)} 
\Phi_{(2)}\; \alpha_s^2 \mu^{4\epsilon}  \,.
\ea
Needless to say, their octet counterparts differ only
by the colour coefficient.
Once the leading order expressions are available, we are ready
to perform the next order corrections to (\ref{amp:ccgg}).
We can write
\be
{\rm Im} {f_1({}^3P_{J} )} =  {\rm Im} {f_1({}^3P_{J} )_0} \Bigg[ 1 + {\alpha_s \over
\pi} V_J \Bigg]\, , 
\label{next}
\ee
where the $ V_J $ are the sum of all the virtual contributions
shown in fig.~\ref{fig:gg} 

Let us consider, for example,  the contribution of graph (e) in 
fig.~\ref{fig:gg}. 
In this case the virtual amplitude reads:
\ba
{\cal A}^{\rm virt}[e]
&\; =  \;&
   g_s^4  \mu^{4 \epsilon}  \;   
\int {d^D Q  \over (2 \pi)^D} 
  \bar v(- \bar p)  
 \; \gamma^\nu {\not \! k_2 \; +  \not \! \bar p \;+ m 
        \over 2 \bar {p} \cdot k_2} \gamma^\rho  {\not \! Q \;+ m 
        \over Q^2-m^2_c} \gamma^\sigma u(-p)  \nn\\ 
&&\times  T^b T^d T^c  \;
\frac {F_{\mu \rho \sigma}^{acd}[k_1,-p-Q-k_1,p+Q]}{(Q+p)^2 (Q+p+k_1)^2}  \;,
\label{vtx} 
\ea
where $Q$ is the loop-momentum and $F_{\mu \rho \sigma}^{acd}$ is the
usual three gluon vertex.        
By making use of the Dirac equation and contracting all
Lorentz indices, it is tedious but straightforward to rewrite (\ref{vtx}) 
in terms of the basic amplitudes in (\ref{bispinors}).
Due to the great number of terms which appear in these expressions,
all the algebraic manipulations have been performed using MATHEMATICA
and the package FEYNCALC ~\cite{feyncalc}.                
Only at this stage, loop tensor integrals of the type
\be
{\cal I}_{1;\alpha;\alpha\beta}= 
\int {d^D Q \over (2 \pi)^{D}} \frac {1; Q^\alpha ; 
Q^\alpha Q^\beta}{ (Q^2-m^2_c) (Q+p)^2 (Q+p+k_1)^2} \;, 
\label{tensor}
\ee
are decomposed with the usual Passarino-Veltman technique, 
which is performed by computer. At
the end of this procedure one is left with a tensor structure based
on the external momenta and a small number of scalar integrals of bosonic type.
We evaluated these integrals
using the results from ref.~\cite{NDE}.
Once all loop integrals are calculated, the remaining
internal Lorentz indices are contracted  and the phase space integration
\be
\;{\rm  Im}\; {\cal M} = 
{1 \over 2}\int_{k_1} \int_{k_2} (2 \pi)^{D} \delta^{D}(P - k_1 - k_2)
\sum ( {\cal A}_{\QQ ' \to g g }^{\rm Born })^*
       {\cal A}_{\QQ  \to g g }^{\rm virt} \;,
\label{match-gg2}
\ee
performed.
The colour structure globally factorizes 
both in the singlet and the octet contributions.
The final results coincide diagram-by-diagram 
with those obtained using the $D$-dimensional
covariant-projector technique, and are shown in Tables 
\ref{tab:virtual3P0} and \ref{tab:virtual3P2}.

\section{Fragmentation functions}
\label{appFF}
We consider here another application of the results
of Section~\ref{sec:opren}. We shall deal with
the cancellation of the infrared singularity in the gluon to $\chi_J$ states
fragmentation functions, which have already been considered in
literature in various occasions \cite{frag,ma}, 
and in one case fully worked out within the matching method with dimensional
regularization \cite{Braaten97}. This is a fourth independent 
calculation: our result coincides with ref.~\cite{Braaten97}.

The general expression for this fragmentation function needs for consistency
both a singlet and an octet part.
Calculations made with the projection method and dimensional regularization
according to the rules given in Section~\ref{sec:projectors} give
the following result in $D=4-2\ep$ dimensions:
\ba
D_{g\to\chi_J}(z) &=&  {{2\assq\mu^{4\ep}}\over{27 m^5}}\left[
    \left(-{1\over\ei} - \ln4\pi +\gamma_E 
 + \hat a_J\right)\delta(1-z) \right. \nn\\
&&+\left.{{2z}\over{(1-z)_+}} + {2\over N_J}P_J(z)\right]
\langle\o_{1}^{\chi_J}(^3P_J)\rangle \\
&&+ {{\pi\as\mu^{2\ep}}\over{8(3-2\ep)m^3}}\delta(1-z)
\langle\o_8^{\chi_J}(^3S_1)\rangle
\ea
The $P_J(z)$ are finite functions which are of no concern to us here, and can 
be found in the literature, for instance in \cite{Braaten97}. $N_J$ is the 
number 
of polarization states of $\threePJ$, in four dimensions. The constants 
$\hat a_J$ take the values
\be
\hat a_0 = -{1\over 6},\qquad\qquad \hat a_1 = -{5\over 12},\qquad\qquad 
\hat a_2 = -{19\over 60}.
\ee
The short distance coefficient of the colour singlet matrix element 
can be seen to be infrared singular.
This singularity is cancelled by substituting, according to eq.~(\ref{3s1full}), 
the bare $D$-dimensional
$\langle\o_8^{\chi_J}(^3S_1)\rangle$ matrix element which appears above. 
The final result then reads
\be
D_{g\to\chi_J}(z) = d_1^J(z,\mul) \langle\o_{1}^{\chi_J}(^3P_J)\rangle +
d_8(z) \langle\o_8^{\chi_J}(^3S_1)\rangle^{(\mul)} \, , 
\label{chifragfin}
\ee
with
\ba
d_1^J(z,\mul) &=& {{4\assq}\over{27 m^5}}\left[
    \left(
a_J - \ln{\mul\over{2m}}\right)\delta(1-z) + {{z}\over{(1-z)_+}} + {1\over
N_J}P_J(z)\right] \, , \\[10pt]
d_8(z) &=& {{\pi\as}\over{24m^3}}\delta(1-z) \, , 
\ea
and
\be
 a_0 = {1\over 4},\qquad\qquad  a_1 = {1\over 8},\qquad\qquad 
 a_2 = {7\over 40}.
\ee

Finally, differentiating (\ref{chifragfin}) with respect to $\ln\mul$ returns 
the evolution equation for the  
$\langle\o_8^{\chi_J}(^3S_1)\rangle^{(\mul)}$ matrix element:
\be
\mul{d\over{d\mul}} \langle\o_8^{\chi_J}(^3S_1)\rangle^{(\mul)} = 
{{32 \as(\mul)}\over{9 \pi m^2}}\langle\o_{1}^{\chi_J}(^3P_J)\rangle \, .
\ee
This corresponds to the standard result~\cite{bbl}, once the difference in the
normalization of the NRQCD operators, eq.~(\ref{eq:opnorm}), 
is taken into account.

\end{document}